\documentclass[12pt]{article}
\pdfoutput=1

\usepackage{color}
\usepackage{amssymb,amsmath,bm}
\usepackage{epsf}
\usepackage{epsfig}
\usepackage{afterpage}
\usepackage{longtable}
\usepackage[dvipsnames]{xcolor}
\usepackage[linktoc=page,bookmarks=false,colorlinks=false,linkbordercolor=RoyalBlue,citebordercolor=ForestGreen,urlbordercolor=CornflowerBlue]{hyperref}
\usepackage{latexsym,mathrsfs}
\usepackage[normalem]{ulem} 
\usepackage[compress]{cite}
\usepackage{graphicx}
\usepackage{url}
\usepackage{paralist}
\usepackage{slashed}
\usepackage{multirow}

\usepackage[hypcap]{caption, subcaption}

\usepackage{booktabs}
\usepackage{listings}

\usepackage{textcomp}

\usepackage[titles]{tocloft}
\renewcommand{\arraystretch}{1.25}

\allowdisplaybreaks[1]

\numberwithin{equation}{section}

\newcounter{MBQ}

\def \refeq#1{(\ref{#1})}
\def \refEq#1{Eq.~(\ref{#1})}
\def \refsec#1{Section~\ref{#1}}

\def \refapp#1{Appendix~\ref{#1}}
\def \reffig#1{Figure~\ref{#1}}
\def \reftab#1{Table~\ref{#1}}

\DeclareMathOperator{\Li2}{Li_2}
\DeclareMathOperator{\re}{Re}
\DeclareMathOperator{\im}{Im}

\newcommand{\MeV}{\,\text{MeV}}

\newcommand{\GeV}{\,\text{GeV}}

\newcommand{\eps}{\epsilon}
\newcommand{\veps}{\varepsilon}

\newcommand{\para}{\parallel}

\def \gf {\gamma_5}

\newcommand \oL[1]{{\overline{#1}}}

\newcommand \bra [1] {\big\langle {#1}\big|}
\newcommand \ket [1] {\big|{#1}\big\rangle}

\newcommand{\Op}{P}

\newcommand{\calA}{\mathcal{A}}

\newcommand{\calL}{\mathcal{L}}

\newcommand{\DelB}{{\Delta B}}

\newcommand{\normEW}{\mathcal{N}_\text{ew}}
\newcommand{\alE}{\alpha_\text{em}}
\newcommand{\alS}{\alpha_s}

\newcommand{\LambQCD}{\Lambda_\text{QCD}}

\newcommand{\lamCKM}[2]{\lambda_{#1}^{(#2)}}

\newcommand{\vb}{v}
\newcommand{\nn}{{n_-}} 
\newcommand{\nb}{{n_+}} 
\newcommand{\nnn}{n_-} 
\newcommand{\nbb}{n_+} 

\newcommand{\nns}{{\slashed n}_-} 
\newcommand{\nbs}{{\slashed n}_+} 

\newcommand{\msbar}{${\overline{\text{MS}}}$}
\newcommand{\SCETI}{SCET$_\text{I}$}
\newcommand{\SCETII}{SCET$_\text{II}$}

\newcommand{\lamBq}{\lambda_{B_q}}
\newcommand{\sigNBq}[1]{\sigma^{(#1)}_{B_q}}
\newcommand{\hsigNBq}[1]{\widehat{\sigma}^{(#1)}_{B_q}}

\newcommand{\lamB}[1]{\lambda_{B_{#1}}}
\newcommand{\sigNB}[2]{\sigma^{(#2)}_{B_{#1}}}
\newcommand{\hsigNB}[2]{\widehat{\sigma}^{(#2)}_{B_{#1}}}

\newcommand{\Bqtoll}{B_q \to \ell\bar\ell}

\newcommand{\Bqtollgamma}{B_q \to \gamma\ell\bar\ell}
\newcommand{\barBqtollgamma}{\oL{B}_q \to \gamma\ell\bar\ell}

\newcommand{\Bstollgamma}{B_s \to \gamma\ell\bar\ell}
\newcommand{\Bdstollgamma}{B_{d,s} \to \gamma\ell\bar\ell}

\newcommand{\Bqtogammagamma}{B_q \to \gamma\gamma}

\newcommand{\Butolvgamma}{B_u \to \gamma\ell\bar\nu_\ell}

\newcommand{\BR}{\mathcal{B}}
\newcommand{\AFB}{\mathcal{A}_\text{FB}}
\newcommand{\ACP}{\mathcal{A}_\text{CP}}

\begin{document}

\allowdisplaybreaks

\begin{titlepage}

\begin{flushright}
{\small
TUM-HEP-1280/20 \\
}
\end{flushright}

\vspace{1cm}
\begin{center}
{\Large\bf\boldmath
$\Bdstollgamma$ decay with an
energetic photon
}
\\[8mm]
{\sc
  Martin~Beneke$^{a}$,
  Christoph~Bobeth$^{a}$ and
  Yu-Ming Wang$^{b}$
}
\\[0.6cm]

$^{a}${\it Physik Department T31,\\
James-Franck-Stra\ss e~1,
Technische Universit\"at M\"unchen,\\
D--85748 Garching, Germany\\[0.2cm]}

$^{b}${\it 
School of Physics, Nankai University,\\ 
Weijin Road 94, 
300071 Tianjin, China\\[0.2cm]}
\end{center}

\vspace{0.6cm}
\begin{abstract}
\vskip0.2cm\noindent
We calculate the differential branching fraction, 
lepton forward-backward asymmetry and direct CP asymmetry for 
$\Bdstollgamma$ decays with an energetic photon. 
We employ factorization methods, which result in rigorous 
next-to-leading order predictions in the strong coupling 
at leading power in the large-energy/heavy-quark expansion, 
together with estimates of power corrections and a resonance 
parameterization of sub-leading power 
form factors in the region of small lepton invariant 
mass $q^2$. The $\Bdstollgamma$ decay shares features of the 
charged-current decay $\Butolvgamma$, and the FCNC decays
$B\to K^{(*)}\ell\bar\ell$. As in the former, the leading-power 
decay rates can be expressed in terms of the $B$-meson 
light-cone distribution amplitude and short-distance 
factors. However, similar to $B\to K^{(*)}\ell\bar\ell$, 
four-quark and dipole operators contribute to the 
$\Bdstollgamma$ decay in an essential way, limiting the 
calculation to $q^2\lesssim 6$~GeV$^2$ below the charmonium 
resonances in the lepton invariant mass spectrum. A detailed 
analysis of the main observables and theoretical uncertainties 
is presented.
\end{abstract}
\end{titlepage}

\newpage

{

}

%
%
%

\section{Introduction}

The semileptonic flavour-changing neutral-current (FCNC) decays
driven by the quark-level transitions $b\to q\,\ell\bar\ell$ with
$q = d,s$ and $\ell = e,\mu,\tau$ are important tests of the
dynamics of the quark and lepton sectors in the framework of the
standard model (SM) and constitute very sensitive probes of
nonstandard effects.

The exclusive $B\to K^{(*)}\ell\bar\ell$ decays are experimentally
most easily accessible due to their comparatively sizeable branching
fractions of $\mathcal{O}(10^{-6})$. The angular distributions of
their three- and four-body final states are among the
prime targets of LHCb and Belle~II for $\ell = e, \mu$, because they
offer many CP-symmetric and CP-asymmetric observables.
However, the hadronic nonperturbative effects entering the
theoretical treatment are complex ranging from form factors to
resonant contributions, preventing currently theoretical
calculations with percent accuracy. In contrast, the purely
leptonic decays $\Bqtoll$ allow theoretical control of hadronic
effects at the percent level \cite{Bobeth:2013uxa}, because they
depend---apart from higher-order QED corrections \cite{Beneke:2017vpq,
Beneke:2019slt}--- only on the $B$-meson
decay constant, which can be computed in QCD lattice calculations
with sub-percent accuracy \cite{Bazavov:2017lyh}. On the other hand
the helicity suppression of the two-body final state leads to tiny
branching fractions of $\mathcal{O}(10^{-9})$ for $\ell = \mu$,
requiring huge data samples at LHC in order to perform measurements
with precision comparable to the prediction. Nevertheless this
decay has been observed by LHCb \cite{Aaij:2017vad},
CMS \cite{Sirunyan:2019xdu} and ATLAS \cite{Aaboud:2018mst} with a
rate compatible with SM predictions.

Compared to the strong research activities on the aforementioned decay
modes, the radiative leptonic decays $\Bqtollgamma$ have received
relatively little attention. The additional photon in the final state
implies suppression by the electromagnetic coupling, but 
lifts the very different helicity suppression for $\ell = e$ and
$\ell = \mu$ present in $\Bqtoll$. The decay rates of $\Bqtollgamma$
become comparable for the electron and muon final states and can be
almost $\mathcal{O}(10^{-8})$ for 
$q=s$ \cite{Eilam:1996vg, Aliev:1996ud,
Guadagnoli:2017quo, Kozachuk:2017mdk}.
Further, the angular distribution offers complementary observables
to test the short-distance couplings. So far there are no experimental studies
of the $\Bqtollgamma$ decays, which for small photon energies also
constitute a background process
in the experimental analyses of $\Bqtoll$ in the dilepton-invariant mass
side-band \cite{Aditya:2012im, Aaij:2017vad, Dettori:2016zff}.

Besides testing the SM, the theoretical interest in $\Bqtollgamma$
also derives from the structure of hadronic effects. The limit of
large photon energy $E_\gamma \gg \LambQCD$, where $\LambQCD$ denotes
the strong interaction scale, allows for a systematic treatment of
nonperturbative effects when adopting the framework of QCD
factorization and
soft-collinear effective theory~(SCET). As pointed out in
\cite{DescotesGenon:2002ja}, in this limit the leading nonperturbative
corrections become universal for the decays $\Butolvgamma$, $\Bqtogammagamma$
and $\Bqtollgamma$, relating thus hadronic effects in $b\to u \ell\bar\nu_\ell$,
$b\to q\gamma$ and $b\to q\ell\bar\ell$, respectively, and offering a programme
for the combined analysis of long- and short-distance quantities in these
decays. Other approaches are available in the literature relying on the
$B_q \to \gamma$ form factor calculation with dispersive methods and quark
models \cite{Eilam:1996vg, Aliev:1996ud, Kozachuk:2017mdk, Dubnicka:2018gqg}.
The case of very soft photons due to initial-state radiation 
has been considered in the framework of heavy-hadron chiral 
perturbation theory in \cite{Aditya:2012im}.

The factorization programme has been carried out for $\Butolvgamma$ 
starting within QCD factorization (QCDF) 
\cite{Korchemsky:1999qb, DescotesGenon:2002mw} and has then been
extended to the two-step matching in SCET 
\cite{Lunghi:2002ju, Bosch:2003fc} at leading power (LP) in 
$\LambQCD/m_b$, including next-to-leading order (NLO) radiative
QCD corrections and resummation of the associated large logarithms.
Next-to-leading power (NLP) corrections at lowest order in QCD have been
considered in \cite{Beneke:2011nf}, studying the consequences for the form
factor symmetry relation. The NLP corrections have been scrutinized further with
sum rule calculations \cite{Braun:2012kp, Wang:2016qii, Wang:2018wfj, Beneke:2018wjp}.

The decay $\Bqtollgamma$ appears as a hybrid of the charged-current
decay $\Butolvgamma$, which is driven by tree-level 
electroweak $b\to u \ell\bar\nu_\ell$ transitions and the FCNC decays
$B\to K^{(*)}\ell\bar\ell$, which receive contributions from semileptonic $b\to s + (\gamma,\, \ell\bar\ell)$ as well as from hadronic
$b\to s + (g,\, q\bar q)$ transitions. When the final-state photon
carries large energy relative to the strong interaction scale,
the non-hadronic final state of the  $\Bqtollgamma$ decay enables
the calculation of the relevant form factors in terms of the
$B$-meson light-cone distribution amplitude (LCDA) at LP
in an expansion in the energy of the photon, similar to the
decay $\Butolvgamma$ \cite{Lunghi:2002ju, Bosch:2003fc}. 
On the other hand, the contribution of
$b\to s q\bar q$ four-quark operators to $\Bqtollgamma$ introduces
issues familiar from $B\to K^{(*)}\ell\bar\ell$ decay, such as
the presence of charmonium resonances in the $\ell\bar\ell$
invariant mass spectrum that limit the applicability of
factorization methods to this decay \cite{Beneke:2001at,Beneke:2004dp}.

Previous work on $\Bqtollgamma$ \cite{Eilam:1996vg, Aliev:1996ud, Wang:2013rfa,
Guadagnoli:2017quo, Kozachuk:2017mdk} has usually focused on providing
theoretical calculations of the (differential) decay rate in the entire
kinematically allowed regions in terms of $B\to\gamma^*$ form factors.
This approach inevitably requires some amount of theoretical modelling of the
form factors, and neglects ``non-factorizable'' $\mathcal{O}(\alS)$
corrections from the $b\to s q\bar q$ four-quark and chromomagnetic dipole
operators, which are known to be sizeable for  $B\to K^{(*)}\ell\bar\ell$
decays \cite{Beneke:2001at}.

In contrast, our focus is on the kinematically more restricted region of
photon energy $E_\gamma \gg \LambQCD$, while making maximal use of
factorization methods applicable in this limit. We perform a SCET analysis
of $\Bqtollgamma$ including $\mathcal{O}(\alS)$ QCD corrections at LP.
We further include local NLP corrections in $\mathcal{O}(\alS^0)$ explicitly,
whereas non-local NLP contributions are parametrized following
\cite{Beneke:2011nf, Wang:2016qii}. The analysis is more involved for the
FCNC $b\to q \ell\bar\ell$ transitions compared to the tree-level decay
$b\to u \ell \bar\nu_\ell$ due to the extended weak operator
basis~\refeq{eq:weak-EFT}. Parts of our analysis are related to earlier
works  on $\Butolvgamma$, $B_q\to\gamma\gamma$ and $B\to K^{(*)}\ell\bar\ell$
within QCDF \cite{Beneke:2001at, DescotesGenon:2002ja, Bosch:2002bv}
or SCET \cite{Lunghi:2002ju, Bosch:2003fc, Beneke:2011nf,Wang:2016qii}.
The factorization approach then allows us to express the decay amplitudes
in terms of only the $B$-meson light-cone distribution amplitude (LCDA)
at LP in the heavy-quark/large-energy expansion. The modelling
of form factors is necessary only at sub-leading power in this expansion. 
The result is expected to be valid for sufficiently broad bins in invariant
mass $q^2$ of the lepton pair below the charmonium resonances, $q^2 \lesssim
6 \GeV^2$, and above the narrow light-meson resonances. The latter restriction
arises, since estimates following \cite{Beneke:2009az} imply that global 
duality is violated not only by charmonium, but also by the light-meson
resonances. In order to extend our calculations to observables local in
$q^2$, we incorporate the light-meson resonances in our model for the
sub-leading power form factors, which confirms the estimates of duality
violation. This together with numerical cancellations of various otherwise
dominant leading-power effects leads us to conclude that theoretical
predictions of $\Bqtollgamma$ observables in the low-$q^2$ region are
plagued by large theoretical uncertainties.

The outline of the paper is as follows. In \refsec{sec:theory} we
detail the conventions for the effective weak interaction Lagrangian
and parametrize the $\Bqtollgamma$ amplitude in terms of hadronic
tensors and the associated form factors. We employ a mix of QCD factorization
and SCET techniques to calculate in \refsec{sec:FF-factorization} the
hadronic tensors at LP including $\mathcal{O}(\alS)$ radiative corrections
and the summation of logarithms, and at NLP in $\mathcal{O}(\alS^0)$.
We next discuss issues related to the interpretation of the rescattering
phase from factorization, duality violation and resonances. We then
parametrize the single NLP form factor, which cannot be computed in
factorization, and discuss the model that will be used for the numerical
analysis. \refsec{sec:phenoanalysis} is devoted to a detailed discussion 
of the various contributions to the $\Bqtollgamma$ decay amplitude,
which exhibits sizeable corrections and cancellations in the $q^2$ region
of interest. We show the differential branching fraction, lepton 
forward-backward asymmetry and direct CP asymmetry as functions of
$q^2$ and integrated in various bins. We conclude in \refsec{sec:concl}.
Further details on definitions, conventions, final-state radiation and
the  $B$-meson LCDA are given in Appendices.

%
%
%

\section{\boldmath
$\Bqtollgamma$ amplitude}
\label{sec:theory}

%
%
%
\subsection[$\Delta B = 1$ effective theory]
{\boldmath  $\Delta B = 1$ effective theory}
\label{sec:def-EFT}

The low-energy effective theory of electroweak interactions in the
SM for $\DelB = 1$
semileptonic $b\to q\ell\bar\ell$ ($q = d, s$) decays is
\begin{align}
  \label{eq:weak-EFT}
  \calL_\text{eff} &
  = \normEW \left[
     \sum_{i=1}^{6} C_i\, \Op_i^{(c)}
   + \frac{\alE}{4\pi} \sum_{i=7}^{10} C_i\, \Op_i
   + \frac{\lamCKM{u}{q}}{\lamCKM{t}{q}} \sum_{i=1}^2
      C_i \left(\Op^{c}_i - \Op^{u}_i\right)
   \right] + \text{h.c.} \,.
\end{align}
We adopt  the basis proposed in \cite{Chetyrkin:1996vx, Bobeth:1999mk} for the
operators $\Op_i$. The normalization factor $\normEW \equiv 2\sqrt{2} \, G_F \,
\lambda_t^{(q)}$ contains products of elements of the quark mixing matrix
$\lamCKM{U}{q} \equiv V_{Ub}^{} V_{Uq}^*$ for $U = u,c,t$. The term
proportional to
$\lamCKM{u}{q}$ leads to tiny doubly Cabibbo-suppressed CP-asymmetries
in $b\to s$ transitions, but is not negligible in $b\to d$ transitions. The
Wilson coefficients $C_i(\nu)$ are evaluated at the hard scale $\nu$
of the order of the $b$-quark mass $m_b$, after renormalization group (RG) evolution \cite{Bobeth:2003at, Huber:2005ig}
from the electroweak scale. There are four-quark operators, the so-called
charged-current operators $\Op_{1,2}^{c, u}$ and the QCD-penguin operators
$\Op_{3,4,5,6}$. The remaining operators in the SM are the semileptonic
operators
\begin{align}
  \Op_{9} &
  = [\bar{q} \gamma^\mu P_L b]  [\bar{\ell} \gamma_\mu \ell] ,
&
  \Op_{10} &
  = [\bar{q} \gamma^\mu P_L b] [\bar{\ell} \gamma_\mu\gamma_5 \ell] ,
\intertext{
with $P_{L/R}=(1\mp\gamma_5)/2$ and the dipole operators
}
  \Op_7 & = - \frac{\oL{m}_b}{e}
            \big[\bar{q} \sigma^{\mu\nu} P_R b \big] F_{\mu\nu} ,
&
  \Op_8 & = - \frac{g_s \oL{m}_b}{e^2}
            \big[\bar{q} \sigma^{\mu\nu} P_R T^a b\big] G^a_{\mu\nu} ,
\end{align}
where $\oL{m}_b$ is the running $b$-quark mass in the \msbar{} scheme at the
scale $\nu$. The QED$\,\times\,$QCD covariant derivative is chosen as
$D_\mu \psi \equiv (\partial_\mu - i e Q_\psi A_\mu - i g_s T^a G^a_\mu ) \psi$,
following \cite{Beneke:2001at},\footnote{The definitions of $\Op_{7,8}$ contain
a minus sign, such that both Wilson coefficients $C_{7,8}< 0$ are negative in the SM.}
and $Q_\ell = -1$ for leptons, $Q_u = +2/3$, $Q_d = -1/3$ for quarks.

%
%
\subsection[$\Bqtollgamma$ form factors]
 {\boldmath $\Bqtollgamma$ form factors}

\label{sec:amp-param}

The transition amplitude for the decay of the $\oL{B}_q$ meson is
\begin{align}
  \label{eq:amplitude}
  \oL{\calA} \equiv \oL{\calA}(\barBqtollgamma) &
  = \bra{\gamma(k, \eps)\, \ell(p_\ell)\, \oL{\ell}(p_\oL{\ell})}
      \calL_\text{eff} \ket{\oL{B}_q(p)} ,
\end{align}
where $\eps$ is the polarization vector of the photon, and $k$ its
momentum, related to the dilepton momentum $q \equiv p_\ell +
p_\oL{\ell} = p - k$. When working to lowest non-vanishing order
in the electromagnetic coupling $\alE=e^2/(4\pi)$, but to all
orders in QCD, the amplitude can be written in the form
\begin{align}
  \oL{\calA} = \, &
  i e \,\frac{\alE}{4\pi} \,\normEW\, \eps^\star_\mu \Bigg\{
    \sum_{i=1}^9 \eta_i C_i \left[
      T^{\mu\nu}_i \,\bra{\ell\oL{\ell}}\oL{\ell}\gamma_\nu\ell\ket{0} +
    S_\nu^{(i)} \int d^4x \,e^{i kx} \,\bra{\ell\oL{\ell}}
      \text{T}\{j_\ell^\mu(x),[\oL{\ell}\gamma^\nu\ell](0)\} \ket{0} \right]
\nonumber \\
  & + \,C_{10} \left[
  T^{\mu\nu}_{10} \, \bra{\ell\oL{\ell}}\oL{\ell}\gamma_\nu\gamma_5
  \ell\ket{0} +
  S_\nu^{(10)} \int d^4x  \,e^{i kx} \, \bra{\ell\oL{\ell}}
  \text{T}\{j_\ell^\mu(x),[\oL{\ell}\gamma^\nu\gamma_5\ell](0)\}
  \ket{0} \right] 
  \Bigg\}
\nonumber \\
  = \, & i e\, \frac{\alE}{4\pi} \,\normEW\, \eps^\star_\mu \left\{
  \sum_{i=1}^9 \eta_i C_i \left[
    T^{\mu\nu}_i L_{V,\nu} + L_V^{\mu\nu} S_\nu^{(i)} \right]
  + C_{10} \left[
    T^{\mu\nu}_{10} L_{A,\nu} + L_A^{\mu\nu} S_\nu^{(10)} \right] \right\}\,,
  \label{eq:ampl-Ti}
\end{align}
where $j_\ell^\mu = Q_\ell \,\oL{\ell}\gamma^\mu\ell$ denotes the
leptonic electromagnetic current. The quantities $T^{\mu\nu}_i$
and $S_\nu^{(i)}$ define hadronic matrix elements, which encode
all QCD effects at lowest non-vanishing order in the electromagnetic
interaction. The calculation of these hadronic tensors is one of
the main purposes of this work.

We briefly consider the terms involving $S_\nu^{(i)}$. The structure
of the purely leptonic tensor multiplying these terms shows
that they correspond to final-state radiation (FSR) of the photon
from the lepton-pair. The hadronic matrix elements  $S_\nu^{(i)}$
are $B$-meson to vacuum matrix elements, which must be proportional
to the $B$-meson momentum $p_\nu$. The contraction of the vectorial
leptonic tensor with $p_\nu$  vanishes, $p_\nu L_V^{\mu\nu}=0$,
while the corresponding contraction of the axi-vectorial leptonic 
tensor is proportional to the lepton
mass. The only non-vanishing FSR contribution is therefore
due to $\Op_{10}$ \cite{Melikhov:2004mk} and helicity-suppressed.
The QCD effects are entirely described by the $B$-meson decay
constant, since
\begin{align}
  \label{eq:def-f_B}
  S_\nu^{(10)} & 
  = \bra{0} \bar q \, \gamma_\nu P_L\, b \ket{\oL{B}_q(p)}
  \equiv -\frac{i}{2} f_{B_q} p_\nu .
\end{align}
The FSR contribution is tiny except in very small phase-space
regions, and can be safely neglected
in the later numerical analysis. For further details on the FSR
contribution we refer to \refapp{app:FSR}.

\begin{figure}
\centering
  \includegraphics[width=0.3\textwidth]{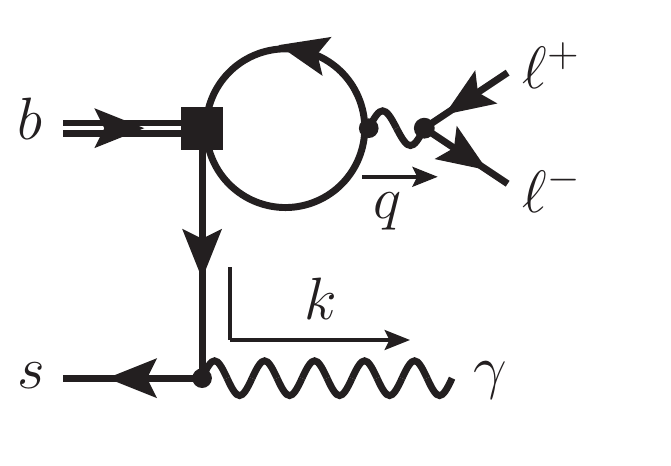}
  \includegraphics[width=0.3\textwidth]{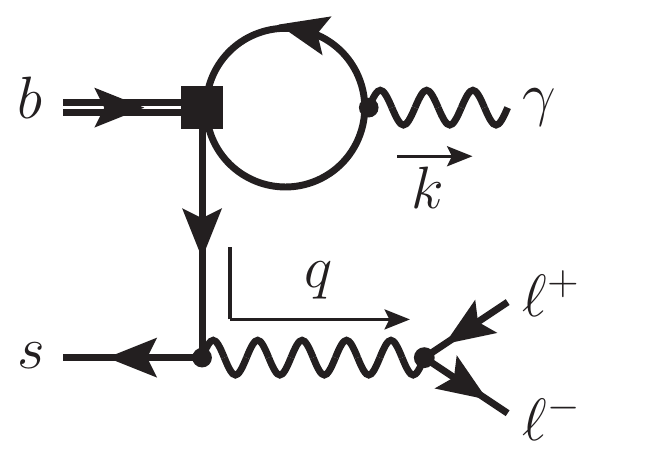}
  \includegraphics[width=0.3\textwidth]{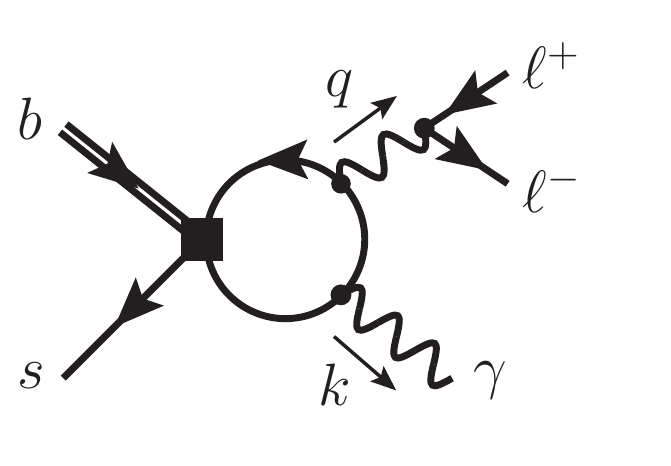}
\caption{\small The three possible contractions ($A$-type left, $B$-type middle)
  of the four-quark operators $\Op_{1,\ldots,6}$ at $\mathcal{O}(\alS^0)$. The
  insertion shown on the right is power-suppressed. The $B$-type insertion is
  power-suppressed when $q^2 \sim m_b^2$ but LP when $q^2 \sim E_\gamma\LambQCD$,
  $m_b\LambQCD$. For the first two diagrams, there exists a counterpart where
  the real/virtual photon is attached to the heavy-quark line (double line),
  which, however, is power-suppressed. The external photon line with momentum
  $k$ symbolizes the electromagnetic current $j^\mu_f(x)$ in the definition
  \refeq{eq:T_16} of the hadronic tensor $T_i^{\mu\nu}$, the virtual photon
  including its decay into the lepton-pair graphically represents the current
  $j^\mu_f(y)$. The $\oL{B}_s$-meson state is represented by its leading
  partonic constituents, a $b$-quark and a light anti-quark.}
  \label{fig:fourquarkLOcontractions}
\end{figure}

Our main concern are therefore the hadronic tensors $T^{\mu\nu}_i$,
sometimes referred to as struc\-ture-dependent (SD) contributions.
The individual SD contributions of each operator are contracted
with the vector or axial-vector lepton currents
\begin{align}
  L_{V(A)}^\alpha &
  = \bra{\ell\oL{\ell}}\oL{\ell}\gamma^\alpha(\gamma_5)\ell\ket{0} = \oL{u}(p_\ell)\, \gamma^\alpha (\gf) \, v(p_{\oL{\ell}})\,,
&
  \eta_i &
  = \left\{ \begin{array}{ll} 1      & \quad i   =  9,10 \\
                              Q_\ell & \quad i \neq 9,10 \end{array} \right. .
\end{align}
The hadronic tensors $T_i^{\mu\nu}(k,q)$ are the
correlation functions
\begin{align}
  \label{eq:T_910}
  T_i^{\mu\nu} &
  = \int \! d^4 x \, e^{i k x} \,
    \bra{0} \text{T} \lbrace j^\mu_f(x), \,
            [\bar{q} \, \gamma^\nu P_L \, b](0) \rbrace \ket{\oL{B}_q} , &
  i & = 9,10,
\\
  \label{eq:T_7A}
  T_{7A}^{\mu\nu} &
  = \frac{2 \oL{m}_b}{q^2} \int \! d^4 x \, e^{i k x} \,
    \bra{0} \text{T} \lbrace j^\mu_f(x), \,
            [\bar{q}\, i \sigma^{\nu\alpha} q_\alpha P_R \, b](0) \rbrace \ket{\oL{B}_q} ,
\\
  \label{eq:T_7B}
  T_{7B}^{\mu\nu} &
  = \frac{2 \oL{m}_b}{q^2} \int \! d^4 x \, e^{i q x} \,
    \bra{0} \text{T} \lbrace j^\nu_f(x), \,
            [\bar{q}\, i \sigma^{\mu\alpha} k_\alpha P_R \, b](0) \rbrace \ket{\oL{B}_q} ,
\\
  T_i^{\mu\nu} &
  = i \,\frac{(4\pi)^2}{q^2} \!\! \int \! d^4 x\, d^4 y\, e^{i k x} e^{i q y} \,
    \bra{0} \text{T} \lbrace j^\mu_f(x),\, j^\nu_f(y),\,
    \Op_i(0) \rbrace \ket{\oL{B}_q} , &
  i & = 1, \ldots, 6, 8
  \label{eq:T_16}
\end{align}
with the electromagnetic quark-current $j^\mu_f = \sum_f Q_f \,
[\bar{f} \gamma^\mu f]$. Here $f$ is summed over all five active 
quark flavours. The hadronic tensors $T_{9,10}^{\mu\nu}$
are time-ordered products of the weak currents with the
electromagnetic current in direct analogy with the charged-current
decay $\Butolvgamma$. The contribution from $\Op_7$,
$T_7^{\mu\nu} = T_{7A}^{\mu\nu} + T_{7B}^{\mu\nu}$, has been split
into the part $T_{7A}^{\mu\nu}$, which accounts for the
emission of the final-state photon from the constituents
of the $B$-meson, and the part $T_{7B}^{\mu\nu}$, where the on-shell
photon is emitted directly from $\Op_7$, while the electromagnetic
current produces the virtual photon with momentum $q$, which
decays into the lepton pair. Quite generally, we refer to
these two different contributions as $A$- and $B$-type,
respectively.\footnote{ In the case of $\Bqtogammagamma$
\cite{Bosch:2002bv} the off-shell photon is replaced by an on-shell
photon and $A$- and $B$-type contributions are equivalent.}
Similarly, the hadronic tensors $T_{1,\ldots 6}^{\mu\nu}$
for the four-quark operators contain $A$-type and $B$-type parts,
depending on which of the two electromagnetic currents
is inserted into a $B$-meson constituent quark line, and
a further annihilation-type contribution, where neither of
the on-shell and virtual photon is emitted from the initial-state
quarks. See \reffig{fig:fourquarkLOcontractions} for
the lowest order diagrams in $\alS$ representing these three
contributions.

%
The hadronic tensors $T_i^{\mu\nu}(k,q)$ can be parametrized in terms of scalar
form factors. Their number is determined by exploiting the Ward identities
(e.g. \cite{Grinstein:2000pc}) that hold upon contraction with $k$ and $q$,
\begin{equation}
  \label{eq:ward-identities}
\begin{aligned}
  k_\mu T_{9,10}^{\mu\nu} =
  k_\mu T_{7A}^{\mu\nu}  =
  q_\nu T_{7B}^{\mu\nu} & = 0 , &
\\
  k_\mu T_i^{\mu\nu} =
  q_\nu T_i^{\mu\nu} & = 0 , &  \quad i & = 1,\ldots,6,8 ,
\end{aligned}
\end{equation}
together with the algebraic relations 
$k_\mu T_{7B}^{\mu\nu} = q_\nu T_{7A}^{\mu\nu} = 0$. The Ward 
identities and the transversality of the on-shell photon, 
$\eps^\star \cdot k = 0$, which implies that $\mu$ is a transverse 
index, lead to the Lorentz decomposition
\begin{align}
  T_i^{\mu\nu} (k, q) &
  = \frac{1}{m_{B_q}} \Big[
    \left[ g^{\mu\nu}(k\cdot q) - q^\mu k^\nu \right]
           \left( F_L^{(i)} - F_R^{(i)} \right)
         + i \veps^{\mu\nu\alpha\beta} q_\alpha k_\beta
           \left( F_L^{(i)} + F_R^{(i)} \right)
  \Big]
\nonumber \\
  \label{eq:Ti-FF-param}
  & = E_\gamma \left[
    g_\perp^{\mu\nu}       \left( F_L^{(i)} - F_R^{(i)} \right)
  + i \veps^{\mu\nu}_\perp \left( F_L^{(i)} + F_R^{(i)} \right)
  \right],
\end{align}
which contains two helicity form factors $F_{L,R}^{(i)}$ for each 
operator $\Op_i$. We use the convention $\veps_{0123} = -1$,\footnote{Note the 
different convention $\veps_{0123} = +1$ in \cite{Beneke:2011nf} for 
$\Butolvgamma$.}
for definitions of $g_\perp^{\mu\nu}$ and $\veps^{\mu\nu}_\perp$ we 
refer to \refapp{app:defs}.

Below we will employ QCD factorization techniques to factorize the hadronic
matrix elements $T^{\mu\nu}_i(k, q)$ and the corresponding form factors
$F_h^{(i)}$ $(h = L,R)$ for large photon energy $E_\gamma \gg \LambQCD$. In this
limit the right-helicity form factors $F_R^{(i)}$ are suppressed by $\LambQCD/E_\gamma$ compared to $F_L^{(i)}$
due to the left-handedness of the weak interaction and helicity-conservation of
QCD at high energy. One often defines vector $F_V^{(i)} = F_L^{(i)} + F_R^{(i)}$
and axial-vector $F_A^{(i)} = F_L^{(i)} - F_R^{(i)}$ form factors, hence the
symmetry $F_V^{(i)} = F_A^{(i)}$ between the transversity form factors holds
at LP in the heavy-quark/large-energy expansion.

%
%
%
\section{Factorization of form factors}
\label{sec:FF-factorization}

QCD factorization can be applied to the hadronic process if the
on-shell photon is very energetic $E_\gamma \gg \LambQCD$. It is
most intuitive to work in the rest frame of the $B_q$ meson, where the
three-vectors $\vec{k}$ and $\vec{q}$ are back-to-back and the
constituent $b$ and $q$ quarks of the $B_q$ meson will have soft
residual momenta of order $\LambQCD$.

We introduce the four velocity $\vb$ ($\vb^2 = 1$) of the
$\oL{B}_q$-meson and a pair of light-like vectors $\nn$ and $\nb$,
with $\nnn^2 = \nbb^2 = 0$, $\nn \nb = 2$, $\vb = (\nn + \nb)/2$,
such that
\begin{align}
  p^\mu & = m_{B_q} \vb^\mu , &
  k^\mu & = \frac{\nb k}{2} \, \nnn^\mu = \frac{m_{B_q}}{2} y \, \nnn^\mu, &
  q^\mu & = \frac{m_{B_q}}{2} \left[ \nbb^\mu + \left(1 - y \right) \nnn^\mu \right] .
\end{align}
The photon energy $E_\gamma = (\nb k)/2 = (m_{B_q}^2-q^2)/
(2m_{B_q})$. For later convenience
we define $y \equiv 2 E_\gamma/m_{B_q}$ with $0 \leq y
\leq 1 - 4 m_\ell^2/m_{B_q}^2$. We refer to momenta $r$ with large
component $\nb r = \mathcal{O}(E_\gamma, m_b)$ in the direction
of $\nn$ and small $r^2$ as collinear. Anti-collinear momenta
have this property with $\nb \leftrightarrow \nn$ interchanged.
Hence the momentum $k^\mu$ of the on-shell final-state photon
is collinear. The nature of the lepton-pair or virtual photon
momentum $q^\mu$ depends on whether the real photon energy
$E_\gamma$ is close to its maximal kinematically allowed value
$m_{B_q}/2$, corresponding to $y=1$. Since
$q^2/m_{B_q}^2 = 1 - 2 E_\gamma/m_{B_q} = 1-y$, we consider two
scalings for~$q^2$:
\begin{itemize}
\item[1)] anti-hard-collinear~($\oL{\text{hc}}$) $q^2 \sim E_\gamma \LambQCD$
  or $m_b \LambQCD$, in which case the momentum $q^\mu$ is dominated by its
  component proportional to the light-like vector $n_+^\mu$, and
\item[2)] hard~(h) $q^2 \sim m_b^2$, in which case the photon energy is large,
  but $y$ is not parametrically close to 1 in the heavy-quark limit.
\end{itemize}
In practice, we will be interested in the region
with the restriction to $q^2 \lesssim 6 \GeV^2$, as originally
introduced for $B\to K^{(*)} \ell\bar\ell$ \cite{Beneke:2001at},
to avoid the charmonium resonance region. This upper limit
lies somewhat in between these two scalings. We will construct
the form factors \refeq{eq:Ti-FF-param} in an expansion in
powers of $\lambda^2 \equiv \LambQCD/E_\gamma$ or $\LambQCD/m_b$,
\begin{align}
  \label{eq:F_h-exp}
  F_h^{(i)} &
  = F_h^{(i,\text{LP})} + F_h^{(i,\text{NLP})}
  + \mathcal{O}(\alS^2, \alS \lambda^2, \lambda^4) ,
&
  h & = L, R
\end{align}
where the LP contribution includes the resummation of NLO radiative
QCD corrections, for which we employ SCET. The  NLP contribution
is included to LO in QCD. Our result will be accurate to
these orders for both possible scalings of $q^2$.

%
%
%
\subsection{Form factors at LP}

The SCET framework allows for a systematic decoupling of fluctuations with
hard virtualities of order $m_b^2$ in the matching QCD$\,\to\,$\SCETI{} and
fluctuations with hard-collinear virtualities of order $E_\gamma \LambQCD$,
$m_b \LambQCD$ in the matching \SCETI{}$\,\to\,$\SCETII{}. The SCET framework
at LP \cite{Lunghi:2002ju, Bosch:2003fc, Beneke:2011nf} can be directly applied
to the semileptonic operators $\Op_{9,10}$ and there is no difference whether
$q^2 \sim \oL{\text{hc}}$ or $q^2 \sim \text{h}$. The factorization of the form
factors of the other operators $i=1,\ldots,8$, which admit $A$- and $B$-type
contributions is more complicated, and we provide the results
below, together with some explanation of their derivation.

\begin{figure}
\centering
  \includegraphics[width=0.3\textwidth]{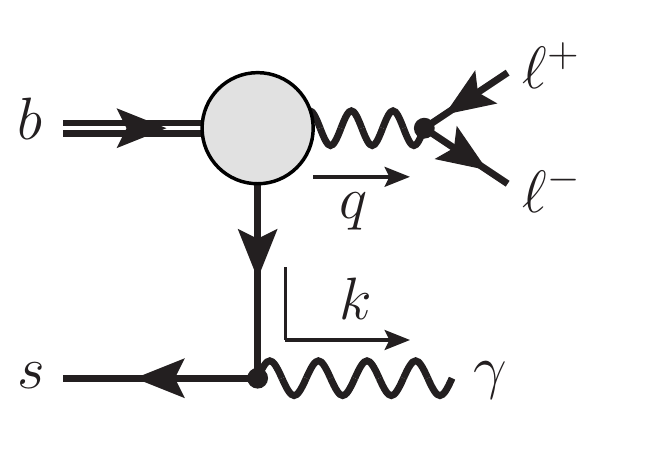}
  \includegraphics[width=0.3\textwidth]{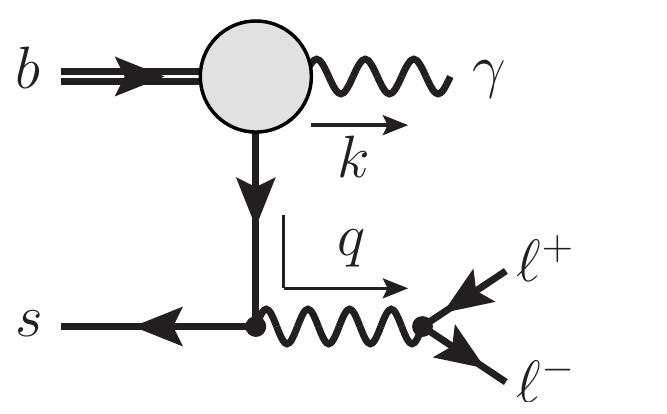}
\caption{\small The LP $A$-type and $B$-type insertions. The grey blob denotes
  the effective hard vertex of the operators $\Op_{1,\ldots,8}$.}
\label{fig:AB-type}
\end{figure}

%
%
\subsubsection{Effective hard vertex}
\label{sec:LP-SCETI}

We begin the discussion of the first matching step to \SCETI{} by
assuming that $q^2$ is hard and the insertion is of
the $A$-type. A LP contribution to the hadronic tensors
$T_i^{\mu\nu}$ is obtained only when the separation $x$ between
the electromagnetic current $j^\mu_f(x)$ and the operator
$P_i(0)$ is of order $1/\sqrt{E_\gamma\Lambda_{\rm QCD}}$. 
In other words, the propagator
joining the two vertices in the left diagram of
\reffig{fig:AB-type} must have hard-collinear virtuality,
since a hard propagator would lead to power-suppression.
Integrating out the hard scale therefore results in an
effective flavour-changing vertex, represented by the grey
circle in \reffig{fig:AB-type}. The hadronic tensors
appearing in \refeq{eq:ampl-Ti} are therefore matched
to \SCETI{} as follows:
\begin{align}
  \label{eq:QCD->SCET1a}
  \sum_{i=1}^{9} \eta_i C_i T_i^{\mu\nu} &
  = \sum_{i=1}^9 C_i \, H_{i}(q^2) \int\! d^4 x \, e^{i k x} \,
    \bra{0} \text{T} \lbrace j^\mu_{f, \,\rm SCET_{\rm I}}(x),\, 
    [\oL{q}_{\rm hc}\gamma^{\nu_\perp} P_L h_\vb](0) \rbrace \ket{\oL{B}_q} \,,
\\
  \label{eq:QCD->SCET1b}
  C_{10} T_{10}^{\mu\nu} &
  = C_{10} \, H_{10}(q^2) \int\! d^4 x \, e^{i k x} \,
    \bra{0} \text{T} \lbrace j^\mu_{f, \,\rm SCET_{\rm I}}(x), \,
    [\oL{q}_{\rm hc}\gamma^{\nu_\perp} P_L h_\vb](0) \rbrace \ket{\oL{B}_q}\,,
\end{align}
where $H_i(q^2)$ denotes the hard matching coefficients, and
the hadronic tensor for $i=7$ is $T_{7A}^{\mu\nu}$.
All operators match to the same correlation function of
the \SCETI{} heavy-to-light current
$\oL{q}_{\rm hc}\gamma^{\nu_\perp} P_L  h_\vb$ and the
representation of the \SCETI{} electromagnetic current.
In the above equation the Wilson coefficients $C_i$ are evaluated
at the scale $\nu\sim m_b$ after evolving them to $\nu$ from the
electroweak scale. The matching coefficients $H_i$ are determined
at the hard scale $\mu_h\sim m_b$ in the
matching QCD$\,\to\,$\SCETI{}.
In addition to $q^2$, they depend on the scale $\nu$ through
the ultraviolet divergences of the QCD diagrams, such that the
dependence on $\nu$ cancels in the product $C_i(\nu)
H_{i}(\nu, \mu_h)$, as well as on the scale $\mu_h$ through the
 infrared divergences of the QCD diagrams. The $\mu_h$-dependence
cancels with the dependence on $\mu_h$ of the \SCETI{}
correlation functions in \refeq{eq:QCD->SCET1a},
\refeq{eq:QCD->SCET1b}.

The $\mathcal{O}(\alS)$ corrections to the hard functions
$H_i(q^2)$ are identical to those for $B\to K^{(*)} \ell\bar\ell$
(see Figure~2 in \cite{Beneke:2001at}). For the four-quark
operators, they arise from two-loop diagrams. It is customary
to absorb the four-quark operator contributions into effective
Wilson coefficients of $\Op_{7-9}$, and we follow this practice
here. Hence we define
\begin{equation}
\begin{aligned}
  \sum_{i=1}^9 C_i \, H_{i}(q^2) &
  \equiv V_9^{\rm eff}(q^2) + \frac{2\, \oL{m}_b\, m_{B_q}}{q^2} \, V^{\rm eff}_7(q^2) ,
\\
  C_{10} \, H_{10}(q^2) &
  \equiv V^{\rm eff}_{10}(q^2) ,
\end{aligned}
\end{equation}
where  $\oL{m}_b$ denotes the \msbar{}-scheme $b$-quark mass at the scale $\nu$.\footnote{Here $q^2$ denotes the momentum-transfer 
squared at the hard vertex, which coincides with the dilepton 
invariant mass squared for the $A$-type contribution. For 
the $B$-type contributions considered below $q^2$
must be substituted by $k^2 = 0$.}
The effective hard functions including the $\mathcal{O}(\alS)$ correction are then given by
\begin{align}
  \label{eq:def-H_7}
  V_7^\text{eff} = \; & 
  C_7^\text{eff} \, C_{T_1}^{(A0)}
  - \frac{\alS}{4\pi} \bigg[
      C_1 \bigg(F_1^{(7c)} + \frac{\lamCKM{u}{q}}{\lamCKM{t}{q}} 
\left[F_1^{(7c)}-F_1^{(7u)}\right] \bigg)
 \nonumber\\
  &    + \,C_2 \bigg(F_2^{(7c)} + \frac{\lamCKM{u}{q}}{\lamCKM{t}{q}} 
\left[F_2^{(7c)}-F_2^{(7u)}\right]  \bigg)
+ C_8^\text{eff} F_8^{(7)} + \text{$C_{3-6}$-terms} \,\bigg]\, ,
\\
  \label{eq:def-H_9}
  V_9^\text{eff} = \; & C_9^\text{eff}(q^2) \, C_V^{(A0)}
  - \frac{\alS}{4\pi} \bigg[
      C_1 \bigg(F_1^{(9c)} + \frac{\lamCKM{u}{q}}{\lamCKM{t}{q}} 
\left[F_1^{(9c)}-F_1^{(9u)}\right] \bigg)
\nonumber\\
  &     +\, C_2 \bigg(F_2^{(9c)} + \frac{\lamCKM{u}{q}}{\lamCKM{t}{q}} 
\left[F_2^{(9c)}-F_2^{(9u)}\right] \bigg) 
+ C_8^\text{eff} F_8^{(9)} + \text{$C_{3-6}$-terms}\, \bigg]\,,
\\
  \label{eq:def-H_10}
  V_{10}^\text{eff} = \; & C_{10} \, C_V^{(A0)} ,
\end{align}
where we have suppressed all arguments $q^2,\mu_h,\nu$. Here $\alS$ is evaluated
at $\mu_h$. The effective Wilson coefficient combinations read as follows
\cite{Chetyrkin:1996vx, Bobeth:1999mk}:
\begin{align}
  \label{eq:def-C7-eff}
  C_7^\text{eff} &
  = C_7 - \frac{C_3}{3}   - \frac{4}{9}  C_4
        - \frac{20}{3} C_5 - \frac{80}{9} C_6 ,
\\
  \label{eq:def-C8-eff}
  C_8^\text{eff} &
  = C_8 + \, C_3 \,  - \frac{1}{6} C_4 + 20 C_5 - \frac{10}{3} C_6 ,
\\
  \label{eq:def-C9-eff}
  C_9^\text{eff} (q^2) &
  = C_9 + Y(q^2)
  - \frac{\lamCKM{u}{q}}{\lamCKM{t}{q}} \left(\frac{4}{3} C_1 + C_2 \right)
    \big[h(q^2, 0) - h(q^2, m_c)\big] .
\end{align}
We use the definition of the function $Y(q^2)$ from~\cite{Beneke:2001at}.
The function $h(q^2, m_q)$~\cite{Bobeth:1999mk} depends on the light quark
masses $m_{u,d}$, which are set to zero, or the charm-quark pole mass $m_c$.

For the semileptonic operators and the parts from the four-quark
operators contained in the effective Wilson coefficients above,
the NLO QCD corrections are contained in the matching coefficients
$C_V^{(A0)}(q^2)$, $C_{T_1}^{(A0)}(q^2)$ of the  heavy-to-light
(axial-) vector and tensor QCD currents to the \SCETI{} current
$\oL{q}_\text{hc}\gamma^{\nu_\perp} P_L h_\vb$ \cite{Bauer:2000yr}.
We use the notation of \cite{Beneke:2005gs} and provide the NLO expression here:
\begin{align}
  \nonumber
  C_V^{(A0)}(q^2) &
  = 1 + \frac{\alS(\mu) C_F}{4 \pi} \,\bigg(
\\ \label{eq:def-CVA0} &
      -2 \ln^2 \frac{m_{B_q}z} {\mu} + 5 \ln \frac{m_{B_q} z}{\mu}
      -\frac{3 - 2 z}{1-z} \ln z - 2 \Li2 (1 - z)
 - \frac{\pi^2}{12} - 6
    \bigg) \,,
\\ \nonumber
  C_{T_1}^{(A0)}(q^2) &
  = 1 + \frac{\alS(\mu) C_F}{4 \pi} \,\bigg(
\\ \label{eq:def-CT1A0} &
      2 \ln \frac{m_b}{\nu}
     -2 \ln^2 \frac{m_{B_q} z}{\mu} + 5 \ln \frac{m_{B_q}z}{\mu}
     -3 \ln z - 2 \Li2 (1 - z) - \frac{\pi^2}{12} - 6
    \bigg) \,,
\end{align}
with $z = 1-q^2/m_{B_q}^2$, $C_F = (N_c^2-1)/(2 N_c)=4/3$, and
$N_c = 3$ in QCD.

The NLO QCD corrections from the charged-current operators $\Op_{1,2}^{u,c}$
and  $\Op_8$ are given by $F^{(7u,7c,9u,9c)}_{1,2}(q^2)$ from \cite{Asatrian:2001de,
Asatryan:2001zw, Seidel:2004jh} and $F_8^{7,9}(q^2)$ from \cite{Asatryan:2001zw,
Beneke:2001at}, respectively. The analogous corrections from $\Op_{3,4,5,6}$
labelled ``$C_{3-6}$-terms'' in \refeq{eq:def-H_7}, \refeq{eq:def-H_9} are
suppressed by the small penguin-operator Wilson coefficients and have been
neglected here.\footnote{They are known only for $q^2 = 0$ \cite{Buras:2002tp}.}
The terms proportional to $\lamCKM{u}{q}$ in $V_{7,9}^\text{eff}$ give rise to
direct CP violation, which is doubly-Cabibbo suppressed for $q = s$ and leads
to tiny CP asymmetries, whereas for $q = d$ CP-violating effects can be larger.

To summarize the discussion up to this point, we note that the $A$-type contribution
to $\Bqtollgamma$ amplitude \refeq{eq:ampl-Ti} after matching to \SCETI{} at LP
is obtained in the form
\begin{align}
  \oL{\calA}_{\text{type}-A} = \; &
  i e \,\frac{\alE}{4\pi} \,\normEW\, \eps^\star_\mu \left\{
  \left( V_9^\text{eff}(q^2) 
       + \frac{2\, \oL{m}_b\, m_{B_q}}{q^2} \, V^\text{eff}_7(q^2)
  \right) L_{V,\nu}
  + V^\text{eff}_{10}(q^2) L_{A,\nu} \right\}
\nonumber\\
  & \times \int\! d^4 x \, e^{i k x} \,
  \bra{0} \text{T} \lbrace j^\mu_{f, \,\rm SCET_{\rm I}}(x), \, 
  [\oL{q}_\text{hc} \gamma^{\nu_\perp} P_L h_\vb](0) \rbrace \ket{\oL{B}_q}\,.
  \label{eq:amplSCETItypeA}
\end{align}
The derivation of this result assumed that the lepton-pair
virtuality is hard. When $q^2$ is anti-hard-collinear, the
structure of the result remains the same. However, the effective
vertices, which are functions of $q^2/m_b^2$, could now
be evaluated at $q^2=0$, since the ratio  $q^2/m_b^2$ is
power-suppressed. More importantly, the effective vertex
$V_7^\text{eff}(q^2)$ becomes power-enhanced due to the photon
pole $1/q^2$, relative to which  $V_{9,10}^\text{eff}(q^2)$ should be dropped as
power-suppressed. This, however, results in a physically
unacceptable approximation. For example, making the same
approximation for the exclusive $B\to K^{(*)} \ell\bar\ell$
decay, the lepton forward-backward asymmetry would disappear and
the predicted branching fractions and angular distributions
in the small-$q^2$ region become unrealistic. The reason is that
the magnitude of the Wilson coefficients $C_{9,10}$ is about
an order of magnitude larger than $C_{7}$, so that it is
more appropriate to count $C_{9,10} \sim m_b^2/q^2\times
C_7$. In other words, the expansion in $1/m_b$ should
be performed for the $C_{9,10}$ and $C_7$ terms separately,
and for each term the LP should be kept.\footnote{In case of
$C_8^\text{eff}$, one then counts $F_8^{(7,9)}(q^2)$ as
$C_7$ and $C_9$ terms, respectively.} 
In the same spirit, no harm is done
by keeping the $q^2$-dependence in the effective vertices
also at small $q^2$. In this way, one obtains a smooth interpolation
between hard and anti-hard-collinear $q^2$, as
\refeq{eq:amplSCETItypeA} applies without modification to the
anti-hard-collinear, small-$q^2$ region. This is also the procedure
that has been adopted for the exclusive $B\to K^{(*)} \ell\bar\ell$
\cite{Beneke:2001at}, which forms the basis of the QCD phenomenology
of this decay.

Another subtlety needs to be mentioned here. When $q^2$ is
anti-hard-collinear, the effective hard vertices are no longer
guaranteed to be dominated by hard virtualities. This is
obvious for the one-loop four-quark operator contribution
shown in the left diagram of \reffig{fig:fourquarkLOcontractions}, 
where the quark-loop is then purely anti-hard-collinear. At $\mathcal{O}(\alS)$
the two-loop diagrams contributing to $F_{1,2}^{(7u,7c,9u,9c)}$
contain hard and hard-collinear regions. It therefore appears
that the previous treatment should be substantially modified,
since only the hard regions are integrated out in the
first matching step to \SCETI. However, the \SCETI{} correlation
function in \refeq{eq:amplSCETItypeA} is governed by 
(hard-)collinear physics. The anti-(hard-)collinear
and the (hard-)collinear sectors of \SCETI{} are already
decoupled after integrating out the hard modes and performing
the soft-decoupling transformation
in \SCETI. The soft Wilson lines from the
decoupling transformation cancel, as the anti-collinear final
state is colour-neutral, hence the existence of the
anti-hard-collinear physics leaves no impact on the remaining
collinear and soft interactions. In consequence, when $q^2$ is
anti-hard-collinear, we may
simply assume that the hard functions $H_i(q^2)$ introduced
above are in fact the hard {\em and} anti-hard-collinear
functions. In principle,
these functions may therefore contain formally large
logarithms $\ln q^2/m_b^2$, which we cannot resum by not
factorizing the hard and anti-hard-collinear physics properly.
However, the existence of the limit $q^2 \to 0$ shows that
such logarithms are absent, at least at LP and
$\mathcal{O}(\alS)$. We note that the same discussion
applies to the standard treatment of the
exclusive $B\to K^{(*)} \ell\bar\ell$ mode, although we are not
aware of its explicit mentioning.

In addition to the $A$-type insertion, there is also a $B$-type
contribution, see the right diagram of \reffig{fig:AB-type},
in which the role of the real and virtual photon is interchanged.
When $q^2$ is anti-hard-collinear, the separation between the
flavour-changing and electromagnetic current is again
of order $1/(E_\gamma\LambQCD)$, and one obtains a LP
contribution. Since the lepton-pair originates from the
electromagnetic current, there is no $B$-type contraction
from the operator $\Op_{9,10}$, and in fact no contribution from
the corresponding effective vertices $V^\text{eff}_{9,10}$.
Hence
\begin{align}
  \oL{\calA}_{\text{type}-B} = \, &
  i e \,\frac{\alE}{4\pi} \,\normEW\, \eps^\star_\mu
  \frac{4\, \oL{m}_b E_\gamma}{q^2} \, V^\text{eff}_7(0) L_{V,\nu}
\nonumber\\
  & \times \int\! d^4 x \, e^{i q x} \,
  \bra{0} \text{T} \lbrace j^\nu_{f, \,\rm SCET_{\rm I}}(x),\, 
  [\oL{q}_{\oL{\rm hc}}\gamma^{\mu_\perp} P_L h_\vb](0) \rbrace \ket{\oL{B}_q} \,.
  \label{eq:amplSCETItypeB}
\end{align}
We note that the effective vertex $V^\text{eff}_7(0)$
is now evaluated for $k^2=0$,\footnote{To be precise, the
definition now contains the hadronic tensor $T_{7B}^{\mu\nu}$
instead of $T_{7A}^{\mu\nu}$, but with the factor $1/q^2$
taken out, the result for $V^\text{eff}_7$ is the same
as \refeq{eq:def-H_7}.}
and the Fourier transform of the \SCETI{} correlation function
is taken with respect to $q$ rather than $k$. The remarks above
concerning the interpretation of the ``hard'' functions $H_i(q^2)$
when the argument $q^2$ is anti-hard-collinear, apply to the present
case $k^2=0$.

When $q^2$ is hard, the \SCETI{} correlation function in
\refeq{eq:amplSCETItypeB} is not the correct expression, since the
distance $x\sim 1/\sqrt{q^2}$ is hard and should already have been
integrated out in the matching to \SCETI. In other words,
the propagator connecting the two currents in the right diagram
of \reffig{fig:AB-type} is far off-shell and should be
contracted to a point, resulting in a local operator rather than
a  \SCETI{} correlation function. Also the vertex
$V^\text{eff}_7(0)$ is not necessarily the correct one in
(\ref{eq:amplSCETItypeB}), since it was obtained by matching the
flavour-changing current
with an on-shell out-going quark. On the other hand, the off-shell
propagator provides power suppression of the $B$-type contraction
in the hard-$q^2$ region relative to the anti-hard-collinear one,
such that $B$-type insertions are NLP effects for hard $q^2$,
for which we aim only at $\mathcal{O}(\alS^0)$ accuracy.

We now argue that within the approximation that we include only LO
$\mathcal{O}(\alS^0)$ contributions at NLP, the expression
\refeq{eq:amplSCETItypeB} can be used even when $q^2$ is hard. The
$B$-type insertion $\mathcal{O}(\alS^0)$ diagram (the diagram corresponding
to the middle diagram of \reffig{fig:fourquarkLOcontractions}
but with the quark-loop replaced by the direct attachment of the
photon to the flavour-changing vertex) contains the expression
\begin{equation}
  \oL{v}_s(l)\, \gamma^\nu \frac{i (\slashed{q}-\slashed{l})}
  {q^2-2 q\cdot l} \left[\gamma^\mu,\slashed{k}\right] P_R u_b(p_b) \,
  \epsilon^*_\mu(k)\,,
\end{equation}
where $l$ is the momentum of the spectator quark.
The LP approximation in both $q^2$ regions is obtained from
keeping only the potentially large momentum components
in the numerator. For hard $q^2$ this includes $(n_+ q)
\frac{\slashed{n}_-}{2}$. The important observation is
that this term vanishes, since $\mu$ is transverse and
$k = E_\gamma n_-$, hence $\slashed{n}_-
\left[\gamma^{\mu_\perp},\slashed{n}_-\right] = 0$.
Thus, the previous equation simplifies to
\begin{equation}
  \frac{1}{2} \, \oL{v}_s(l)\, \gamma^\nu \frac{i (n_- q )\slashed{n}_+}
  {q^2-2 q\cdot l} \left[\gamma^\mu,\slashed{k}\right] P_R u_b(p_b) \,
  \epsilon^*_\mu(k)
  \label{eq:typeBstring}
\end{equation}
in both $q^2$-regions. \refEq{eq:typeBstring} would be obtained
from \refeq{eq:amplSCETItypeB} in the anti-hard-collinear $q^2$
region, which proves that we can smoothly extrapolate
\refeq{eq:amplSCETItypeB} into the hard region at
$\mathcal{O}(\alpha_s^0)$, in which case
\begin{align}
  \frac{n_- q } {q^2-2 q\cdot l} & \to \frac{1}{n_+ q} \,.
  \label{eq:typeBstringlimit}
\end{align}
Since we do not aim at $\mathcal{O}(\alS)$ accuracy at NLP,
we can use \refeq{eq:amplSCETItypeB} in the hard-collinear
and hard $q^2$ region, even if in the hard region
the  $\mathcal{O}(\alS)$
correction to $V_7^\text{eff}(0)$ is not the complete one at
NLO in $\alS$.

To summarize the first matching step, we obtain the LP
$\Bqtollgamma$ amplitude \refeq{eq:ampl-Ti} in the form
\begin{align}
  \oL{\calA}_\text{LP} &
  = \oL{\calA}_{\text{type}-A} + \oL{\calA}_{\text{type}-B} \,,
  \label{eq:amplSCETILP}
\end{align}
where the two terms are given by \refeq{eq:amplSCETItypeA},
\refeq{eq:amplSCETItypeB}. The second step consists of matching
the \SCETI{} correlation function in these terms to \SCETII{}
by integrating out the hard-collinear modes. Before turning
to this task in the following subsection, we mention that
the same correlation function appears for all operators $\Op_i$
from the electroweak effective Lagrangian. The RG evolution
from the hard to the hard-collinear scale is therefore universally
related to the anomalous dimension of the LP A0-type
\SCETI{} heavy-to-light current
$\oL{q}_\text{hc}\gamma^{\nu_\perp} P_L h_\vb$ \cite{Bauer:2000yr}
and equals the one that appears in  the $\Butolvgamma$ decay.
We therefore have
\begin{align}
  \label{eq:H-RGE}
  V_i^\text{eff}(q^2, \mu_{hc}, \nu) &
  = U_H(q^2, \mu_{hc}, \mu_h) \, V_i^\text{eff}(q^2, \mu_h, \nu)\,.
\end{align}
The RG equation for $U_H(q^2,\mu_{hc}, \mu_h)$ and its solution
to NLL\footnote{The RG solution sums Sudakov double logarithms.
In the literature on Sudakov resummation, the approximation would
be called ``NNLL''.} can be found in Eqs.~(A.1) and (A.3)
of~\cite{Beneke:2011nf} with $U_H (q^2,\mu_{hc}, \mu_h) =
U_1((m_{B_q}^2-q^2)/(2 m_{B_q}),\mu_h, \mu_{hc})$ in terms of the
evolution factor defined there.

%
%
\subsubsection{Hard-collinear function}
\label{sec:LP-SCETII}

The matching of \SCETI{} to \SCETII{} amounts to integrating
out the hard-collinear modes in the \SCETI{} correlation
function
\begin{equation}
  \mathcal{T}^{\mu\nu}(r)
  \equiv \int\! d^4 x \, e^{i r x} \,
  \bra{0} \text{T} \lbrace j^\mu_{f, \,\text{\SCETI{}}}(x), \, 
  [\oL{q}_\text{hc}\gamma^{\nu_\perp} P_L h_\vb](0) \rbrace \ket{\oL{B}_q}\,.
  \label{eq:SCETIcorrelation}
\end{equation}
No new calculation is required in this step. For $r^2=0$,
it has first been explained and computed to the one-loop order
for  $\Butolvgamma$ in \cite{Lunghi:2002ju,Bosch:2003fc}.
The generalization to $r^2 \neq 0$ needed here was worked
out in \cite{Wang:2016qii}.

We recall from \cite{Lunghi:2002ju} that the \SCETI{}
electromagnetic current $j^\mu_{q, \text{\SCETI{}}}(x)$
relevant here consists of two power-suppressed pieces
\begin{align}
  j^\mu_{q, \text{\SCETI{}}}(x) = \; &
  j^{(0)\mu}_{q, \text{\SCETI{}}}(x) + j^{(1)\mu}_{q, \text{\SCETI{}}}(x)
\nonumber\\
  = \; & \sum_q Q_{q} \left[ \oL{q}_\text{hc}
    \left(\gamma_\perp^\mu \frac{1}{i\nb D_\text{hc}} i \slashed{D}_{\text{hc}\perp}
      + i \slashed{D}_{\text{hc}\perp} \frac{1}{i\nb D_\text{hc}} \gamma_\perp^\mu
    \right) \frac{\nbs}{2} q_\text{hc} \right] (x)
\nonumber\\
  \label{eq:SCET-QED}
  & + \sum_q Q_{q} \left[\oL{q}_s(x_-) \, \gamma^\mu_\perp \, q_\text{hc}(x) \right] 
\end{align}
with $x_-^\mu \equiv (\nb x) \,n_-^\mu/2$. Here (and above) 
$q_\text{hc} = W^\dagger \xi_\text{hc}$ is the hard-collinear
quark field multiplied by the QCD hard-collinear Wilson line. 
The second term on the
right-hand side describes the conversion of the soft spectator
quark into a hard-collinear quark at the photon vertex and
counts as $\mathcal{O}(\lambda^2)$. The first term is
only $\mathcal{O}(\lambda)$ suppressed, but contributes to
the amplitude with an external soft quark only through
the  $\mathcal{O}(\lambda)$ suppressed quark-gluon vertex of
the SCET Lagrangian, that converts a soft quark into a
hard-collinear quark through interaction with a hard-collinear
gluon. This part of the current contributes to the matching
only from NLO in the $\alS$ expansion through hard-collinear
gluon corrections to the photon vertex.

With the hard-collinear physics at the scale $E_\gamma \LambQCD$,
$m_b\LambQCD$ integrated out, the remaining
nonperturbative soft physics is parametrized at LP in the
matching by the leading-twist light-cone distribution
amplitude of the $B$ meson, $\phi_+(\omega)$. The matching equation
reads
\begin{equation}
  \mathcal{T}^{\mu\nu}(r) 
  = (g_\perp^{\mu\nu} + i \veps_\perp^{\mu\nu}) \,
    \frac{Q_q F_{B_q} m_{B_q}}{4} \,
    \int_0^\infty \!\! d\omega\, \phi_+(\omega) \,
    \frac{J(n\cdot r,r^2,\omega)}{\omega - r^2/n\cdot r - i0^+}\,.
  \label{eq:matchtoSCETII}
\end{equation}
Since both cases will be needed, we employ the convention
that $n\cdot r$ denotes the large component $n_+ r$, when
$r$ is a collinear momentum and  $n_- r$ when it is
anti-collinear.\footnote{In the latter case, which 
applies to the $B$-type contribution, exchange 
$\text{hc} \to \oL{\text{hc}}$ and $\nb \leftrightarrow \nn$ in 
\refeq{eq:SCET-QED}.} Concretely, for the $A$-type contribution,
we need $J(n_+ k, 0, \omega)$ with $n_+ k = 2 E_\gamma$, while
for the $B$-type one the relevant function is $J(n_- q, q^2, \omega)$
with $n_-q = m_{B_q}$ and $q^2 = m_{B_q} (m_{B_q}-2 E_\gamma)$.

The scale-dependent quantities $F_{B_q}$, $J$, $\phi_+$ are
assumed to be evaluated at the hard-collinear scale $\mu_{hc}$,
and the scale argument has been omitted in \refeq{eq:matchtoSCETII}.
Their definitions are as follows. The $B$-meson LCDA
$\phi_+(\omega)$ \cite{Grozin:1996pq, Beneke:2000wa} is
the Fourier transform of the HQET matrix element
\begin{align}
  \label{eq:def-B-LCDA}
  \bra{0} \oL{q}_s (t \nn)\, [t\nn, 0] \, \nns \!\gamma_5 \, h_\vb(0) \,
  \ket{\oL{B}_q (p)}  &
  = i m_{B_q} F_{B_q} \int_0^\infty \!\!\! d\omega \,
    e^{-i\omega t} \,\phi_+ (\omega) \,,
\end{align}
where $ [t\nn, 0]$ denotes a straight soft Wilson
line connecting the light-like separated points 0 and $t \nn$.
It is customary to relate the HQET $B$-meson decay constant
$F_{B_q}$ to the scale-independent decay constant of full
QCD $f_{B_q}$, introduced in \refeq{eq:def-f_B}, and use
the latter as an input. The relation is
\begin{align}
  \label{eq:rel-FB-and-fB}
  F_{B_q}(\mu_{hc}) &
  = U_F^{-1}(\mu_{hc},\mu_h)\, K^{-1}(\mu_h)\, f_{B_q}\,
\end{align}
with
\begin{align}
  \label{eq:def-f_B-K}
  K^{-1}(\mu) &
  = 1 +
    \frac{\alS(\mu) C_F}{4\pi} \left(\frac{3}{2} \ln \frac{\mu^2}{m_b^2} + 2 \right) .
\end{align}
The RG evolution factor $U_F^{-1}$ from $\mu_{hc}$ to $\mu_h$ can be
obtained from the Appendix of~\cite{Beneke:2011nf} with the identification
$U_F^{-1}(\mu_{hc},\mu_h) = U_2^{-1}(\mu_h,\mu_{hc})$ in terms of the
evolution factor defined there.

Finally, the hard-collinear matching function (``jet function'') reads
\cite{Wang:2016qii}
\begin{align}
  J(n\cdot r, r^2, \omega; \mu) = \; &
  1 + \frac{\alS C_F}{4\pi}\, \Bigg\{
    \ln^2 \frac{\mu^2}{n\cdot r \, (\omega - \oL{n}\cdot r)}
    - \frac{\pi^2}{6} - 1
\nonumber\\
  & - \,\frac{\oL{n}\cdot r}{\omega} \ln \frac{\oL{n}\cdot r - \omega}{\oL{n}\cdot r}
      \left[ \ln \frac{\mu^2}{-r^2}
           + \ln \frac{\mu^2}{n\cdot r \, (\omega - \oL{n}\cdot r)}
           + 3 \right]
  \Bigg\}
  \label{eq:def-jet-func}
\end{align}
with $\oL{n}\cdot r = r^2/n\cdot r$. We note that the second line vanishes
for $r^2=0$. Hence for the $A$-type insertions, the convolution $J(\omega)
\otimes \phi_+(\omega)$ in \refeq{eq:matchtoSCETII} can be expressed as
\begin{align}
  & \int_0^\infty \frac{d\omega}{\omega} \, 
  J(2 E_\gamma, 0, \omega; \mu)\, \phi_+(\omega; \mu)
\nonumber \\
  \label{eq:jet-func-conv}
  & = \, \frac{1}{\lamBq(\mu)} \Bigg[
    1 + \frac{\alS(\mu) C_F}{4 \pi} \left(
       \ln^2 \frac{2 E_\gamma \mu_0}{\mu^2}
     - 2 \sigNBq{1}(\mu) \ln \frac{2 E_\gamma \mu_0}{\mu^2}
     + \sigNBq{2}(\mu) - \frac{\pi^2}{6} - 1
  \right) \Bigg] \qquad
\end{align}
in terms of the inverse ($\lamBq$) and the first two inverse-logarithmic
moments ($\sigNBq{1,2}$)
\begin{align}
  \label{eq:def-B-LCDA-moments}
  \sigNBq{n}(\mu) &
  = \int_0^\infty \!\! d\omega \, \frac{\lamBq(\mu)}{\omega}
    \ln^n \frac{\mu_0}{\omega} \, \phi_+(\omega,\mu) ,
&
  \sigNBq{0} & \equiv 1
\end{align}
of the $B$-meson LCDA \cite{Beneke:2011nf}.
However, for the $B$-type insertions the expression is more complicated and
the entire function $\phi_+(\omega)$ must be known to evaluate
the convolution integral.

%
%
\subsubsection{Final factorized form}
\label{sec:finalff}

Putting together \refeq{eq:amplSCETItypeA}, \refeq{eq:amplSCETItypeB} and
\refeq{eq:matchtoSCETII}, we obtain the following compact result for the
LP $\Bqtollgamma$ amplitude \refeq{eq:ampl-Ti}:
\begin{eqnarray}
\oL{\calA}_{{\rm LP}} &=&
i e \,\frac{\alE}{4\pi} \,\normEW\, \eps^\star_\mu \,
(g_\perp^{\mu\nu}+i\varepsilon_\perp^{\mu\nu}) \,
\frac{Q_q F_{B_q} E_\gamma}{2}\,
\Bigg\{
\nonumber\\
&&
\left\{
\left(V_9^{\rm eff}(q^2) +
\frac{2\, \oL{m}_b\, m_{B_q}}{q^2} \, V^{\rm eff}_7(q^2)\right)
 L_{V,\nu} +
V^{\rm eff}_{10}(q^2) L_{A,\nu}\right\}
\frac{ m_{B_q}}{2E_\gamma} \,
\int_0^\infty \!\frac{d\omega}{\omega}\, \phi_+(\omega)
\,J(2 E_\gamma,0,\omega)
\nonumber\\
&& + \, \frac{2\, \oL{m}_b\, m_{B_q}}{q^2} \,
V^{\rm eff}_7(0) L_{V,\nu}
\,
\int_0^\infty \!\!d\omega\, \phi_+(\omega)
\frac{J(m_{B_q},q^2,\omega)}{\omega - q^2/m_{B_q} - i0^+}
\Bigg\}\,.
\label{eq:ampLPfinal}
\end{eqnarray}
Note that the amplitude contains
$g_\perp^{\mu\nu} + i\veps_\perp^{\mu\nu}$ and not
$g_\perp^{\mu\nu} - i\veps_\perp^{\mu\nu}$. From \refeq{eq:Ti-FF-param}
this implies
\begin{equation}
  \label{eq:F_R-LP}
  F_R^{(i,\text{LP})} = 0 ,  \qquad
  i = 1, \ldots, 10 ,
\end{equation}
and the so-called form-factor-symmetry relation 
$F_V^{(i)} = F_A^{(i)}$  as a consequence of helicity conservation
in the heavy-quark and large-energy limit.

For completeness, we also give the left-handed form factors.
For this purpose we define
\begin{align}
  \label{eq:effFF-LP}
  \sum_{i=7,9} F_L^{(i\text{-eff}, \text{LP})} &
  \equiv \sum_{i=1}^9 \eta_i C_i F_L^{(i,\text{LP})} ,
&
  F_L^{(10\text{-eff}, \text{LP})} &
  \equiv \eta_{10} C_{10} F_L^{(10,\text{LP})}
\end{align}
and find
\begin{align}
  F_L^{(7\text{-eff},\text{LP})} = \; &
  \frac{Q_q F_{B_q}}{2}\,\frac{2\, \oL{m}_b\, m_{B_q}}{q^2} \,
  \Bigg\{ V^\text{eff}_7(q^2) \frac{m_{B_q}}{2 E_\gamma} \,
  \int_0^\infty \! d\omega \, \phi_+(\omega) \, \frac{J(2 E_\gamma, 0, \omega)}{\omega}
\nonumber\\
  & + V^\text{eff}_7(0) \, \int_0^\infty \! d\omega \, \phi_+(\omega)
  \frac{J(m_{B_q}, q^2, \omega)}{\omega - q^2/m_{B_q} - i0^+}
  \Bigg\} \,,
\label{eq:finalFF-7eff}
\\[0.2cm]
  F_L^{(i\text{-eff},\text{LP})} = \; &
  \frac{Q_q F_{B_q}}{2}\,V^{\rm eff}_{i}(q^2)\,
  \frac{ m_{B_q}}{2E_\gamma} \,
  \int_0^\infty \! d\omega \, \phi_+(\omega) \, \frac{J(2 E_\gamma, 0, \omega)}{\omega}
  \qquad i = 9,10\,.
\label{eq:finalFF-9eff}
\end{align}
We recover the result for $\Butolvgamma$ \cite{Beneke:2011nf} from
\eqref{eq:finalFF-9eff}, by setting $C_9 = 1$ and $C_i=0$ otherwise.

%
%
%
\subsection{Form factors at NLP}
\label{sec:FF-NLP}

As for the case of $\Butolvgamma$ \cite{Beneke:2011nf} we include NLP
$\LambQCD/E_\gamma$, $\LambQCD/m_b$ corrections to the above
$\Bqtollgamma$ form factors, but we aim only at $\mathcal{O}(\alS^0)$
accuracy at NLP. Such power corrections arise from three sources:
\begin{itemize}
\item The coupling of the real or virtual photon to the heavy quark.
\item Power corrections to the (anti-) hard-collinear light-quark propagator
  in the LP  $\mathcal{O} (\alS^0)$ contributions, that is, power corrections to
  the \SCETI{} correlation function (\ref{eq:SCETIcorrelation}) at tree level.
\item Annihilation-type insertions of the four-quark operators, such that the
  real and the virtual photon are attached to the quark loop, see the right
  diagram in \reffig{fig:fourquarkLOcontractions}.
\end{itemize}
The first two effects are also present in $\Butolvgamma$. The
chromomagnetic dipole operator $\Op_8$ can be ignored, since
its insertions involve at least one power of $\alS$.

The NLP contributions from the semileptonic operators $\Op_{i}$ ($i=9,10$)
can be taken directly from the $\Butolvgamma$ calculation \cite{Beneke:2011nf}:
\begin{align}
  \label{eq:FL-9_10-NLP}
  F_L^{(i,\text{NLP})} & = \frac{\xi_{B_q}(E_\gamma)}{2} ,
\\
  F_R^{(i,\text{NLP})} & = \frac{f_{B_q}}{4 E_\gamma} \left(
    Q_b \frac{m_{B_q}}{m_b} + Q_q \frac{m_{B_q}}{2 E_\gamma}
  \right) .
\end{align}
At NLP the right-helicity form factor is non-vanishing, but this
``symmetry-breaking'' contribution (as it implies $F_V \neq F_A$)
is local at  $\mathcal{O}(\alS^0)$. By this we mean that 
at this order they can be expressed in terms
of the $B$-meson decay constant. On the other hand, the power
correction to the left-helicity form factor cannot be factorized
and is parametrized by an unknown function, the
``symmetry-preserving'' soft form factor. We also refer to 
these contributions as ``non-local'' power corrections. 
For $q=u$ relevant
to  $\Butolvgamma$ it has been calculated with QCD sum
rules~\cite{Braun:2012kp, Wang:2016qii, Beneke:2018wjp}. The
definition in \refeq{eq:FL-9_10-NLP} is such that in the SU(3)-flavour
symmetry limit of QCD  $\xi_{B_q} = Q_q/Q_u \times \xi_{B_u}$
($q = d, s$) is related to the one introduced in
\cite{Beneke:2011nf} for $\Butolvgamma$ simply by the ratio of the
electric charges of the spectator quarks. We remark that we
consistently set the strange-quark mass to zero in our analysis,
which would otherwise give $m_s/(2 E_\gamma)$ corrections to the
above expressions.

The NLP contributions of $\Op_7$ from the $A$-type insertion are
\begin{align}
  \label{eq:FL-7A-NLP}
  F_L^{(7A,\text{NLP})} &
  = - \frac{\oL{m}_b \, m_{B_q}}{q^2} \left( 
      \xi_{B_q}(E_\gamma) + Q_b \frac{f_{B_q}}{2 E_\gamma} 
    \right) ,
\\
  \label{eq:FR-7A-NLP}
  F_R^{(7A,\text{NLP})} & 
  = - Q_q \frac{\oL{m}_b f_{B_q}}{(2 E_\gamma)^2} \, .
\end{align}
We note that the emission of the photon from the heavy quark
is symmetry-preserving for $\Op_7$. For the $B$-type insertion of
$\Op_7$ we find
\begin{align}
  \label{eq:FL-7B-NLP}
  F_L^{(7B,\text{NLP})} &
  = - \frac{\oL{m}_b \, m_{B_q}}{q^2} \left(
    \widetilde{\xi}_{B_q}(E_\gamma) + Q_b \frac{f_{B_q}}{2 E_\gamma}
  \right),
\\
  F_R^{(7B,\text{NLP})} & = 0 \,.
\end{align}
The parameterization of the NLP $B$-type insertion requires
another soft form factor $\widetilde{\xi}_{B_q}(E_\gamma)$.
The \SCETI{} correlation function \refeq{eq:SCETIcorrelation}
can be considered as a function of $n\cdot r$ and $r^2$, where
$n\cdot r$ is the large component of the (anti-)collinear momentum
$r$. To parametrize the NLP correction, we may introduce
the soft form factor $\zeta_{B_q}(n\cdot r, r^2)$ of two variables.
Similar to the general hard-collinear function 
$J(n\cdot r, r^2, \omega)$ in \refeq{eq:matchtoSCETII}, the 
soft-form factors in \refeq{eq:FL-7A-NLP},
\refeq{eq:FL-7B-NLP} are then the special cases
\begin{align}
  \label{eq:xi-zeta}
  \xi_{B_q}(E_\gamma) &
  = \zeta_{B_q}(2 E_\gamma, 0),
&
  \widetilde{\xi}_{B_q}(E_\gamma) &
  = \zeta_{B_q}(m_{B_q}, m_{B_q} (m_{B_q}-2 E_\gamma)) \,.
\end{align}

\begin{figure}
\centering
  \includegraphics[width=0.24\textwidth]{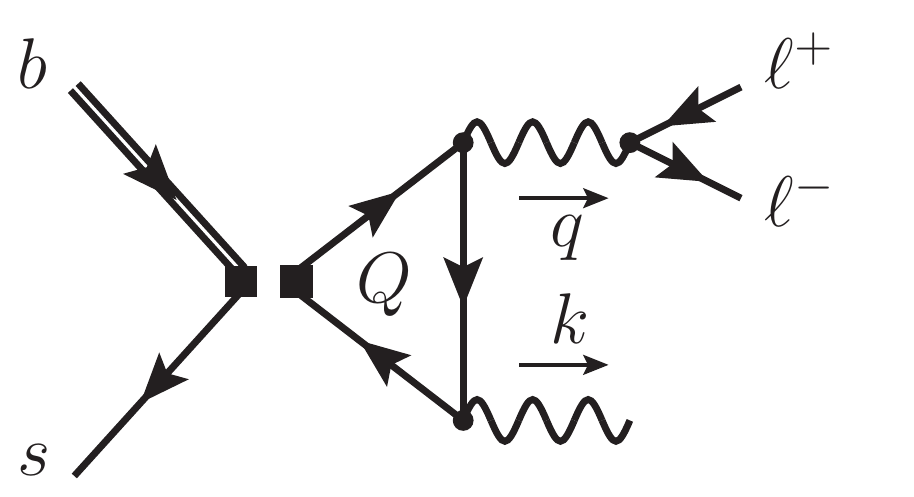}
  \includegraphics[width=0.24\textwidth]{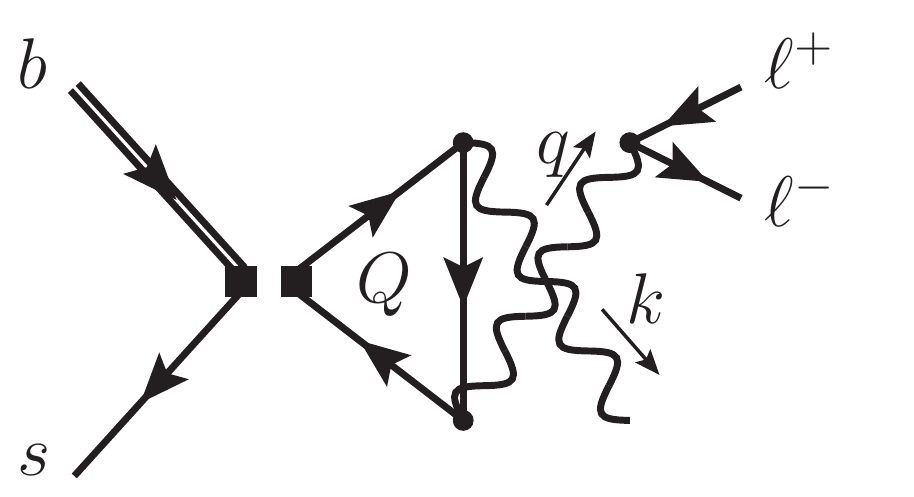}
  \includegraphics[width=0.24\textwidth]{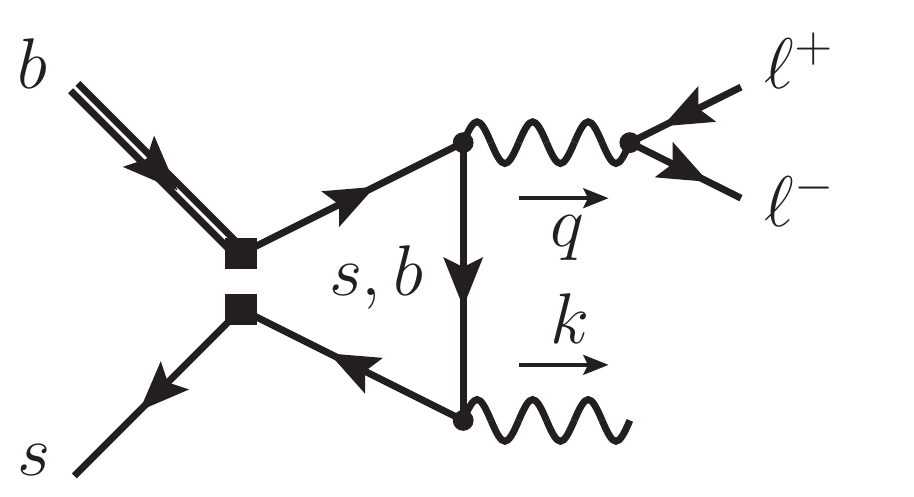}
  \includegraphics[width=0.24\textwidth]{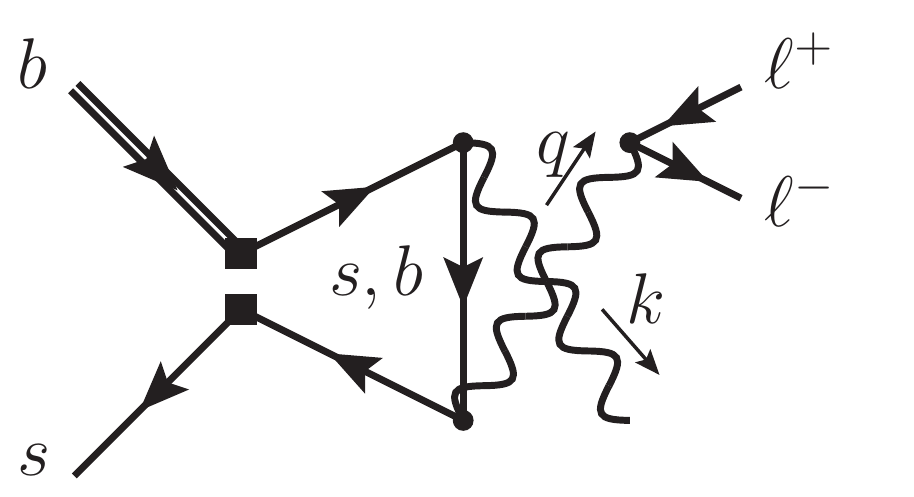}
\caption{\small Annihilation-type one-loop diagrams with insertions
  of the four-quark operators $\Op_{1,2}^{(u,c)}, \Op_{3-6}$.
  The operator insertion is depicted by black squares.
  The $Q$ stands for the five quark flavours $u,d,s,c,b$.}
\label{fig:ME-Op_3456}
\end{figure}

The effect of the four-quark operators at $\mathcal{O}(\alS^0)$
is two-fold. First, the NLP terms from the left and middle diagram
of \reffig{fig:fourquarkLOcontractions} replace
$C_7$ and $C_9$ multiplying $F_{L,R}^{(7A,\text{NLP})}$,
$F_{L,R}^{(7B,\text{NLP})}$,  $F_{L,R}^{(9,\text{NLP})}$ above
by $C_7^\text{eff}$ and $C_9^\text{eff}$. Second, they allow
annihilation-type contractions (right diagram in
\reffig{fig:fourquarkLOcontractions}), which appear only
at NLP, and must be computed separately. The complete set of
such diagrams is depicted in \reffig{fig:ME-Op_3456}.
Contributions of this type were computed for the case of
two real photons, $\Bqtogammagamma$ in \cite{Bosch:2002bv}.
The authors of this paper also showed that no infrared singularities
appear in the two-loop $\mathcal{O}(\alS)$ corrections to these
diagrams, hence allowing for a quantitative interpretation of
the lowest-order one-loop diagrams.

We computed the one-loop annihilation contribution for the
$\Bqtollgamma$ situation of one virtual and one real photon
attached to the quark loop. As expected, the result is local
such that the nonperturbative hadronic physics can be expressed
in terms of $f_{B_q}$. We express the result in the form
\begin{align}
  \label{eq:ampl-NLP-4qu}
  F_{L,R}^{(\text{4q,NLP})} \equiv
  \sum_{i = 1}^6 \eta_i C_i F_{L,R}^{(i,\text{NLP})} &
  = \frac{m_{B_q}^2}{q^2} \frac{f_{B_q}}{2 \, E_\gamma }
    \big[f_V(y) \mp f_A(y) \big] \,,
\end{align}
where the minus (plus) sign refers to $L$ ($R$). The functions $f_{V,A}(y)$
read
\begin{align}
  f_V(y) = \; & 
  Q_u^2 \, \left(C_F C_1 + C_2 \right)
    \left[ \left( 1 + \frac{\lamCKM{u}{q}}{\lamCKM{t}{q}} \right) 
\left[y + 2 z_c \, \Delta C_0(z_c, y)\right] -  \frac{\lamCKM{u}{q}}{\lamCKM{t}{q}} \,y \right]
\nonumber \\
  & + \,2 \, Q_d^2 \left(C_F C_4 + C_3 \right)
    \big[ \; y \, + \sum_{q=b,s} (z_q - \sqrt{z_q}) \, \Delta C_0(z_q, y)\big]
\nonumber \\
  & + \, 8 \, Q_d^2 \left(C_F C_6 + C_5 \right)
    \big[ 4 y + \sum_{q=b,s}  (4 z_q - \sqrt{z_q})  \Delta C_0(z_q, y) \big]
\nonumber \\
  \label{eq:def-A_perp}
  & + \, 12 N_c C_5 \sum_Q Q_Q^2  \big[ y + 2 z_Q \, \Delta C_0(z_Q, y) \big] \,,
\\
  f_A(y) = \; &
  Q_d^2 \; \big[C_F (C_4 + 4 C_6) + C_3 + 4 C_5 \big]
\nonumber \\
  \label{eq:def-A_para}
  & \times \,
  \sum_{q=b,s} \frac{\sqrt{z_q}}{y} \big[
    4 y
  + 2 (4 z_q - y) \Delta C_0(z_q, y)
  - 4 (1 - y) \Delta B_0 (z_q, y) \big] \,.
\end{align}
Here $y = 2 E_\gamma/m_{B_q}$ as before.  We further introduced the 
mass ratio $z_Q = m_Q^2/m_{B_q}^2$, where $Q = u,d,s,c,b$ denotes 
the various quark flavours. In the numerical
evaluation we set $m_u = m_d = m_s = 0$. The contribution from
$\Op_{1,2}^u$ is included in $f_V(y)$ through the second 
term proportional to $\lamCKM{u}{q}$. The loop functions are
\begin{align}
  \label{eq:def-DeltaB0,C0}
  \Delta B_0(z, y) &
  \equiv B_0\left(\frac{1}{z} \right) - B_0\left(\frac{1-y}{z} \right), 
&
  \Delta C_0(z, y) &
  \equiv C_0\left(\frac{1}{z} \right) - C_0\left(\frac{1-y}{z} \right) ,
\end{align}
with
\begin{align}
  \label{eq:def-B0-C0}
  B_0(x) &
  \equiv 2 \sqrt{\frac{4}{x}-1} \arctan \left(\frac{1}{\sqrt{\frac{4}{x}-1}}\right) - 2 ,
&
  C_0(x) &
  \equiv - 2 \arctan^2 \left(\frac{1}{\sqrt{\frac{4}{x}-1}}\right) .
\end{align}
Numerically, we find $|f_A| < 0.05 |f_V|$ for $q^2 < 8\GeV^2$, such that
approximately $F_L^{(\text{4q,NLP})} \approx F_R^{(\text{4q,NLP})}$.

%
%
%
\subsection{Discontinuity, duality and validity of 
the factorization approach}
\label{sec:duality}

To proceed to the numerical predictions for the $\Bqtollgamma$ decay rates, we 
need a model for the generalized soft form factor $\zeta_{B_q}(n\cdot r, r^2)$,
or the two single-variable form factors in~\eqref{eq:xi-zeta}. The form factor 
$\xi_{B_q}(E_\gamma) = \zeta_{B_q}(2 E_\gamma, 0)$ that appears in $A$-type
contributions could be computed with QCD sum rules as for $\Butolvgamma$, for
which sophisticated results including radiative corrections and higher-twist
effects already exist \cite{Braun:2012kp, Wang:2016qii, Wang:2018wfj,
Beneke:2018wjp}. This method applies to the transition to a real photon or a
photon with Euclidean virtuality $k^2<0$, but not to the case $q^2 = m_{B_q}
(m_{B_q}-2 E_\gamma)$ relevant to $\widetilde{\xi}_{B_q}(E_\gamma) = 
\zeta_{B_q}(m_{B_q}, m_{B_q} (m_{B_q}-2 E_\gamma))$ in the $B$-type contribution.
This form factor develops an imaginary part and resonances, which cannot be
fully described with large-energy factorization methods. In the following we
provide some general considerations on the factorization calculation of $B\to
\gamma^*$ form factors in the physical region $q^2>0$. We take note of a recent 
computation of these form factors with QCD sum rules \cite{Albrecht:2019zul},
which however applies to larger $q^2$ than of interest here. 

We first note that the $\Bqtollgamma$ decay amplitude contains a discontinuity
already at LP from two sources. First, in the $A$-type contribution from
$V_{7,9}^{\rm eff}(q^2)$, in lowest order given by the cut through the quark
loop in the left-most diagram of \reffig{fig:fourquarkLOcontractions}. This
discontinuity is similar to the one for $B_q\to V\ell\bar\ell$ and its implications
for short-distance and factorization calculations are well understood. For the
$b\to s$ transition it leads to prominent charmonium resonances in the
$q^2$-spectrum and large global parton-hadron duality violation for integrated
or binned spectra, which limit the short-distance calculation to $q^2 \lesssim
6 \GeV^2$. On the other hand, the light-meson resonances related to light-quark
loops cause negligible amounts of parton-hadron duality violation when
sufficiently wide bins in $q^2$ are considered, as explained in
\cite{Beneke:2009az}. The second source of a discontinuity appears in the 
$B$-type contribution from real intermediate states in the correlation function
$\mathcal{T}^{\mu\nu}(q)$ \eqref{eq:SCETIcorrelation} in the physical region
$q^2 > 0$ (see right diagram in \reffig{fig:AB-type}).
In the following we are concerned with this discontinuity.

From \eqref{eq:ampLPfinal} or \eqref{eq:finalFF-7eff} we obtain 
\begin{eqnarray}
\frac{1}{2i}\,\mbox{disc}\,F_L^{(7\text{-eff},\text{LP})} &=& 
\frac{Q_q F_{B_q}}{2}\,\frac{2\, \oL{m}_b\, m_{B_q}}{q^2} \,
  V^\text{eff}_7(0) \, \int_0^{\,q^2/m_{B_q}} 
  \hskip -0.7cm d\omega \, \phi_+(\omega) \, 
  \frac{1}{2i}\, \mbox{disc}
  \frac{J(m_{B_q}, q^2, \omega)}{\omega - q^2/m_{B_q} - i0^+}
\nonumber\\
&\approx&
\pi\,\frac{Q_q F_{B_q}}{2}\,\frac{2\, \oL{m}_b\, m_{B_q}}{q^2} \,
  V^\text{eff}_7(0) \,\phi_+(\omega_*)_{|\omega_* = q^2/m_{B_q}}\,,
\label{eq:Btypedisc}
\end{eqnarray}
where we used that the discontinuity is restricted to $\omega < q^2/m_{B_q}$
to set the upper integration limit. The second line holds in the tree-level
approximation $J(m_{B_q}, q^2, \omega) = 1$.  Since $\phi_+(\omega) \propto
\omega$ for $\omega\to 0$, the discontinuity survives as $q^2\to 0$, but is
negligible relative to the real part of $F_L^{(7\text{-eff},\text{LP})}$,
which develops the photon pole. This partonic discontinuity must be interpreted
as dual to a large number of continuum hadronic intermediate states in the
sense of parton-hadron duality. When $q^2 \sim E_\gamma \LambQCD$ is
hard-collinear or larger, this is certainly the case, since the invariant
mass of the hadronic intermediate state becomes parametrically large in the
heavy-quark/large-energy limit. On the other hand, for small $q^2 \sim 
\LambQCD^2$, the partonic discontinuity becomes unreliable. Note however,
that it also becomes power-suppressed, since $\omega_* \sim \LambQCD^2/m_b$
when $q^2\sim\LambQCD^2$, hence $\phi_+(\omega_*) \sim 1/m_b$ as opposed to
$\phi_+(\omega_*) \sim 1/\LambQCD$ for hard-collinear $q^2$. In reality,
these parametric estimates do not work well. For example, when $q^2 \sim
m_\phi^2$ is near the $\phi$ meson resonance, $m_\phi$ formally counts as
$\LambQCD$, but $\omega_* \sim 200\,$MeV is closer to the QCD scale $\LambQCD$
than to $\LambQCD^2/m_b$.

We next compare the leading-power $B$-type contribution $\oL{\calA}_{\text{type}-B}$
defined in \refeq{eq:amplSCETItypeB} to the amplitude generated by saturating
the hadronic tensor with a single vector resonance $V$ of mass $m_V$ and width
$\Gamma_V$. Employing a standard spectral representation of $T^{\mu\nu}_{7B}(k,q)$
\eqref{eq:T_7B}, we obtain 
\begin{eqnarray}
  \oL{\calA}_\text{res} &=&
  i e \,\frac{\alE}{4\pi} \,\normEW\, \eps^\star_\mu \,
  (g_\perp^{\mu\nu} + i \veps_\perp^{\mu\nu}) \,
  \frac{m_{B_q}}{2} \,
  \frac{4 \oL{m}_b E_\gamma}{q^2} \,
  V^\text{eff}_7(0) L_{V,\nu}\,
  \frac{c_V f_V m_V T_1^{B_q\to V}(0)}{m_V^2 - i m_V\Gamma_V - q^2}
  \,.\quad
  \label{eq:ampRes}
\end{eqnarray}
The vector meson decay constant and $B\to V$ transition form factors 
are defined by 
\begin{equation}
  \bra{0} \bar q \gamma^\mu q \ket{V(p^\prime, \varepsilon)}
  = i a_V^{(q)} f_V^{(q)} m_V \,\varepsilon^\mu(p^\prime)
\end{equation}
and
\begin{align}
  \bra{V(p',\veps^\ast)} \bar q \sigma^{\mu\nu}q_\nu b \ket{\bar{B}_q(p)} &
  = -2\,a_V^{(q)} \, T^{B_q\to V}_1(q^2)\,\veps^{\mu\nu\rho\sigma}
    \veps^{\ast}_\nu\, p_\rho p^{\prime}_\sigma,
\nonumber\\
  \bra{V(p',\veps^\ast)} \bar q \sigma^{\mu\nu} \gamma_5 q_\nu b \ket{\bar{B}_q(p)} &
  = (-i)\,a_V^{(q)} \, T^{B_q\to V}_2(q^2) \left[
      (m_{B_q}^2-m_V^2) \,\veps^{\ast\mu}
     -(\veps^\ast\cdot q)\,(p^\mu + p^{\prime\,\mu}) \right]
\nonumber\\
  \hspace*{2cm} + \, (-i)\, a_V^{(q)} & \, T^{B_q\to V}_3(q^2) \,
  (\veps^\ast\cdot q)
  \left[q^\mu - \frac{q^2}{m_{B_q}^2 - m_V^2}(p^\mu + p^{\prime\,\mu}) \right],
\label{eq:ffdef}
\end{align}
respectively, with $q=p-p^\prime$. The factor $a_V^{(q)}$ arises 
from the quark flavour wave function of the resonance. For 
the cases of interest below---the $\phi$, $\rho$ and $\omega$ 
mesons---the non-vanishing constants are $a_\phi^{(s)}=1$, 
$a_\rho^{(u)} = -a_\rho^{(d)} = 1/\sqrt{2}$, and 
$a_\omega^{(u)} = a_\omega^{(d)} = 1/\sqrt{2}$. The constant 
$c_V$ in (\ref{eq:ampRes}) collects these flavour factors 
as well as the electric charges from the matrix element 
$\bra{0} j^\mu_f \ket{V(q, \veps)}$ of the electromagnetic current, 
\begin{equation}
  c_V = a_V^{(q)} \sum_{f=u,d,s} a_V^{(f)} Q_f \,,
\end{equation}
with values $-1/3, -1/2, 1/6$ for $V = \phi, \rho, \omega$. We further 
used $T_2(0) = T_1(0)$ to obtain  \refeq{eq:ampRes}. 
Eq.~(\ref{eq:ampRes}) should be compared to the first and last 
line of \eqref{eq:ampLPfinal}. Hence, we obtain the resonant 
amplitude from the LP $B$-type amplitude by the substitution
\begin{eqnarray}
\frac{Q_q F_{B_q}}{2}\,
\int_0^\infty \!\!d\omega\, \phi_+(\omega)
\frac{J(m_{B_q},q^2,\omega)}{\omega - q^2/m_{B_q} - i0^+}
\quad\to\quad 
  \frac{c_V f_V m_V T_1^{B_q\to V}(0)}{m_V^2-i m_V\Gamma_V - q^2}\,.
\label{eq:LPtoRes}
\end{eqnarray}
With $F_{B_q} \sim \LambQCD^{3/2}/m_b^{1/2}$ and 
$T_1^{B_q\to V}(0) \sim (\LambQCD/m_b)^{3/2}$, we find that 
the resonant amplitude is suppressed by 
$(\LambQCD/m_b)^2$ in the heavy-quark limit when $q^2$ is 
hard-collinear and by $\LambQCD^2/(m_b \Gamma_V)$ in the 
resonance region. This suggests that we can add the resonant 
amplitude to the parameterization of the soft form factor 
$\zeta_{B_q}(n\cdot q, q^2)$ without double counting 
of part of the short-distance contributions, since they 
are formally of lower order in the heavy-quark 
expansion.\footnote{The counting for the ratio of the imaginary parts is 
$\LambQCD/\Gamma_V$ in the resonance region. As mentioned 
above, the imaginary part of the left-hand side of 
(\ref{eq:LPtoRes}) is suppressed for small $q^2$. The 
real part is dominated by the subtraction constant in a 
once-subtracted dispersion relation.}

In order to address the question whether global duality 
is violated by the presence of resonances, we consider the 
ratio
\begin{equation}
R\equiv \int_{q_\text{min}^2}^{q_\text{max}^2} dq^2\,
\frac{d\Gamma_{\rm res}}{dq^2}\,
\mbox{\Huge$\mathbin{/}$}
\int_{q_\text{min}^2}^{q_\text{max}^2} dq^2\,
\frac{d\Gamma_{\rm LP}^{\text{type}-B}}{dq^2}
\end{equation}
of differential decay rates in a $q^2$-bin. Since we are 
only interested in an $\mathcal{O}(1)$ estimate, we evaluate 
the resonance contribution in the narrow-width approximation, 
and the short-distance contribution in the tree-level 
approximation $J=1$ for the hard-collinear function. 
We define the complex, $q^2$-dependent inverse moment 
\begin{equation}
\frac{1}{\lamB{q}(q^2)}\equiv 
\int_0^\infty \!\!d\omega\, \phi_+(\omega)
\frac{1}{\omega - q^2/m_{B_q} - i0^+}
\end{equation}
of the $B$-meson LCDA, and obtain 
\begin{align} 
  R &
  = 4\pi\,\frac{(c_V f_V  T_1^{B_q\to V}(0))^2 } {m_V\Gamma_V Q_q^2 F_{B_q}^2} 
  \left( \int_{q_\text{min}^2}^{q_\text{max}^2} 
    \frac{dq^2}{q^2} \,
    \left(1 - \frac{q^2}{m_{B_q}^2}\right)^{\!\!3/2} \!\!\!
    \frac{1}{\left|\lamB{q}(q^2)\right|^2} \,
    \right)^{\!\!-1}
\nonumber \\ &
  \approx 4\pi\,
  \left(\frac{c_V \lambda_{B_q}  T_1^{B_q\to V}(0)} {Q_q F_{B_q}}\right)^{\!\!2} 
  \!\times \frac{f_V^2}{m_V\Gamma_V}
  \times {\displaystyle \frac{1}{\ln\frac{q_\text{max}^2}{q_\text{min}^2}}}\,
\label{eq:Restimate}
\end{align}
provided $m_V^2\in [q_\text{min}^2,\, q_\text{max}^2]$. The second line
is obtained upon neglecting the $q^2$-dependence of $\lamB{q}(q^2)$, and
neglecting $q^2/m_{B_q}^2$ in the integrand. The above estimate allows us
to draw a number of important conclusions:
\begin{itemize}
\item 
Since the resonance is localized while the short-distance amplitude is
smooth, we would have expected the ratio to decrease as $1/q_\text{max}^2$
with the width of the integration interval for $q_\text{max}^2 \gg
q_\text{min}^2$. Instead the impact of the resonance on the integrated
decay spectrum decreases only logarithmically. This behaviour appears
because the amplitude shows the photon-pole $1/q^2$ enhancement. 
\item 
In the large-energy/heavy-quark limit $R\sim (\LambQCD/m_b)^2$, counting
$\Gamma_V\sim \LambQCD$; as expected the resonance is a sub-leading power
correction. However, $R$ can be strongly enhanced for narrow resonances.  
In \cite{Beneke:2009az} this mechanism has been identified as the source
of large global duality violation for the inclusive $B\to X_s\ell\bar\ell$
decay. Here we find (with parameters as specified in \refsec{sec:phenoanalysis}
and omitting the bin-size dependent logarithm in this estimate)
$R\approx 57$ for the $\phi$ resonance contribution to $B_s\to \gamma\ell
\bar\ell$, and $R \approx 2.8$ and $3.9$ for the $\rho$ and $\omega$ meson
contribution, respectively, to $B_d\to \gamma\ell\bar\ell$. Hence we
conclude that global duality is violated by a large factor in $B_s$ decay
due to the narrow width of the $\phi$ meson, and there is still an
$\mathcal{O}(1)$ contribution from the $\rho$ and $\omega$ resonance for
the $B_d$ case. The short-distance contribution is only dominant, if the
$q^2$ bin of $d\Gamma/dq^2$ does not include the resonance and is
sufficiently in the hard-collinear region.
\item 
Eq.~(\ref{eq:Restimate}) is similar to Eq.~(44) of \cite{Beneke:2009az},
except that the factor $f_V^2/m_b^2$ there is replaced by $(\lamB{q} 
T_1^{B_q\to V}(0)/F_{B_q})^2$ here. The first factor arises from the
production of the resonance from a local current, while here the resonance 
arises from the transition from an extended $B_q$ meson. Although both
ratios scale as $(\LambQCD/m_b)^2$ in the heavy-quark limit, the second
is larger by nearly two orders of magnitude, since the $B_q$-meson form
factors and decay constant do not satisfy the heavy-quark mass scaling 
at $m_b\approx 5\,$GeV. This explains why the $\rho$ resonance makes a
negligible contribution for the inclusive $B\to X_s\ell\bar\ell$
transition discussed in \cite{Beneke:2009az}, but is $\mathcal{O}(1)$
for  $B_d\to \gamma\ell\bar\ell$.
\end{itemize}
Since the prominent lowest-mass resonance(s) may dominate any 
bin of the $q^2$-distribution, which contains them, in particular 
for  $B_s\to \gamma\ell\bar\ell$, the 
short-distance calculation is valid only in a small $q^2$ 
region from approximately 2~GeV$^2$ to about 6~GeV$^2$ below 
the charmonium resonances. To extend the theoretical prediction 
to smaller $q^2$, we propose an ansatz for the soft NLP form 
factor that includes the 
lowest resonance(s). From the above discussion we deduce 
that this can be done without double counting, but the ansatz 
departs from the rigour of the short-distance calculation. 

%
%
%
\subsection{Ansatz for the soft NLP form factor}
\label{sec:FF-NLP-model}

The form factors $\xi_{B_q}(E_\gamma)$, 
$\widetilde{\xi}_{B_q}(E_\gamma)$ are suppressed by a single 
power of $\LambQCD/E_\gamma$ 
in the large-energy/heavy-quark limit. As mentioned 
above we add to these form factors a resonance contribution, 
which is formally suppressed by another power, 
to extend the local description of the $q^2$ spectrum into 
the low-$q^2$ region. We further draw on the observation from 
\cite{Beneke:2018wjp} that the power-suppressed soft 
form-factor contribution tends to have opposite sign to the 
leading-power contribution. We therefore subtract from 
$\xi_{B_q}(E_\gamma)$, $\widetilde{\xi}_{B_q}(E_\gamma)$
the leading-power contribution in the tree approximation 
multiplied by a parameter $r_\text{LP}$, which formally scales as 
$\LambQCD/E_\gamma$ and for which we adopt the value 
$r_\text{LP}=0.2\pm 0.2$. This leads to 
the ansatz
\begin{align}
  \label{eq:soft-FF-A-model}
  \xi_{B_q}(E_\gamma) &
  = \sum_V \frac{2 c_V f_V }{m_V} \, T_1^{B_q \to V}(q^2)
  - r_\text{LP} \times \frac{Q_q F_{B_q}}{\lamB{q}} 
    \frac{m_{B_q}}{2 E_\gamma} 
\,,
\\
  \label{eq:soft-FF-B-model}
  \widetilde{\xi}_{B_q}(E_\gamma) &
  = \sum_V \frac{2 c_V m_V f_V}{m_V^2-i m_V\Gamma_V-q^2}\, T_1^{B_q \to V}(0)
  - r_\text{LP} \times\frac{Q_q F_{B_q}}{\lamB{q}(q^2)}\,,
\end{align}
where $\lamB{q} = \lamB{q}(q^2=0)$ is the standard 
$q^2$-independent inverse moment of the $B$-meson LCDA. The 
first term in each expression involves the tensor form factor 
$T_1^{B_q \to V}(q^2)$ of $B_q \to V$ transitions, for which 
we will employ the combination of LCSR and lattice results 
from~\cite{Straub:2015ica} in the numerical evaluation below. 
For the form factor $\xi_{B_q}(E_\gamma)$, which appears in 
the $A$-type contribution, QCD sum rule calculations
\cite{Braun:2012kp, Wang:2016qii, Wang:2018wfj, Beneke:2018wjp} 
exist including radiative corrections and higher-twist 
effects. However, for a coherent approximation of both 
types of form factors, we choose the same ansatz here. The 
two expressions above can be considered as special cases  
of the same resonance+factorization ansatz for the 
generalized soft form factor $\zeta_{B_q}(n\cdot r, r^2)$. 
We also remark that the tensor form factor literally appears only 
for the $7A$ contribution to $F_L$. For $i=9,10$, the vector 
and axial-vector form factors would appear. However, all these 
form factors reduce to the universal, transverse vector 
meson form factor $\xi^{B_q \to V}_\perp(q^2)$ when radiative 
and power corrections are neglected
\cite{Charles:1998dr, Beneke:2000wa}. Within this approximation, 
it is consistent to use $T_1^{B_q \to V}(q^2)$ everywhere 
instead of $\xi^{B_q \to V}_\perp(q^2)$.

By default we only include the $\phi(1020)$ resonance in case of the
$B_s\to \gamma\ell\bar\ell$ decay. The higher $s\bar s$ resonances have large
masses and we assume that they can be subsumed in binned averages of the
short-distance contribution.\footnote{A comparison to including the $\phi(1680)$
and $\phi(2170)$ resonances explicitly will be done in the analysis section.}
For the $B_d\to \gamma\ell\bar\ell$ decay, we include $V = \rho, \omega$.

%
%
%
\section{Phenomenological analysis}
\label{sec:phenoanalysis}

The phenomenological analysis of $\Bqtollgamma$ for $q = s, d$ and $\ell = e, \mu$
will use the results of the form factors presented in the previous section. The
systematic factorization leads to a reduction of renormalization scheme dependencies 
when including NLO QCD corrections to the LP contribution. Further, the NLP
contributions allow to study the breaking of the form factor symmetry and provide
estimates of nonperturbative resonant and nonresonant contributions in the large-energy limit, i.e. the low-$q^2$ region. Many features of the factorization
approach can be investigated at the level of amplitudes, for which we refer to
\refsec{sec:Bqllgam-amp}. The observables of phenomenological interest and our
final results will be presented in \refsec{sec:Bqllgam-obs}. 

The numerical values of the SM and hadronic parameters, which will be used in
the analysis, are collected in \reftab{tab:num-input}. The strong coupling
$\alS(\mu)$ in the \msbar{} scheme is calculated from $\alS(m_Z)$ with $n_f = 5$
using three-loop evolution, including quark flavour threshold crossings at the
scale $\mu_4 = \mu_h$ (transition to $n_f = 4$) and $\mu_3 = 1.2\GeV$ ($n_f = 3$).
The Wilson coefficients $C_i$ of the weak EFT are calculated at the electroweak
matching scale  $\mu_W = 160\GeV$ and then evolved with the required accuracy
(following \cite{Bobeth:2003at, Huber:2005ig}) to the scale $\nu$ that we 
equate to the hard factorization scale $\mu_h$ in the $n_f = 5$ theory. Their
values at the central scale $\nu = \mu_h = 5.0\GeV$ are $C_i^\text{NLL} =
(-0.294,\, 1.004,\, -0.004,\, -0.081,\, 0.0003,\, 0.0009)_i$ for
$i = 1,\ldots,6$, $C_7^\text{eff,NLL} = -0.303$, $C_8^\text{eff,LL} = -0.136$,
$C_9^\text{NNLL} = 4.327$ and $C_{10}^\text{NNLL} = -4.262$.
The central value of the hard-collinear scale is set to $\mu_{hc} = 1.5\GeV$.
The RG resummation in \SCETI{} between $\mu_h$ and $\mu_{hc}$ is done
in the $n_f = 4$ theory, see $U_H$ in \refeq{eq:H-RGE} and $U_F$ in 
\refeq{eq:rel-FB-and-fB}. The same holds for the factor $\alS(\mu_{hc})$ 
in the NLO QCD correction to the jet function in \refeq{eq:def-jet-func}.

\begin{table}
\centering
\renewcommand{\arraystretch}{1.5}
\resizebox{\columnwidth}{!}{
\begin{tabular}{|lll|lll|}
\hline
  Parameter
& Value
& Ref.
&  Parameter
& Value
& Ref.
\\
\hline \hline
  $G_F$                            & $1.166379 \cdot 10^{-5}$ GeV$^{-2}$  & \cite{Zyla:2020zbs} 
& $m_Z$                            & $91.1876(21)$ GeV                    & \cite{Zyla:2020zbs} 
\\
  $\alS^{(5)}(m_Z)$                & $0.1181(11)$                         & \cite{Zyla:2020zbs} 
& $m_\mu$                          & $105.658\ldots$ MeV                  & \cite{Zyla:2020zbs} 
\\
$\alE^{(5)}(\nu=5.0\GeV)$          & $1/132.18$
  & 
& $m_t^\text{OS}$                  & $172.4(7)$ GeV                       & \cite{Zyla:2020zbs}
\\
\hline
  $\oL{m}_b(\oL{m}_b)$             & $4.198(12)$ GeV                      & \cite{Aoki:2019cca}
& $\oL{m}_c(3\GeV)$                & $0.988(7)$ GeV                       & \cite{Aoki:2019cca}
\\
  $m_b^\text{PS}(\mu_f= 2.0\GeV)$  & $4.52^{+0.01}_{-0.04}$ GeV           & \cite{Beneke:2014pta, Beneke:2016oox}
& $m_c^\text{PS}(\mu_f= 1.0\GeV)$  & $1.39(5)$ GeV                        &
\\
\hline
  $m_{B_s}$                        & $5366.88(17)$ MeV                    & \cite{Zyla:2020zbs} 
& $m_{B_d}$                        & $5279.64(13)$ MeV                    & \cite{Zyla:2020zbs} 
\\
  $f_{B_s}$                        & $230.3(1.3)$ MeV                     & \cite{Aoki:2019cca} 
& $f_{B_d}$                        & $190.0(1.3)$ MeV                     & \cite{Aoki:2019cca} 
\\
  $\tau_{B_s}$                     & $1.527(11)$ ps                       & \cite{Zyla:2020zbs} 
& $\tau_{B_d}$                     & $1.519(4)$ ps                        & \cite{Zyla:2020zbs} 
\\
  $\lamB{s}(\mu_0)$                & $400(150)$ MeV                       &
& $\lamB{d}(\mu_0)$                & $350(150)$ MeV                       &
\\
  $\hsigNB{s}{1}(\mu_0)$           & $0.0(0.7)$                           & \cite{Beneke:2018wjp}
& $\hsigNB{d}{1}(\mu_0)$           & $0.0(0.7)$                           & \cite{Beneke:2018wjp}
\\
\hline
  $\lambda$                        & $0.22650(48)$                   & \cite{Zyla:2020zbs}
& $\oL{\rho}$                      & $0.141^{+0.016}_{-0.017}$       &
\\ 
  $A$                              & $0.790^{+0.017}_{-0.012}$       &
& $\oL{\eta}$                      & $0.357(11)$                    &
\\
\hline
\end{tabular}
}
\renewcommand{\arraystretch}{1.0}
\caption{\label{tab:num-input}
  \small
  Numerical input values for parameters. The values of the bottom- and charm-quark
  masses in the \msbar{} scheme are averages of $N_f = 2+1+1$ lattice determinations
  from the FLAG group from \cite{Chakraborty:2014aca, Colquhoun:2014ica, Bussone:2016iua,
  Gambino:2017vkx, Bazavov:2018omf} and \cite{Carrasco:2014cwa, Alexandrou:2014sha,
  Chakraborty:2014aca, Bazavov:2018omf, Lytle:2018evc}.
  The $B_q$-meson decay constants $f_{B_q}$ are averages from the FLAG group for
  $N_f = 2+1+1$ from \cite{Bazavov:2017lyh, Bussone:2016iua, Dowdall:2013tga,
  Hughes:2017spc}. The combinations of CKM parameters have been determined from
  \cite{Charles:2015gya} for PDG 2020, which are very similar to \cite{Bona:2016dys}.
  The reference scale for the $B$-meson LCDA parameters $\lamB{q}$, $\hsigNB{q}{1}$
  is $\mu_0 = 1$~GeV. 
}
\end{table}

The bottom- and charm-quark masses, which enter the one- and 
two-loop matrix elements of the four-quark operators 
$\Op_{1,\ldots,6}$ through the hard functions in \refeq{eq:def-H_7}, 
\refeq{eq:def-H_9} and \refeq{eq:def-C9-eff}, are usually chosen 
to be the pole masses. 
The pole masses suffer from large uncertainties because the 
conversion from the very precisely known \msbar{} masses to the 
pole scheme does not converge when using one-, two, or three-loop 
expressions. For example we obtain 
$m_b^\text{OS} = 4.80(^{+14}_{-20})\GeV$ and $m_c^\text{OS} = 
1.67(^{+27}_{-20})\GeV$ from the corresponding \msbar{}-values in
\reftab{tab:num-input} with two-loop expressions and an uncertainty 
spanned by using one- and three-loop
ones. On the other hand, the \msbar{} masses are also not 
appropriate here, at least for the charm quark, since the 
hard functions then exhibit imaginary parts from unphysically 
small values of $4 \overline{m}_c^2$. A good compromise is the 
potential-subtracted~(PS) renormalization 
scheme~\cite{Beneke:1998rk}, since the PS mass has a well-behaved 
relation to the \msbar{} mass while being numerically closer 
to the physical thresholds. We therefore use the PS masses 
as parameters and perform the scheme conversion of the hard 
functions from the pole to the PS scheme for the 
masses. In practice, this has to be done only for the charm mass, 
such that the NLO corrections $F_{1,2}^{(9c)}$ 
\cite{Asatrian:2001de} acquire an additional term from the 
change of scheme of $m_c$ in $C_9^{\rm eff}(q^2)$. The bottom 
PS mass is then also used in the matching coefficient 
\refeq{eq:def-f_B-K}, \refeq{eq:def-CVA0}, \refeq{eq:def-CT1A0}, 
\refeq{eq:def-A_perp}, and \refeq{eq:def-A_para}. 

We shall see below that the parameters of the $B$-meson LCDA 
cause by far the largest theoretical uncertainty from 
input parameters. This is not surprising, since the related charged
current process $\Butolvgamma$ is usually advocated as a 
measurement of $\lamB{u} \approx \lamB{d}$. While  $\lamB{d}$ 
may therefore be known more precisely in the future, there is 
no obvious measurement of $\lamB{s}$. There also do not exist studies 
of SU(3) breaking for this quantity, which requires us to 
make an educated guess. The value assumed 
in \reftab{tab:num-input} is obtained from the fact that the 
$B$-meson LCDA represents the distribution of light-cone 
momentum of the light-quark in the meson. A quark with larger mass 
is expected to have a larger light-cone momentum on average. 
More details on the treatment of the $B$-meson LCDA in this 
analysis are provided in \refapp{app:B-LCDA}.

If not stated otherwise, below we provide results for 
$q = s$ and $\ell = \mu$, because $B_s\to\gamma\mu\bar\mu$ 
is experimentally most easily accessible at LHCb. 

%
%
%
\subsection[$\Bqtollgamma$ amplitude]
{\boldmath $\Bqtollgamma$ amplitude}
\label{sec:Bqllgam-amp}

The $\Bqtollgamma$ amplitude can be the parametrized in terms of four 
helicity amplitudes, 
\begin{equation}
  \label{eq:ampl-FFs}
\begin{aligned}
  \oL{\calA}(\barBqtollgamma)
  = i e \frac{\alE}{4\pi} \normEW \, E_\gamma \eps^\star_\mu & \,
  \Big[ \; (g_\perp^{\mu\nu} + i \veps^{\mu\nu}_\perp)
        \left( \oL{\calA}_{L V} [\bar{u} \gamma_\nu v] 
             + \oL{\calA}_{L A} [\bar{u} \gamma_\nu \gf v] \right)
\\ & \!\! - (g_\perp^{\mu\nu} - i \veps^{\mu\nu}_\perp)
        \left( \oL{\calA}_{R V} [\bar{u} \gamma_\nu v] 
             + \oL{\calA}_{R A} [\bar{u} \gamma_\nu \gf v] \right)
  \Big] \,,
\end{aligned}
\end{equation}
where $V$ and $A$ refer to the vector and axial-vector chirality structure 
of the lepton currents, respectively. With the aid of \refeq{eq:ampl-Ti}
and \refeq{eq:Ti-FF-param} the helicity amplitudes are given by
\begin{align}
  \label{eq:F_LR^i}
  \oL{\calA}_{h V} & = \sum_{i=1}^9 \eta_i C_i F_h^{(i)} ,
&
  \oL{\calA}_{h A} & = C_{10} F_h^{(10)} ,
& 
  h & = L,R 
\end{align}
with LP and NLP contributions according to \refeq{eq:F_h-exp}.
The axial-vector amplitudes $\oL{\calA}_{h A}$ depend only on a  
single Wilson coefficient, $C_{10}$.
The transversity amplitudes
\begin{align}
  \oL{\calA}_{\perp \chi} & = \frac{1}{\sqrt{2}}
    \left(\oL{\calA}_{L \chi} + \oL{\calA}_{R \chi} \right) , &
  \oL{\calA}_{\parallel \chi} & = \frac{1}{\sqrt{2}}
    \left(\oL{\calA}_{L \chi} - \oL{\calA}_{R \chi} \right) , &
  \chi = V, A,
\end{align}
have definite CP transformation properties facilitating the analysis 
of the time-dependence in neutral $B$-meson decays. 
The amplitude of the CP-conjugated decay $\calA(\Bqtollgamma)$ 
is given by replacing in \refeq{eq:ampl-FFs} 
\begin{align}
  \label{eq:ampl-CP}
  \oL{\calA}_{\perp \chi} & \to - e^{i \xi_{B_q}} \calA_{\perp \chi}[g \to g^*] , &
  \oL{\calA}_{\para \chi} & \to + e^{i \xi_{B_q}} \calA_{\para \chi}[g \to g^*] ,
\end{align}
with all complex-valued fundamental couplings $g$ (Wilson coefficients and
CKM elements) complex conjugated. The dependence on the phase of the 
CP transformation  $\mathcal{CP} \ket{\oL{B}_q} = 
e^{i \xi_{B_q}} \ket{B_q}$ of the  $B_q$-meson state will cancel in observables.

%
%
\subsubsection{Amplitudes at LP}
\label{sec:num-LP-amps}

We start with the LP contributions to the amplitudes $\oL{\calA}_{LV}$ and
$\oL{\calA}_{LA}$. (Recall that $\oL{\calA}_{RV} = \oL{\calA}_{RA} = 0$ at
LP.) In the evaluation of the LP amplitude \refeq{eq:ampLPfinal} we drop
systematically higher-order terms in $\alS$ than NLO. This concerns the
weak-EFT Wilson coefficients, the hard functions and the \SCETI{} evolution
matrix $U_H(\mu_h, \mu_{hc})$ in \refeq{eq:H-RGE}, the static HQET decay
constant $F_{B_q}(\mu_{hc})$ in \refeq{eq:rel-FB-and-fB}, and the convolution
of the jet function with the $B$-meson LCDA $\phi_+(\omega; \mu_{hc})$. 
Schematically, the NLO correction reads
\begin{align}
  F \cdot V \cdot \mathcal{J}\, \big|_\text{NLO} &
  \;\sim\; F^{(1)} \cdot V^{(0)} \cdot \mathcal{J}^{(0)} 
     + F^{(0)} \cdot V^{(1)} \cdot \mathcal{J}^{(0)} 
     + F^{(0)} \cdot V^{(0)} \cdot \mathcal{J}^{(1)}\,,
\end{align}
where each quantity $X = X^{(0)} + X^{(1)} + \mathcal{O}(\alS^2)$ 
is expanded in $\alS$ at its corresponding scale. In case of the 
evolution factors $U$, ``${}^{(0)}$'' means LL instead of NLL.

Previous studies of $B\to (X_s, K, K^*) + \ell\bar\ell$ \cite{Asatrian:2001de,
Beneke:2001at, Asatryan:2001zw, Beneke:2004dp} have shown that NLO QCD
corrections to these decays are sizeable, especially from the four-quark
operators $\Op_{1,2}^{c}$. Including them leads to a significant reduction
of the hard renormalization scale uncertainty. We therefore first look 
at the scale uncertainties at LO and NLO of our calculation of the 
$\Bqtollgamma$ mode. Here the LO approximation also implies only LL evolution
in $U_H$ and $U_F$, and the omission of $\mathcal{O}(\alpha_s)$ corrections
to the hard and jet functions. 

\begin{figure}
\centering
  \includegraphics[width=0.41\textwidth]{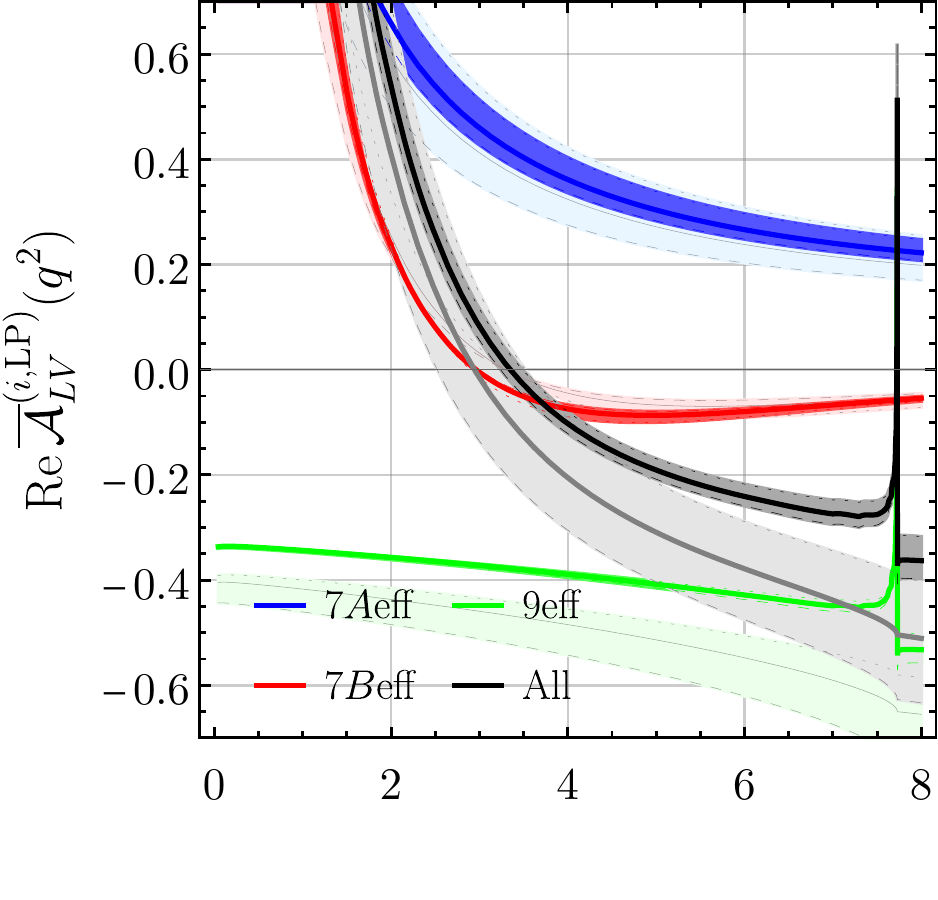}
  \hskip 0.05\textwidth
  \includegraphics[width=0.41\textwidth]{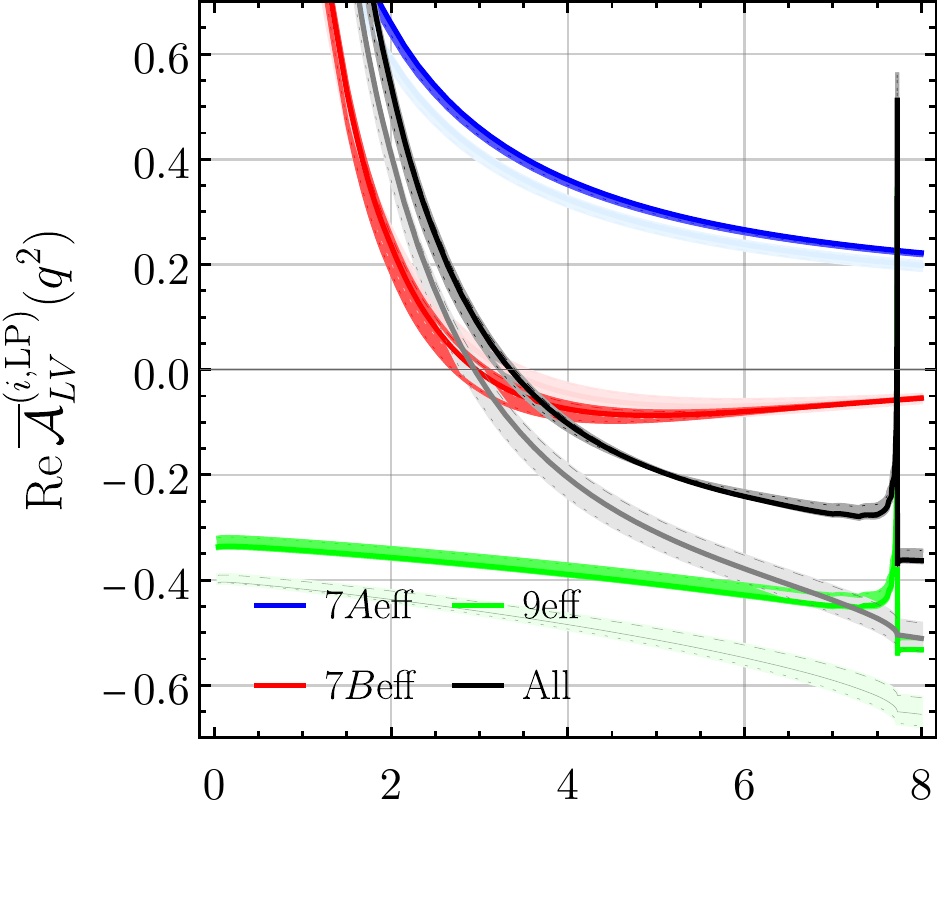}
\\[-0.3cm]
  \includegraphics[width=0.41\textwidth]{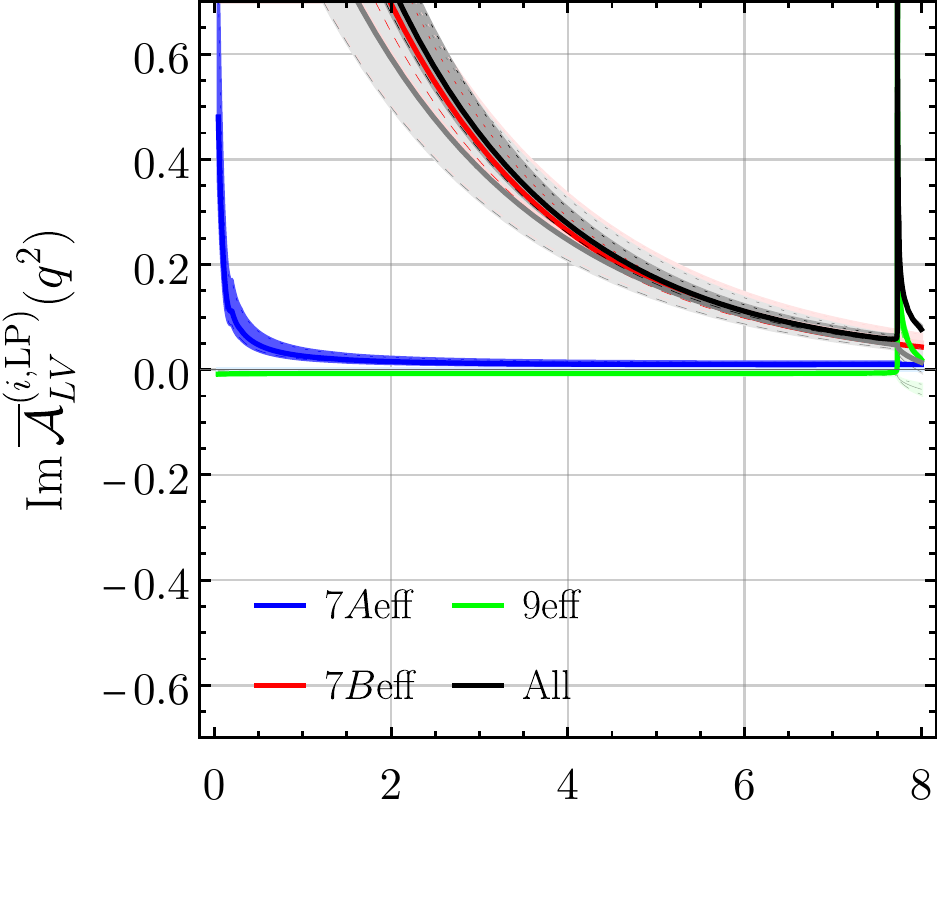}
  \hskip 0.05\textwidth
  \includegraphics[width=0.41\textwidth]{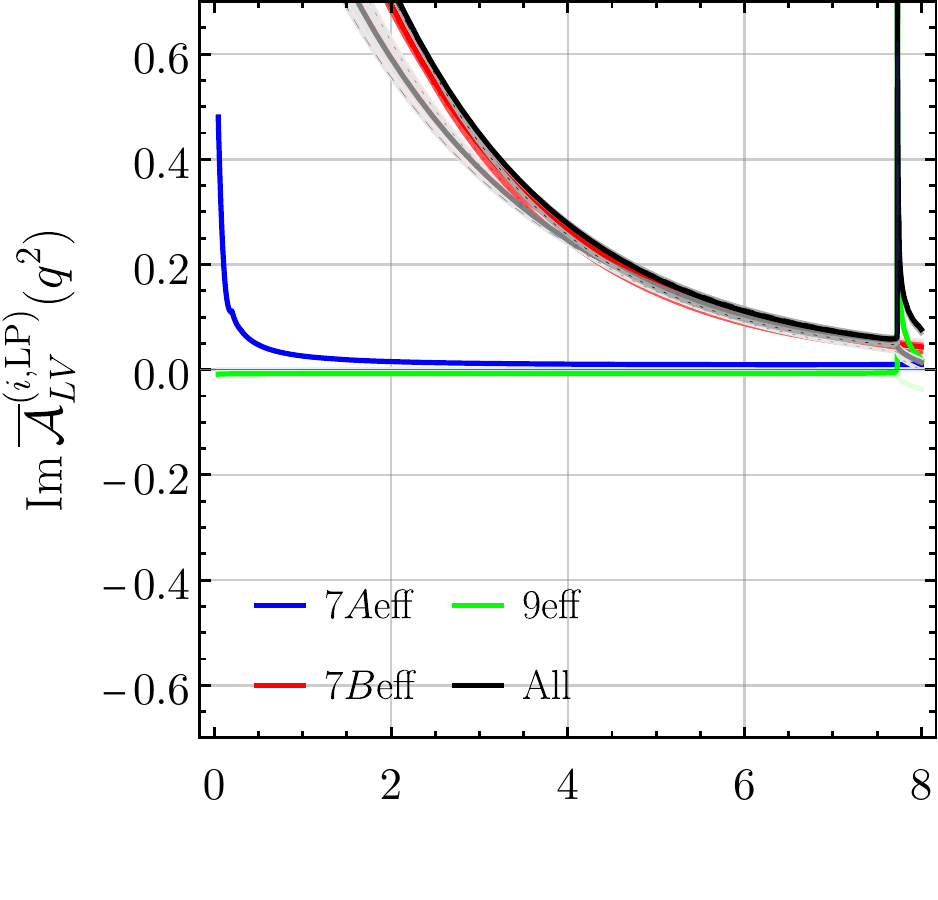}
\\ [-0.3cm] 
  \includegraphics[width=0.41\textwidth]{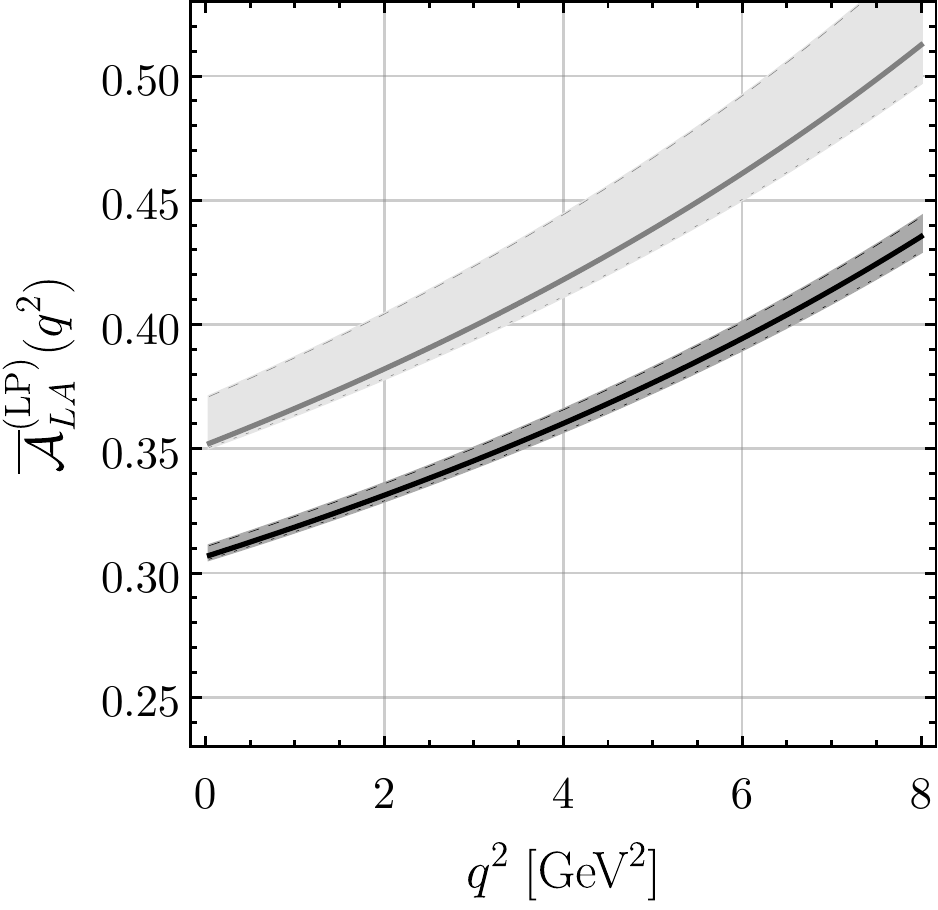}
  \hskip 0.05\textwidth
  \includegraphics[width=0.41\textwidth]{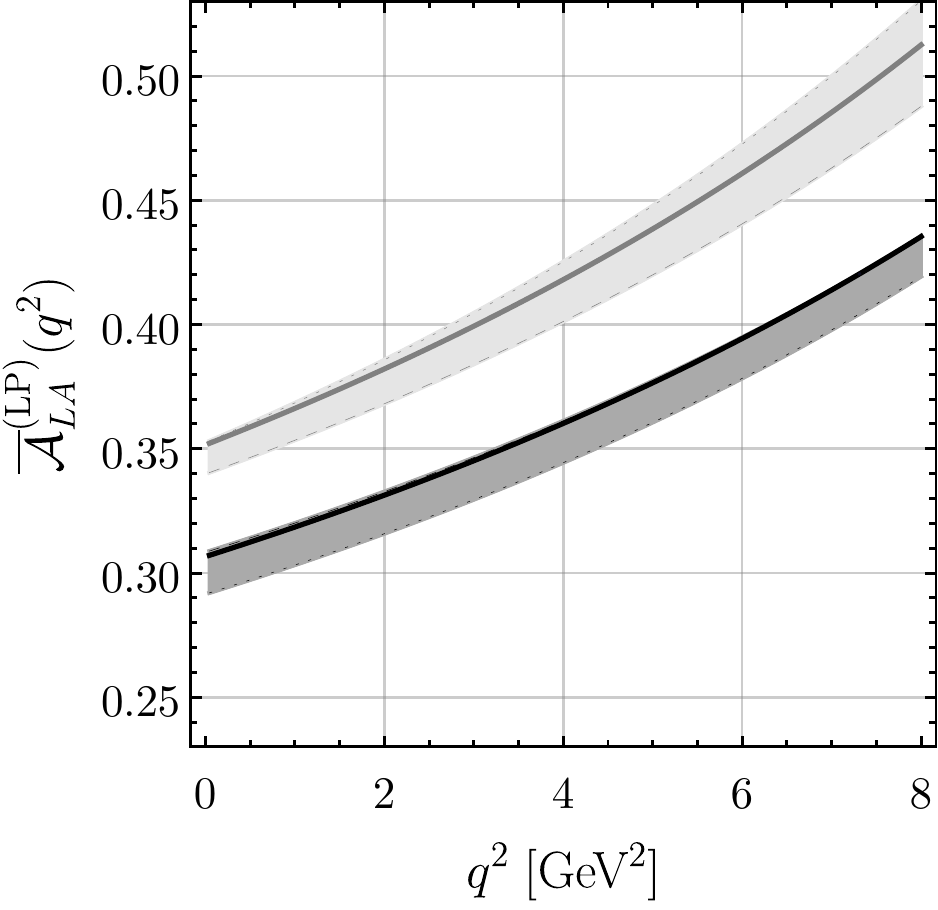}
\caption{\small The $\mu_h$ [left] and $\mu_{hc}$ [right] dependence of the
  helicity amplitudes $\re \oL{\calA}_{LV}$ [upper], $\im \oL{\calA}_{LV}$ 
  [middle] and $\oL{\calA}_{LA}$ [lower] at LP. The variation of 
  $\mu_h = \{2.5,\, 5.0,\, 10.0\} \GeV$ and  $\mu_{hc} = \{1.0,\, 1.5,\, 2.0\}\GeV$
  [dotted, solid, dashed] leads to the grey bands at LO QCD [lighter] and
  NLO QCD [darker]. The individual contributions $i = 7A, 7B, 9$ 
  [blue, red, green] are shown for $\oL{\calA}_{LV}$.
}
\label{fig:LP-amp-mu-dep}
\end{figure}

The reduction of the dependence on $\mu_h = \{2.5,\, 5.0,\, 10.0\}
\GeV$ upon going from LO to NLO is shown in \reffig{fig:LP-amp-mu-dep} for
$\oL{\calA}_{LV}$ and $\oL{\calA}_{LA}$. The NLO QCD corrections are sizeable
for $V_9^\text{eff}$ and $V_{10}^\text{eff}$, where LO and
NLO scale uncertainty bands do not overlap. These corrections 
have been neglected in previous predictions
of $\Bqtollgamma$. Whereas $\oL{\calA}_{LA}$ is
real-valued and slowly varying over $q^2 \in [4 m_\ell^2, \, 8.0]\GeV^2$, 
$\oL{\calA}_{LV}$ has a sizeable imaginary part already at LO QCD because
of the $B$-type contribution $i = 7B\text{eff}$, in contradistinction
to the decays $B\to (X_s, K, K^*) + \ell\bar\ell$, where such contributions are
suppressed by the small QCD-penguin coefficients.\footnote{In the 
case of charged $b\to q \ell\bar\ell$ decays also charged-current 
operators $\Op_{1,2}^u$ contribute to these so-called weak 
annihilation contributions \cite{Beneke:2001at, Beneke:2004dp}. 
They are CKM suppressed for $q=s$, but not negligible for
$q = d$, as for example in $B^+ \to M^+\ell\bar\ell$, 
$M=\rho,\pi$ \cite{Beneke:2004dp,Hambrock:2015wka}.} 
Further, $\re(\oL{\calA}_{LV})$ is dominated for 
$q^2 \lesssim 2\GeV^2$ by the photon pole in $i = 7A\text{eff},\, 7B\text{eff}$,
which interferes destructively with the $i = 9\text{eff}$ contribution for
$q^2 \gtrsim 2\GeV^2$ and leads to a zero crossing around $q^2_0 \approx 
(3.0 - 3.5)\GeV^2$. The zero crossing is responsible for the sign flip of
the lepton forward-backward asymmetry $\AFB(q^2)$, see \refsec{sec:Bqllgam-obs}.
NLO QCD corrections shift $q_0^2$ to slightly
larger values and reduce the $\mu_h$-dependence. Moreover, in the
$q^2$-region of the zero-crossing of the LP contribution, the total amplitude
is also very sensitive to NLP corrections, see \refsec{sec:num-NLP-amps}. 

The dependence of $\oL{\calA}_{L\chi}$ on the hard-collinear scale 
$\mu_{hc}$  is also shown in \reffig{fig:LP-amp-mu-dep}, varying
$\mu_{hc} = \{1.0,\, 1.5,\, 2.0\} \GeV$. Also here a reduction
of the scale uncertainty can be observed after including the 
NLO QCD corrections.
Compared to the $\mu_h$ dependence, the residual $\mu_{hc}$ 
dependence at NLO is smaller in $\oL{\calA}_{LV}$ and 
slightly larger in $\oL{\calA}_{LA}$, but both scale dependencies 
are small compared to other theoretical uncertainties.

\begin{figure}
\centering
  \includegraphics[width=0.41\textwidth]{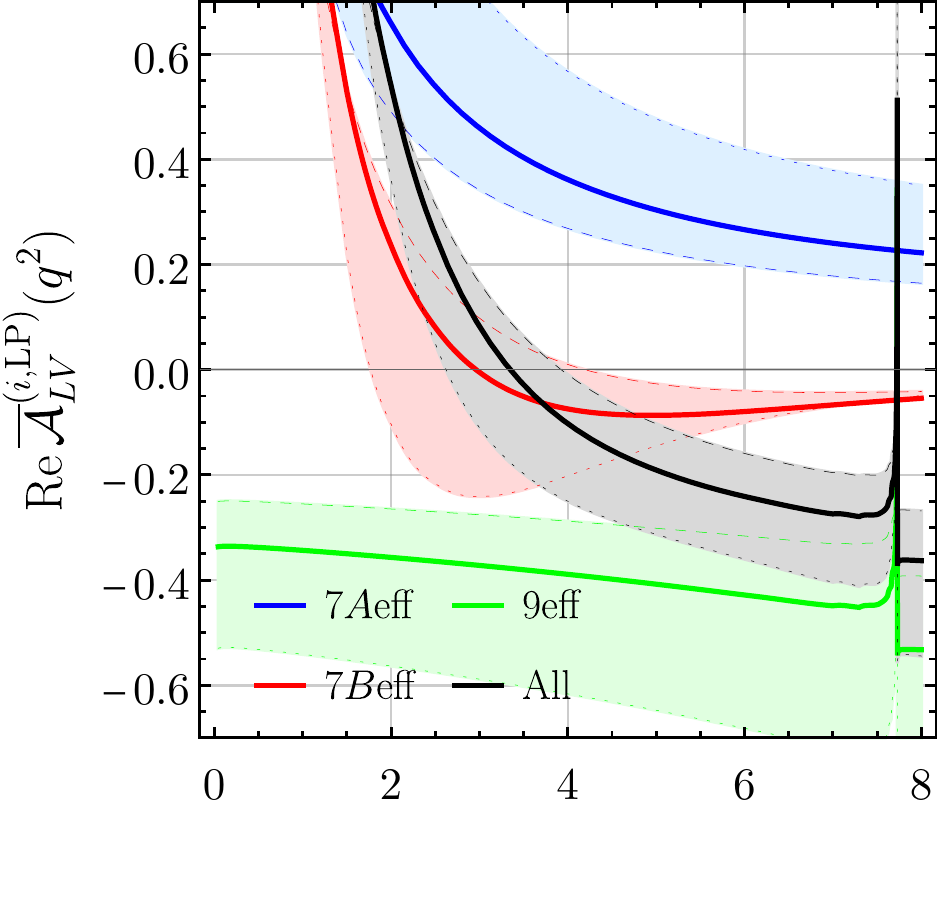}
  \hskip 0.05\textwidth
  \includegraphics[width=0.41\textwidth]{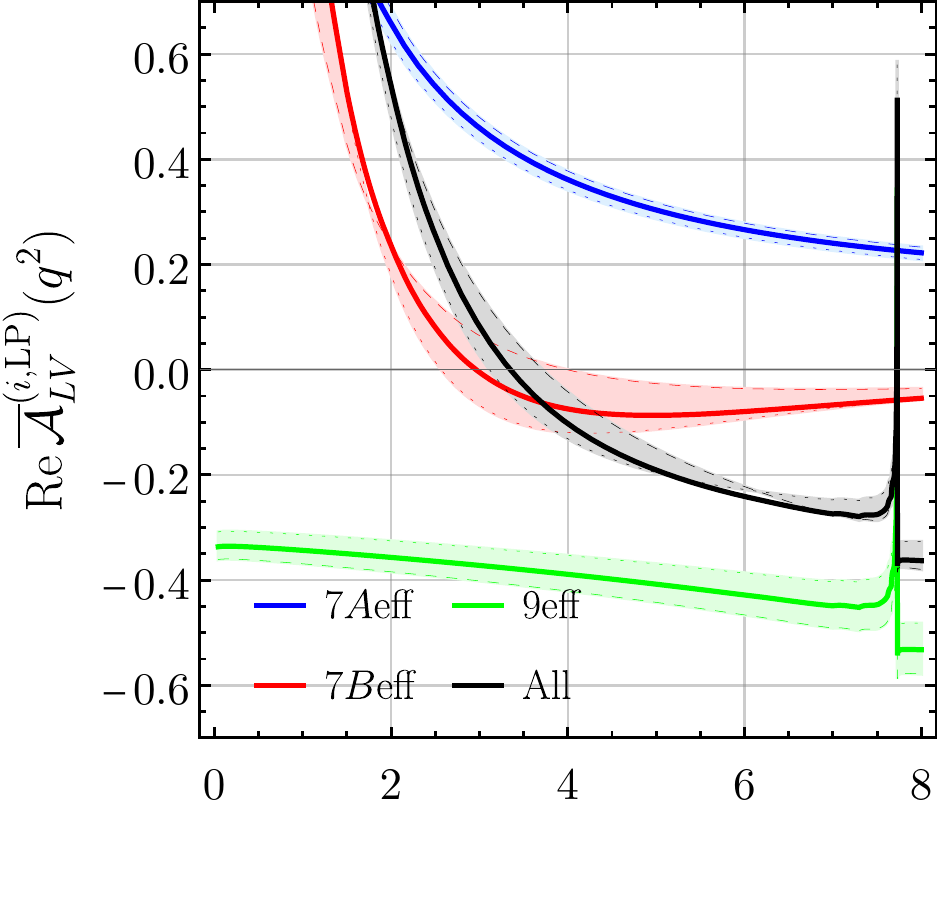}
\\[-0.3cm]
  \includegraphics[width=0.41\textwidth]{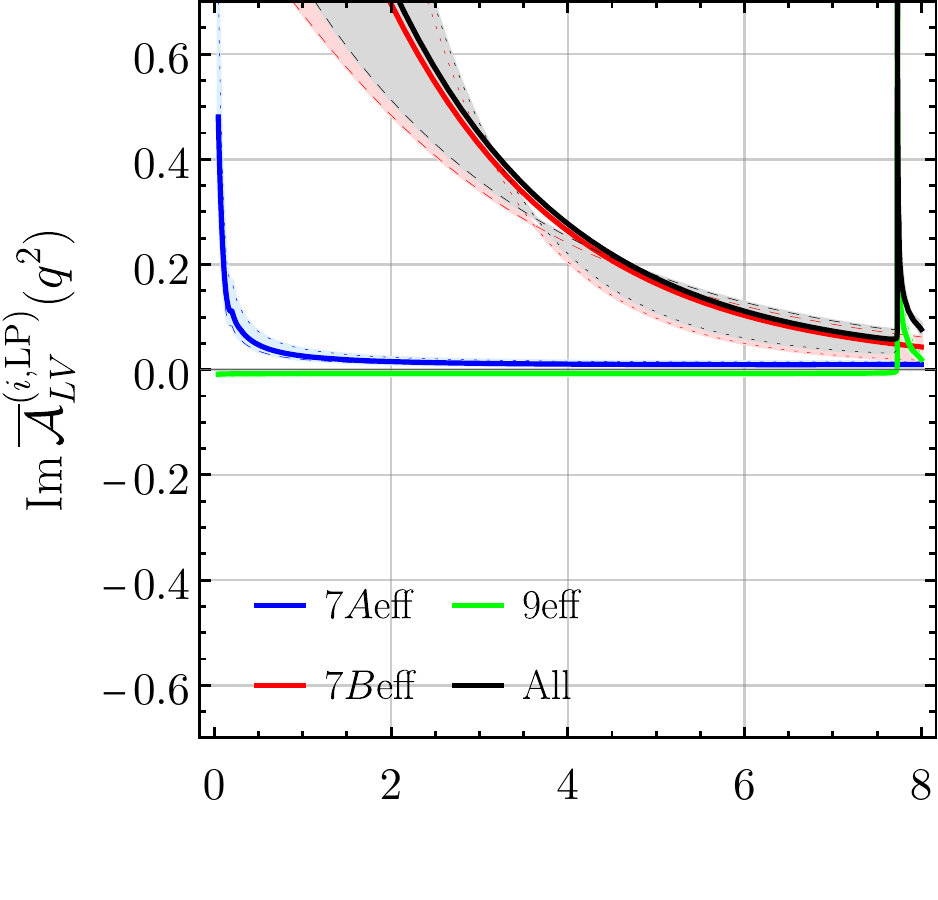}
  \hskip 0.05\textwidth
  \includegraphics[width=0.41\textwidth]{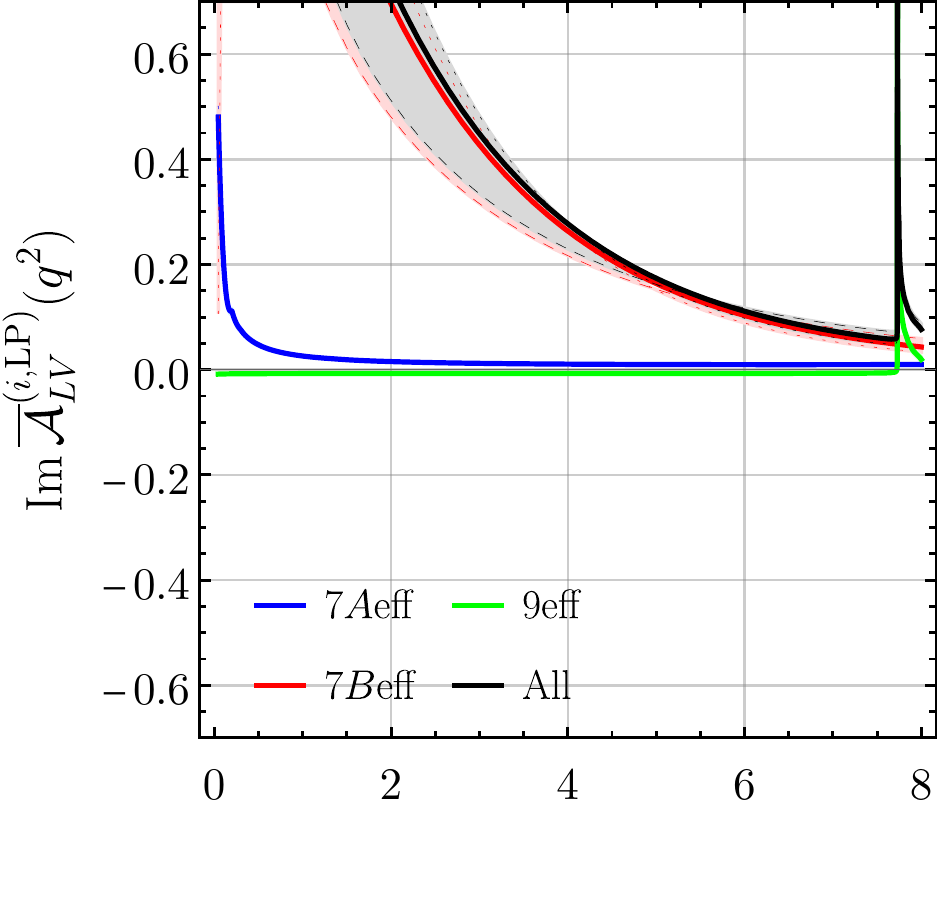}
\\[-0.3cm]
  \includegraphics[width=0.41\textwidth]{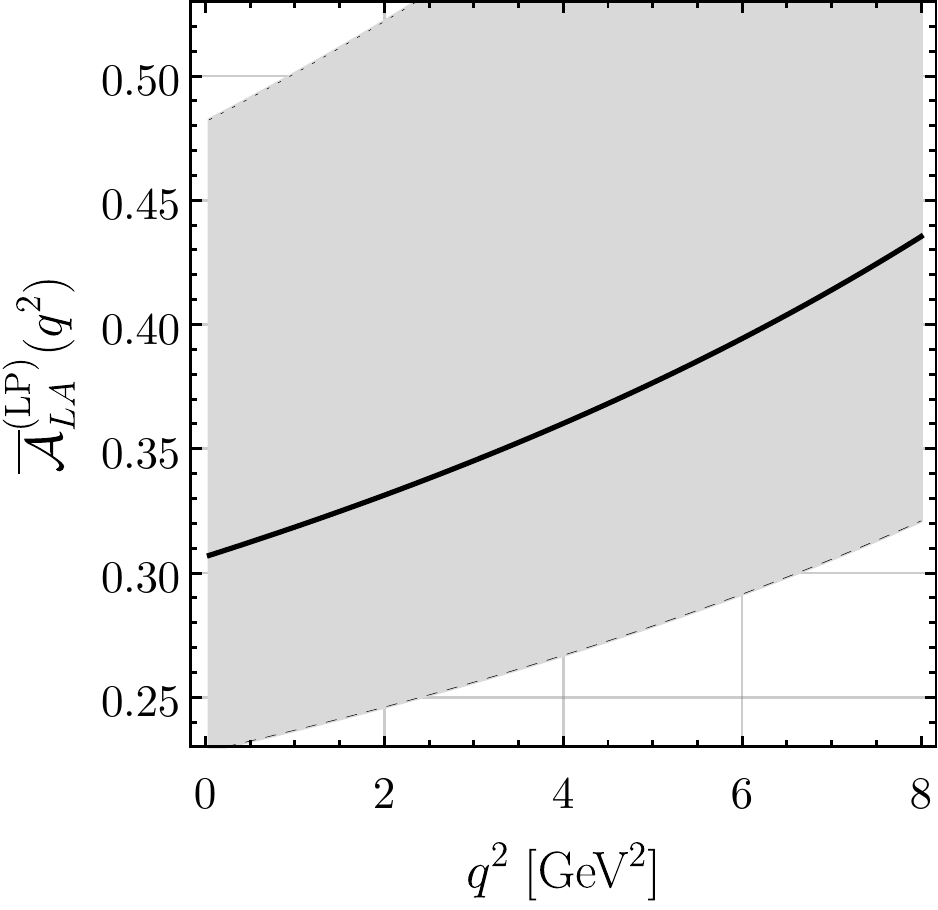}
  \hskip 0.05\textwidth
  \includegraphics[width=0.41\textwidth]{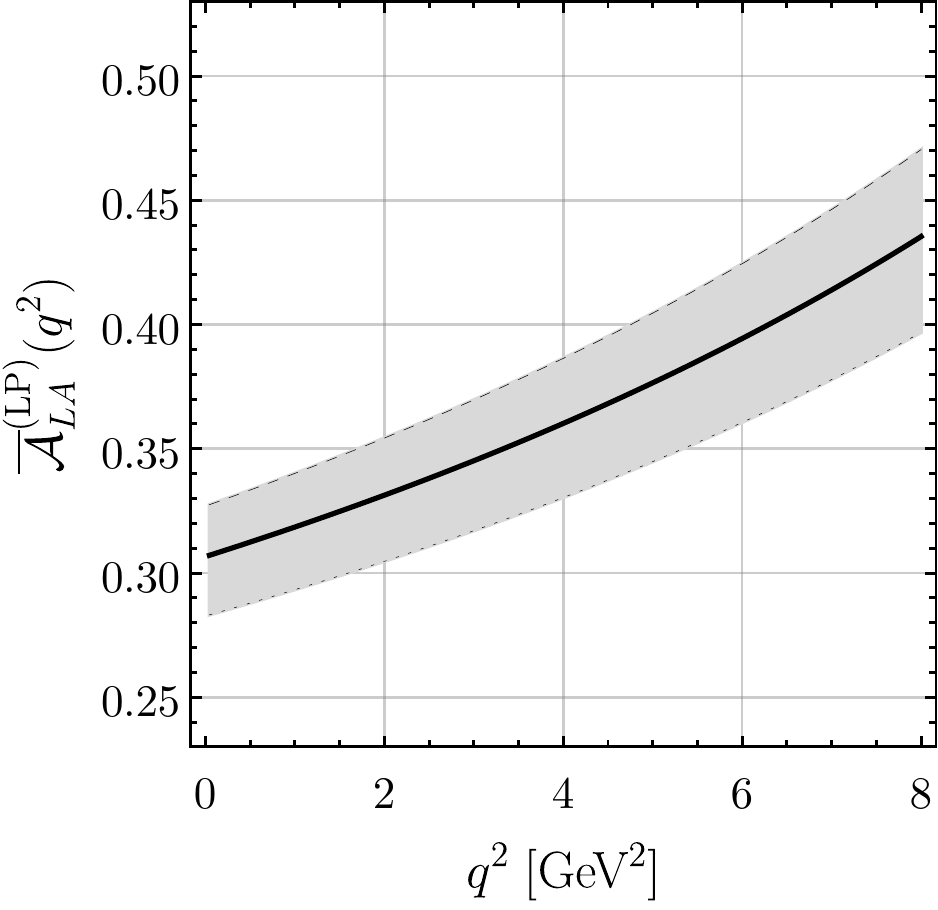}
\caption{\small The $\lamB{s}$ [left] and $\hsigNB{s}{1}$ [right] dependence
  of the helicity amplitudes $\re \oL{\calA}_{LV}$ [upper], $\im \oL{\calA}_{LV}$ 
  [middle] and $\oL{\calA}_{LA}$ [lower] at LP. The variation of 
  $\lamB{s} = (0.40 \pm 0.15) \GeV$ and  $\hsigNB{s}{1} = (0.0 \pm 0.7)$
  [dotted, solid, dashed] leads to the grey bands. The individual contributions 
  $i = 7A, 7B, 9$ [blue, red, green] are also shown for $\oL{\calA}_{LV}$.
}
\label{fig:LP-amp-B-LCDA-dep}
\end{figure}

The charm-quark pair threshold causes a divergence of $\oL{\calA}_{LV}$
in \reffig{fig:LP-amp-mu-dep} at $q^2 = 4 (m_c^\text{PS})^2 \approx
7.73 \GeV^2$ from the two-loop matrix elements of $\Op_{1,2}^{c}$, which
signals a breakdown of factorization and restricts us to  $q^2 \lesssim
6 \GeV^2$. The uncertainties due to the charm-quark mass on $\re \oL{\calA}_{LV}$ 
($\im \oL{\calA}_{LV}$) for $q^2 \lesssim 6 \GeV^2$ are at most $1.0\%$
($2.0\%$) in the PS scheme. Note that the physical threshold is at
$q^2 = m_{J/\psi}^2 > 4 (m_c^\text{PS})^2$. One might therefore be tempted
to adopt the pole scheme for the charm-quark mass, in which case the partonic
threshold divergence occurs closer to the physical threshold. However, we
find that  the uncertainty on $\re \oL{\calA}_{LV}$ ($\im \oL{\calA}_{LV}$)
from $m_c$ for $q^2 \lesssim 6 \GeV^2$ is $5.0\%$ ($5.0\%$) in the pole
scheme, much larger than in the PS scheme, due to the larger uncertainty
in the mass value. 

The variation of the bottom-quark mass shows that for $q^2 \lesssim 6 \GeV^2$
the uncertainty on $\re \oL{\calA}_{LV}$ ($\im \oL{\calA}_{LV}$) is less
than $0.1\%$ ($0.3\%$) in the PS scheme and less than $0.3\%$ ($1.0\%$)
in the pole scheme, and similarly for $\re \oL{\calA}_{LA}$. 

As already mentioned, a strong dependence of $\oL{\calA}_{L\chi}$ ($\chi = V, A$)
on the first inverse moment $\lamB{s}$ of the $B$-meson LCDA $\phi_+(\omega)$
should be expected, which is shown in \reffig{fig:LP-amp-B-LCDA-dep}.
It represents the largest uncertainty compared to the tiny $\mu_h$ and 
the $\mu_{hc}$ uncertainties, and also affects the location of the
zero-crossing $q_0^2$ of $\re \oL{\calA}_{LV}$. To estimate the further
model-dependence on the $B$-meson LCDA, we vary also the first (hatted)
logarithmic moment $\hsigNB{s}{1} = (0.0 \pm 0.7)$, 
as described in more detail in \refapp{app:B-LCDA}. The uncertainty due to
$\hsigNB{s}{1}$ is rather large for the contribution of $i = 7B\text{eff}$,
but overall small compared to the $\lamB{s}$ uncertainty. 

%
%
\subsubsection{Amplitudes at NLP}
\label{sec:num-NLP-amps}

The NLP corrections to the amplitudes are due to local and 
nonlocal $A$-type and $B$-type, as well
as the local four-quark contributions in \refeq{eq:ampl-NLP-4qu}, see 
\refsec{sec:FF-NLP} and \reffig{fig:ME-Op_3456}. 

\begin{figure}
\centering
  \includegraphics[width=0.41\textwidth]{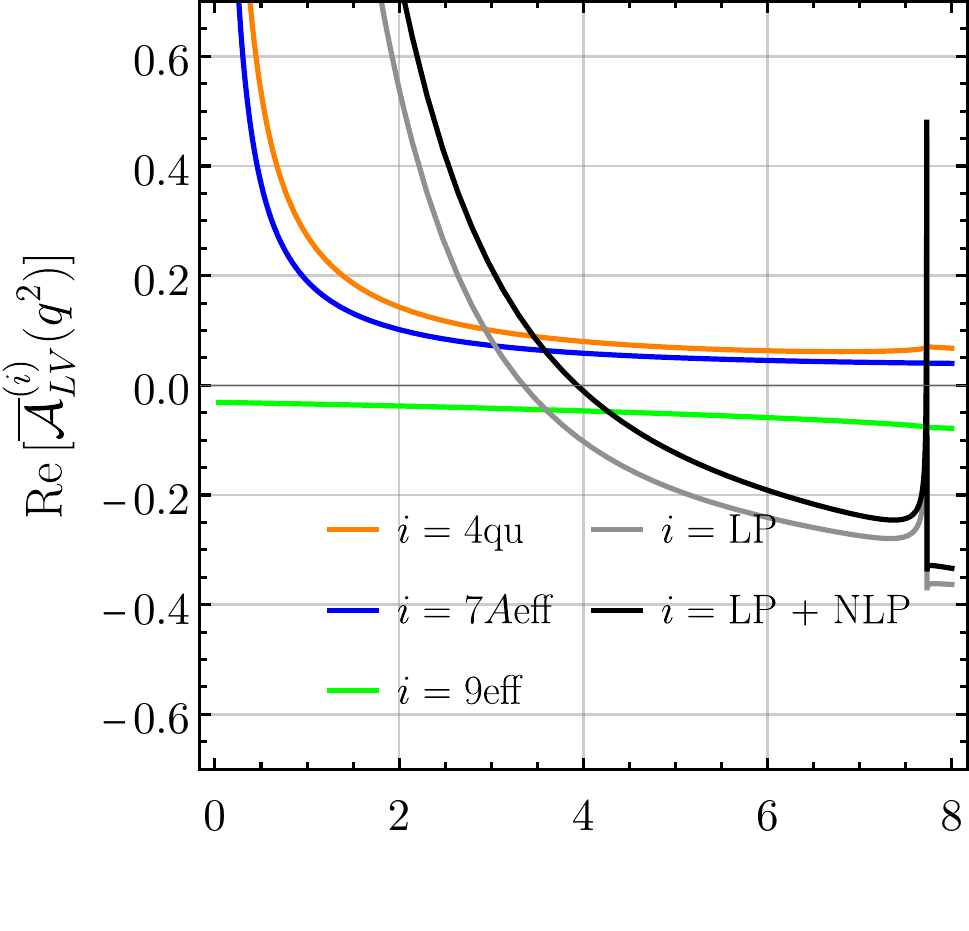}
  \hskip 0.05\textwidth
  \includegraphics[width=0.41\textwidth]{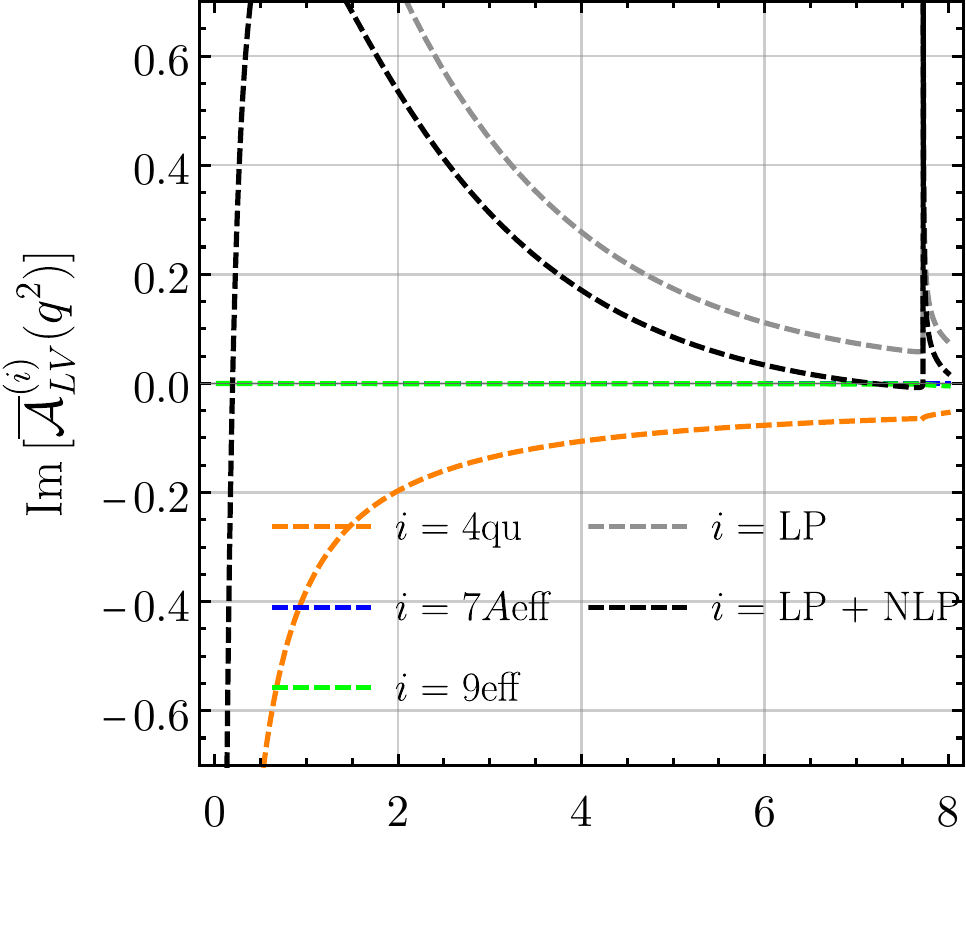}
\\[-0.3cm]
  \includegraphics[width=0.41\textwidth]{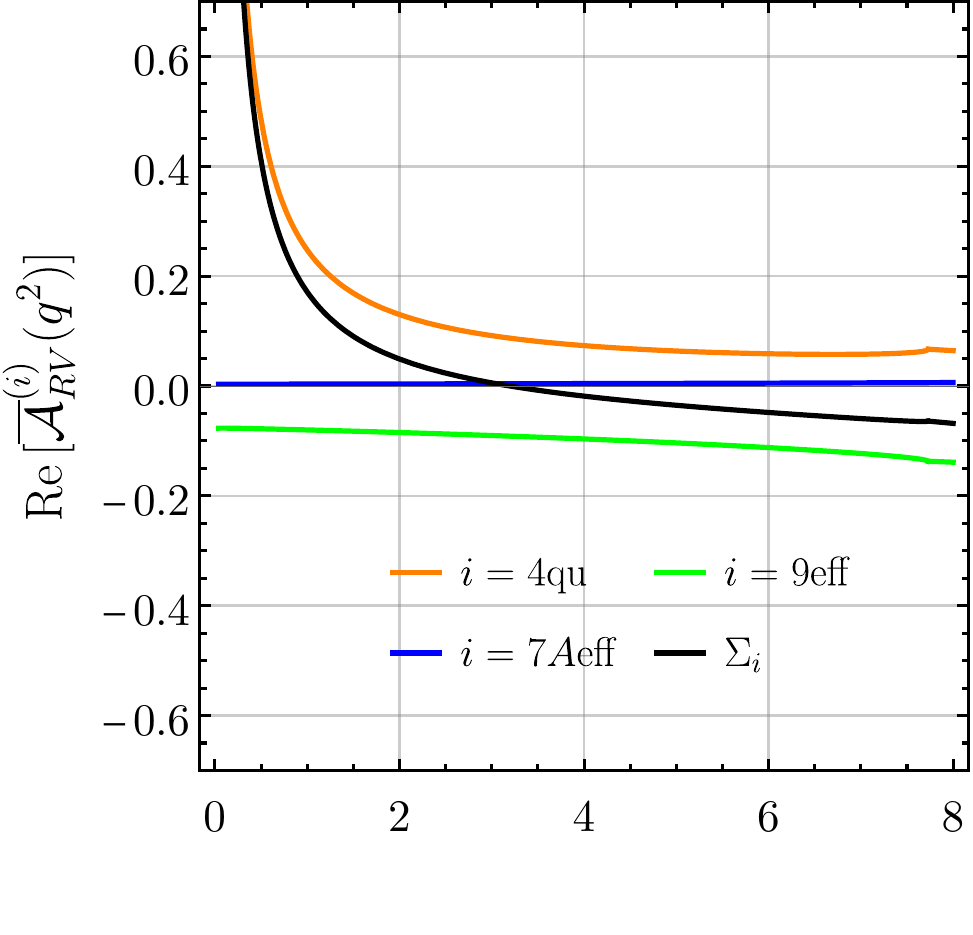}
  \hskip 0.05\textwidth
  \includegraphics[width=0.41\textwidth]{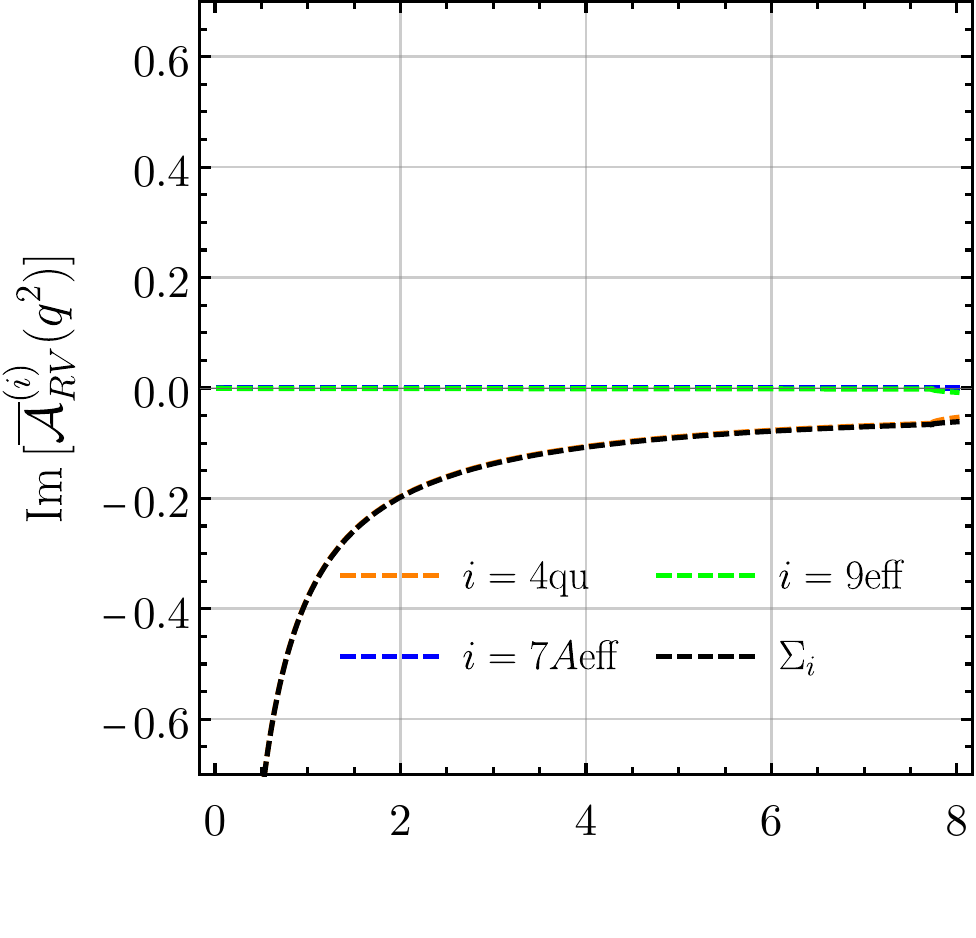}
\\[-0.3cm]
  \includegraphics[width=0.41\textwidth]{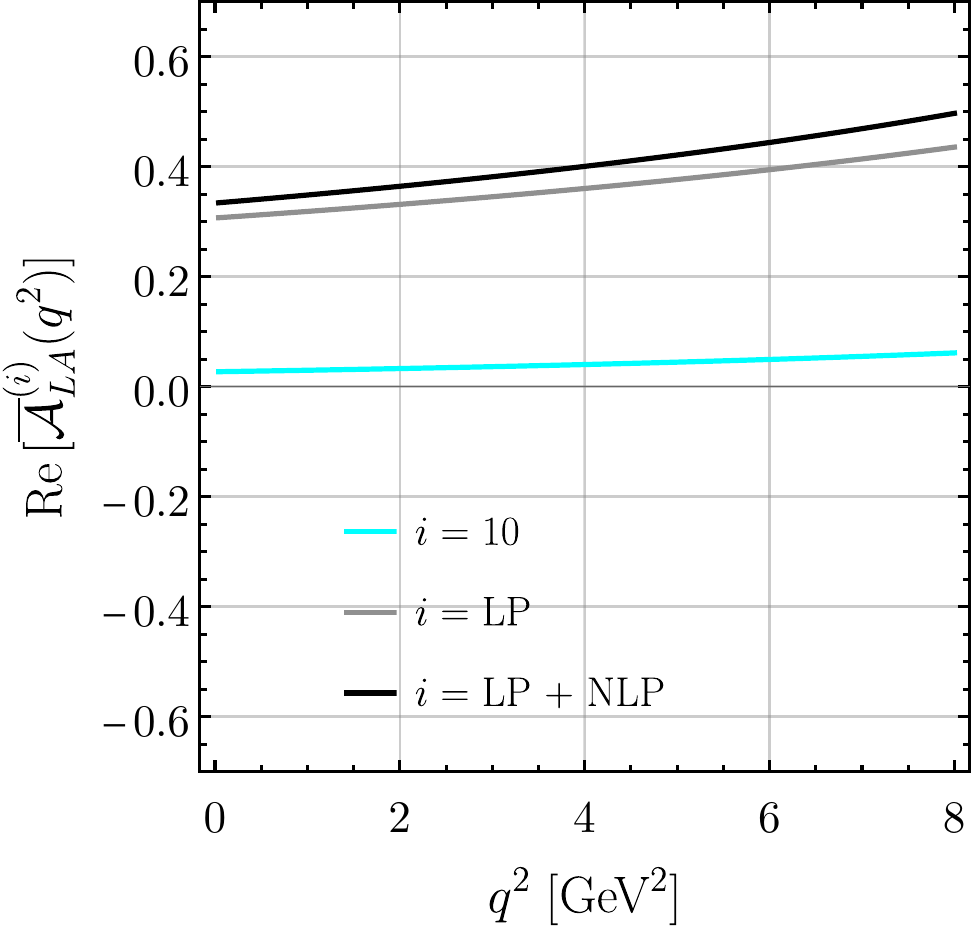}
  \hskip 0.05\textwidth
  \includegraphics[width=0.41\textwidth]{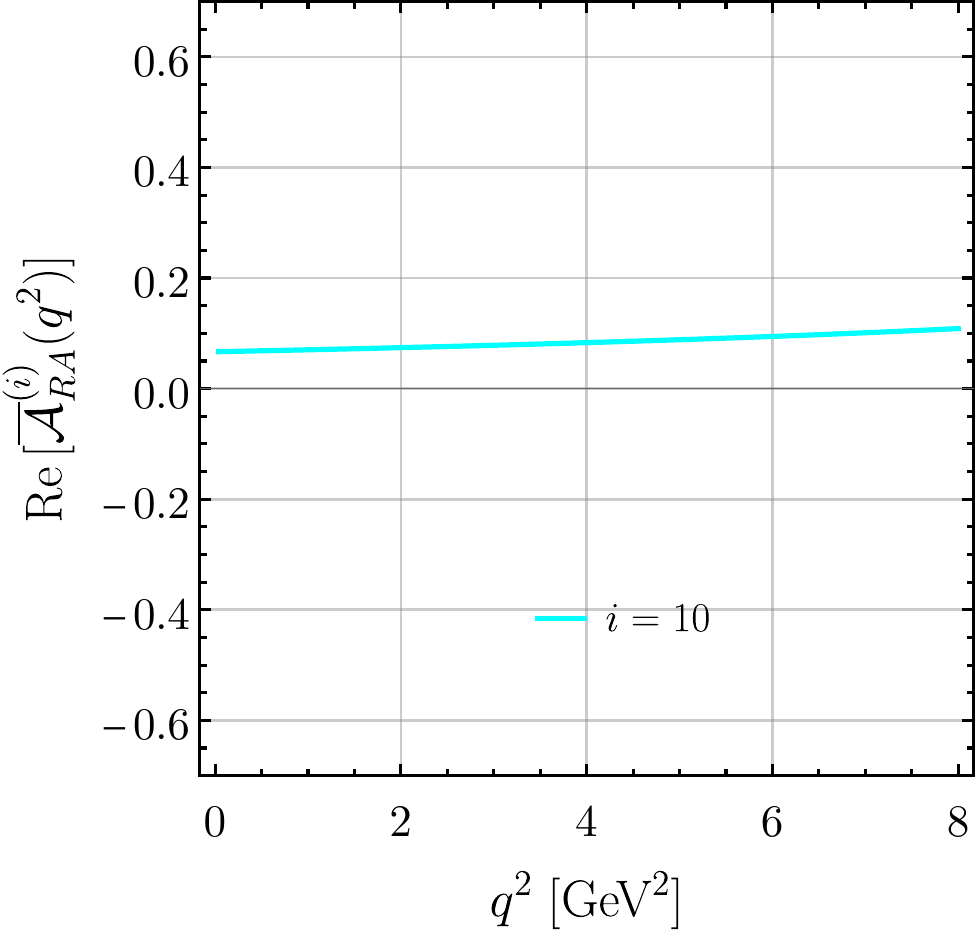}
\caption{\small The local and nonlocal $A$-type NLP contributions to the
  real [solid] and imaginary [dashed] parts of the helicity amplitudes
  $\oL{\calA}_{LV}$ [upper], $\oL{\calA}_{RV}$ [middle]
  and $\oL{\calA}_{hA}$ with $h = L,R$ [lower]. For comparison the
  LP [gray] and the total LP + NLP [black] results are also shown.
}
\label{fig:NLP-amp}
\end{figure}

The local and nonlocal $A$-type NLP contributions to $\oL{\calA}_{hV}$ ($h=L,R$)
and $\oL{\calA}_{LA}$ in \refeq{eq:FL-9_10-NLP}--\refeq{eq:FR-7A-NLP} due to the
semileptonic ($i = 9\text{eff},\, 10$) and dipole ($i = 7A\text{eff}$) operators
as well as the local contributions from the four-quark (4qu) operators in
\refeq{eq:ampl-NLP-4qu} are compared in \reffig{fig:NLP-amp} 
separately to the LP results. For the non-local form factor 
parameterization we use the central value $r_\text{LP} = 0.2$.
It can be seen that the contributions $i = 7A\text{eff},\, \text{4qu}$ have a
photon pole for $q^2 \lesssim 2 \GeV^2$ and that $i=\text{4qu}$ contributes an imaginary part to the left- and right-helicity  amplitudes. The NLP corrections to $\oL{\calA}_{LA}$ are within the
expected size of a $\LambQCD/m_b \sim (10 - 15)\%$ correction 
relative to LP. 
In the case of $\oL{\calA}_{LV}$, this applies individually to 
the real part of $i = 7A\text{eff}$ and $i = 9\text{eff}$,
amounting to $15\%$ and $(10-15)\%$ relative to their respective 
LP contributions, whereas their imaginary parts of the
LP and NLP contributions are tiny ($9\text{eff}$) or vanish altogether ($7A\text{eff}$). 
The real and imaginary parts of the local NLP contribution $i=\text{4qu}$
turn out to be rather large, constituting $+(20 - 25)\%$ and $-30\%$
of the real part of the LP $i = 7A\text{eff}$ contribution, which has been
chosen for comparison because of a similar photon pole.
The NLP corrections are most important in the $q^2$ region
where the LP contributions cancel and give rise to the zero crossing of
$\re(\oL{\calA}_{LV})$. In fact, the location of the zero, 
$q^2_0 \approx 4 \GeV^2$ is significantly
shifted and the sum of these NLP corrections is sizeable  
compared to the LP part for $q^2 \in [3,\, 5] \GeV^2$
around the zero crossing $q^2 \in [3,\, 5] \GeV^2$, especially for 
$\re(\oL{\calA}_{LV})$. 

The right-handed helicity amplitudes $\oL{\calA}_{R\chi}$ 
($\chi = V,A$) are entirely NLP. $\oL{\calA}_{RV}$ exhibits a
zero at $q^2_0 \approx 3\GeV^2$ due to the 
interference of the $i = \text{4qu}, 9\text{eff}$ contributions, 
whereas the $i=7A\text{eff}$ part is not
enhanced at $q^2 \to 0$, see \refeq{eq:FR-7A-NLP}. 
The amplitude $\oL{\calA}_{RA}$ is small.

\begin{table}
\centering
\renewcommand{\arraystretch}{1.4}
\begin{tabular}{|c|cccc|}
\hline\hline
                      & $m_V$           & $\Gamma_V$     & $f^\parallel_V$   & $T_1^{B_q\to V}(0)$ \\
  $V$                 & $[\MeV]$        & $[\MeV]$       & $[\MeV]$          &                     \\
\hline\hline
  $\phi(1020)$        & $1019.461(16)$  & $4.249(13)$    & $233(4)$          & $0.309(27)$         \\
  $\phi(1680)$        & $1680(20)$      & $150(50)$      & ---               & ---                 \\
  $\phi(2170)$        & $2170(15)$      & $104(20)$      & ---               & ---                 \\
\hline
  $\rho^0(770)$       & $775.26(25)$    & $147.8(9)$     & $213(5)$          & $0.272(26)$         \\
  $\omega(782)$       & $782.65(12)$    & $8.49(8)$      & $197(8)$          & $0.251(31)$         \\
  $\omega(1420)$      & $1410(60)$      & $290(190)$      & ---               & ---                 \\
  $\rho(1450)$        & $1465(25)$      & $400(60)$      & ---               & ---                 \\
\hline\hline
\end{tabular}
\renewcommand{\arraystretch}{1.0}
\caption{\label{tab:num-input-res}
  \small
  Numerical input values for the parameters of the resonances entering
  the $b\to s$ and $b\to d$ transitions. Masses and decay widths are 
  from \cite{Zyla:2020zbs} and decay constants and form factors 
  from \cite{Straub:2015ica}, with the $q^2$ dependence of $T_1^{B_q\to V}$
  in the simplified series expansion (SSE) in Table 14 
  from light-cone sum rules only.}
\end{table}

\begin{figure}
\centering
  \includegraphics[width=0.41\textwidth]{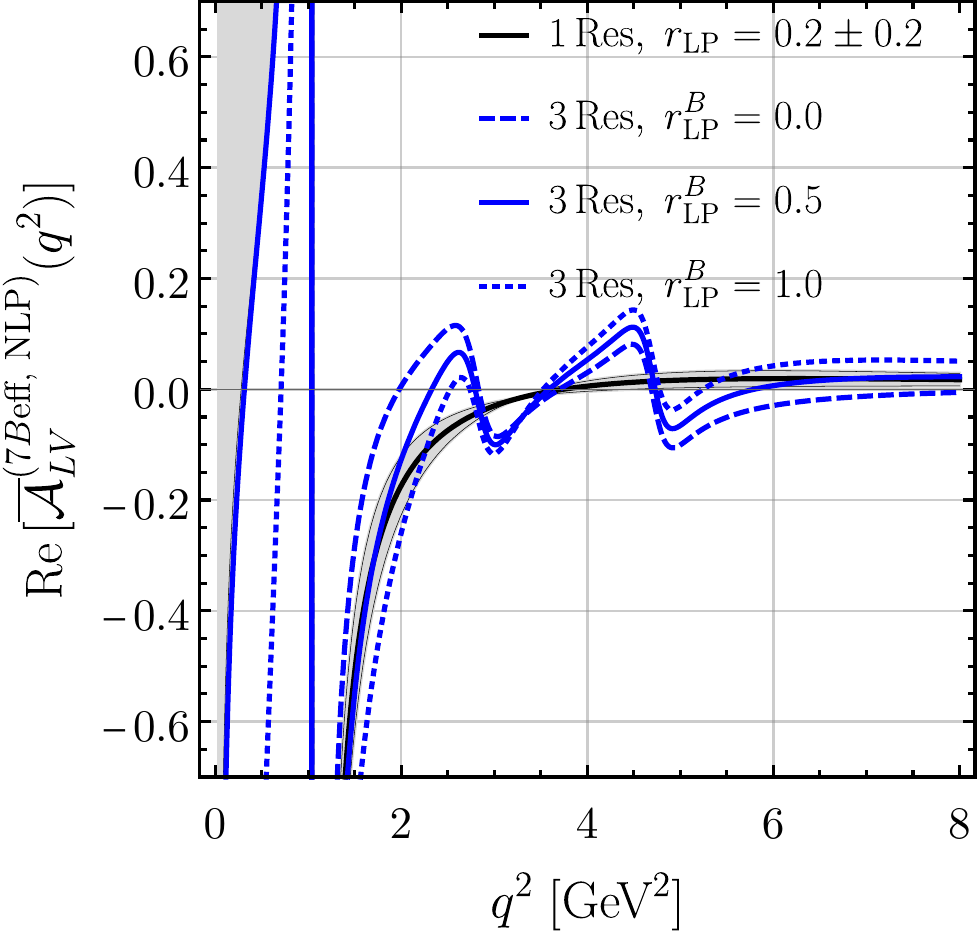}
  \hskip 0.05\textwidth
  \includegraphics[width=0.41\textwidth]{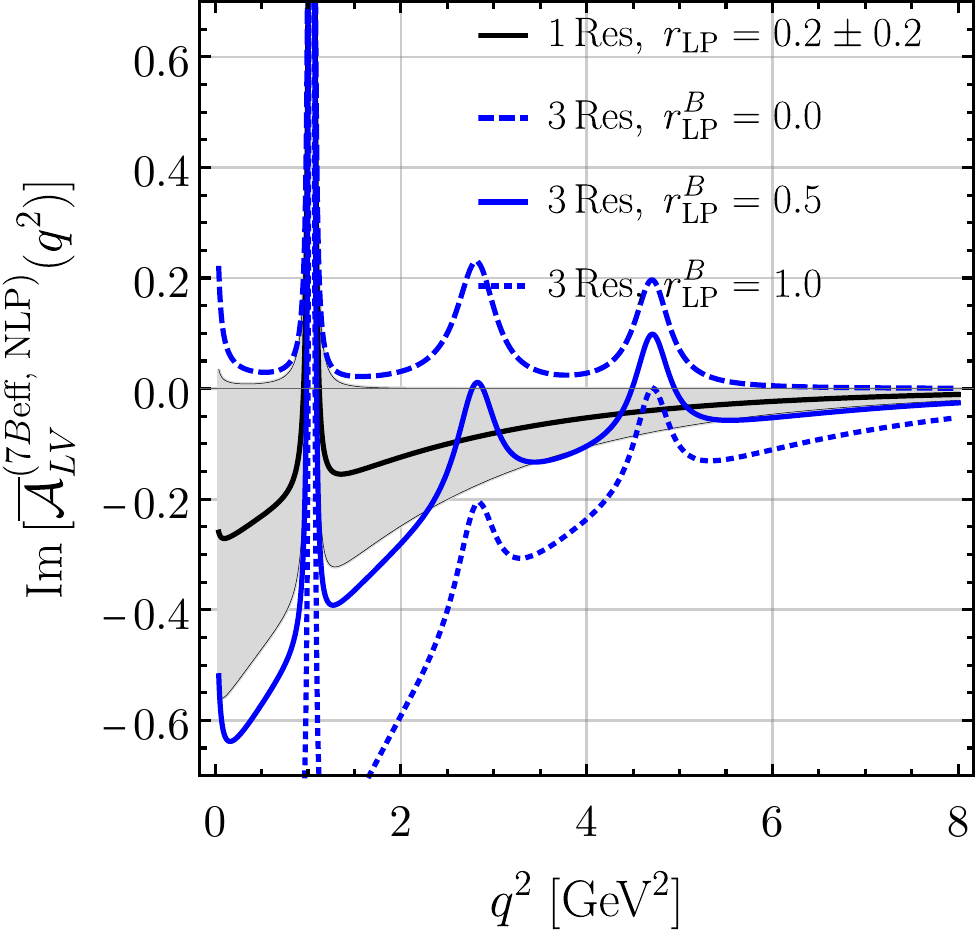}
\\[2mm]
  \includegraphics[width=0.41\textwidth]{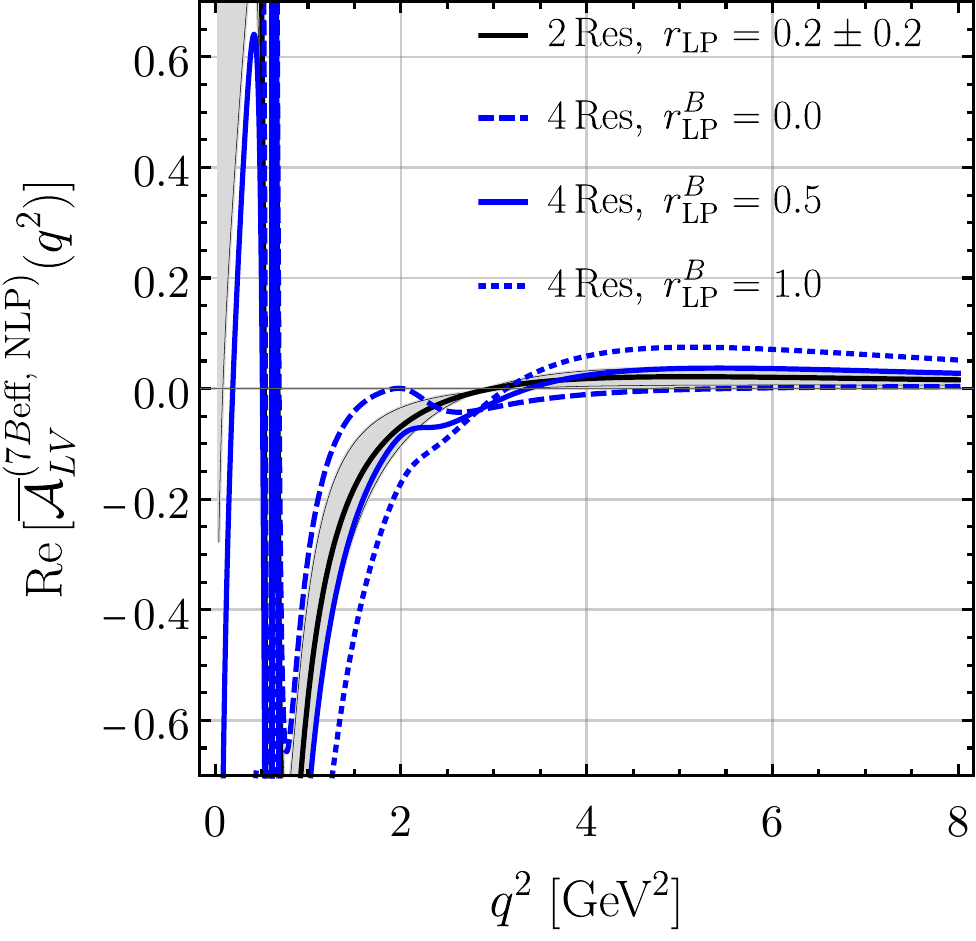}
  \hskip 0.05\textwidth
  \includegraphics[width=0.41\textwidth]{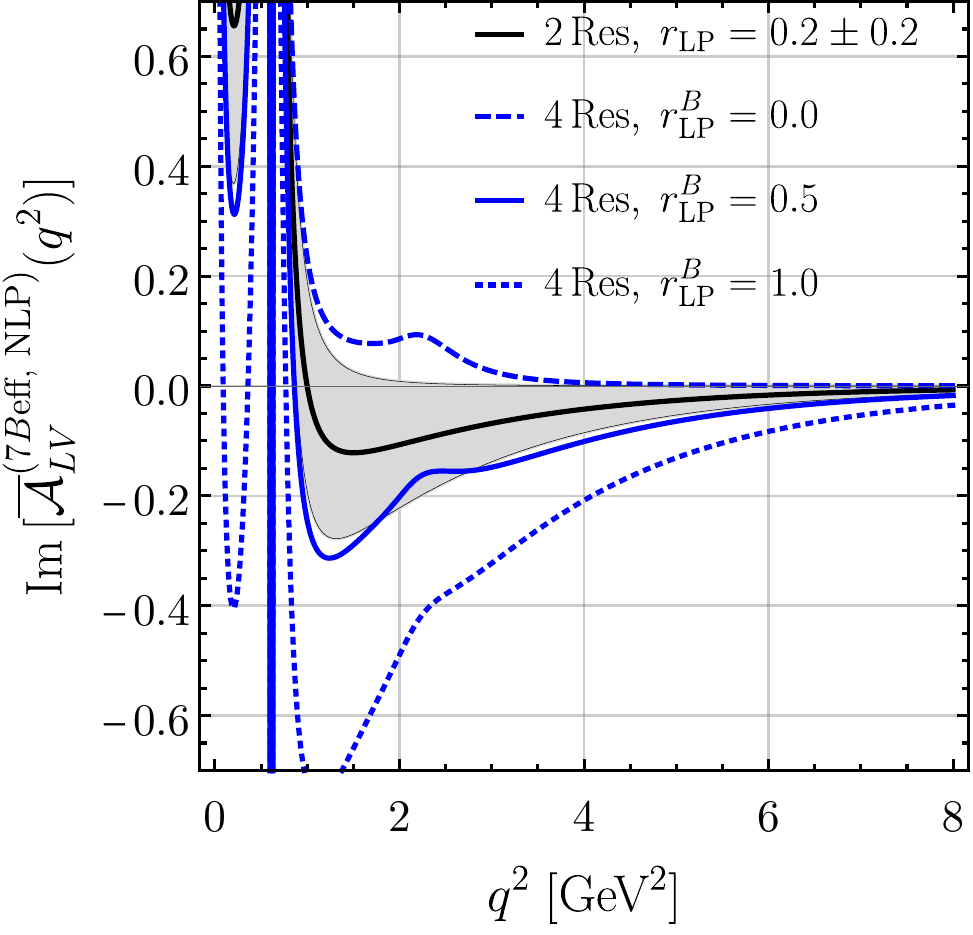}
\caption{\small The real [left] and imaginary [right] parts of the NLP 
  local + nonlocal $7B\text{eff}$-type contributions $i = 7B\text{eff}$
  to $\oL{\calA}_{LV}$ for $q = s$ [upper] and $q = d$ [lower]. The default
  1(2)-Res model for $r_\text{LP} = 0.2 \pm 0.2$ [grey band] is compared
  with the 3(4)-Res model for $r^B_\text{LP} = \{0.0,\, 0.5,\, 1.0\}$
  [blue dashed, solid, dotted].
}
\label{fig:NLP-B-type}
\end{figure}

The $B$-type NLP contributions ($i = 7B\text{eff}$)
enter only in $\oL{\calA}_{LV}$. The nonlocal form factors 
are modelled according to 
\refsec{sec:FF-NLP-model} with parameters listed in \reftab{tab:num-input-res}.
The lowest resonances affect the very-low $q^2$ region 
$q^2 \lesssim 1\,\GeV$: $\phi(1020)$ for $q = s$, and 
$\rho^0(770)$ and $\omega(782)$  for $q = d$. We include these 
lowest resonances in our default model (``1(2) Res'') in both 
$A$- and $B$-type contributions. Further we describe the
continuum part of the nonlocal NLP $B$-type contribution by the 
fraction $r_\text{LP} = 0.2 \pm 0.2$ of the LP $B$-type 
amplitude \refeq{eq:amplSCETItypeB}, omitting NLO QCD corrections. 
The positive central value of $r_\text{LP}$ implies destructive
interference with the LP contribution, 
see \refeq{eq:soft-FF-B-model}, in agreement with sum
rule calculations of the corresponding $A$-type form factor 
\cite{Beneke:2018wjp}.
The resulting $B$-type NLP contributions 
to the helicity amplitudes are shown in \reffig{fig:NLP-B-type} 
as solid black line and grey band.

The $B$-type contribution 
needs the form factor in the time-like region. 
In order to investigate the 
impact of higher-mass resonances on this form factor, 
we define an alternative model (``3(4) Res''), which includes 
the next two resonances listed in 
\reftab{tab:num-input-res} in the 
sum over $V$ in (\ref{eq:soft-FF-B-model}).
We assume the unknown decay constants and form factors of the higher
$\phi$, $\rho$ and $\omega$ resonances to be a fraction
\begin{equation}
  r^q_{fT} 
  \equiv \Big(f_{V'} T_1^{B_q \to V'}\Big) 
   \Big/ \Big(f_V T_1^{B_q \to V}\Big)
\hskip2cm
  (q  = d,s)
\end{equation}
of those of the lowest ones and use the same value $r^q_{fT}=0.3$ 
for all higher resonances $V' > V$. Since the higher resonances 
lie in the range of $q^2$, where we might expect the continuum 
contribution calculated in factorization to already be dual to 
the contribution from the relatively broad resonances, the value 
of $r_\text{LP}$ should be increased (implying a smaller 
continuum part in the sum of LP and NLP according to 
\refeq{eq:soft-FF-B-model}). Hence, we introduce separate parameters 
$r_\text{LP}^A$ and $r_\text{LP}^B$, where the first is 
always kept at $r_\text{LP}^A = 0.2\pm 0.2$, but the second 
should depart from this value in the ``3(4) Res'' model, at 
least for $q=s$, where the effect of resonances is more 
pronounced.

In \reffig{fig:NLP-B-type} the default model is compared
to the one including the higher resonances. In the case $q=s$ 
(upper plots), the higher $\phi$ resonances are clearly visible
despite their large widths of about $(100 - 150) \MeV$.
The ``3-Res'' model is shown 
for three different values of  $r_\text{LP}^B$ 
and oscillates around the default 1-Res model for the real 
part independently of the choice of $r_\text{LP}$. On the other 
hand, the imaginary part is very sensitive to this choice. 
For $r_\text{LP}^B \approx 0.5$ the local effects of the
resonances would be included globally in the 1-Res model with 
$r_\text{LP}^B = 0.2$, in agreement with the qualitative expectation 
above. In the case of $q = d$ (lower plots) the effect of the 
excited resonances $\rho(1450)$ and $\omega(1420)$ is much smaller 
in $\oL{\calA}_{LV}$, such that local duality above 
$q^2 \gtrsim 2.0\GeV^2$ works better compared to $q = s$. Also 
we see that with $r_\text{LP}^B \approx 0.3$ the 
4-Res description is close to the 2-Res one, again in agreement 
with expectations.

\begin{figure}
\centering
  \includegraphics[width=0.42\textwidth]{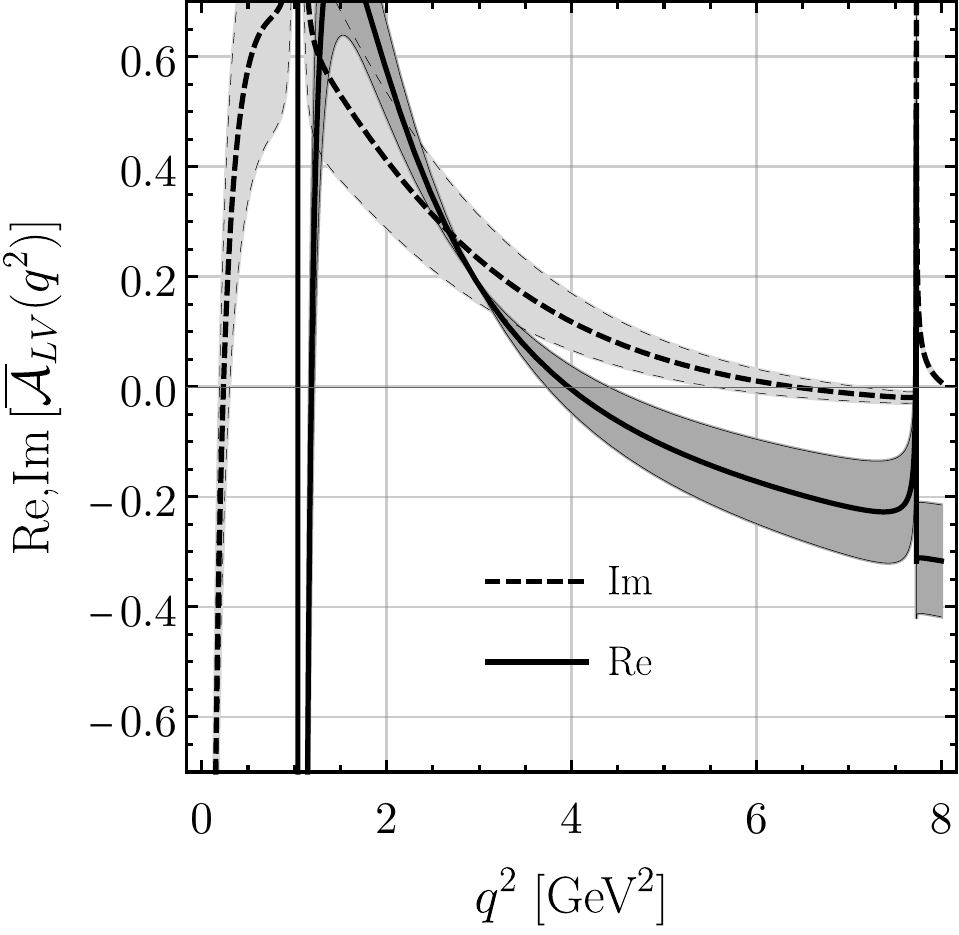}
  \hskip 0.05\textwidth
  \includegraphics[width=0.42\textwidth]{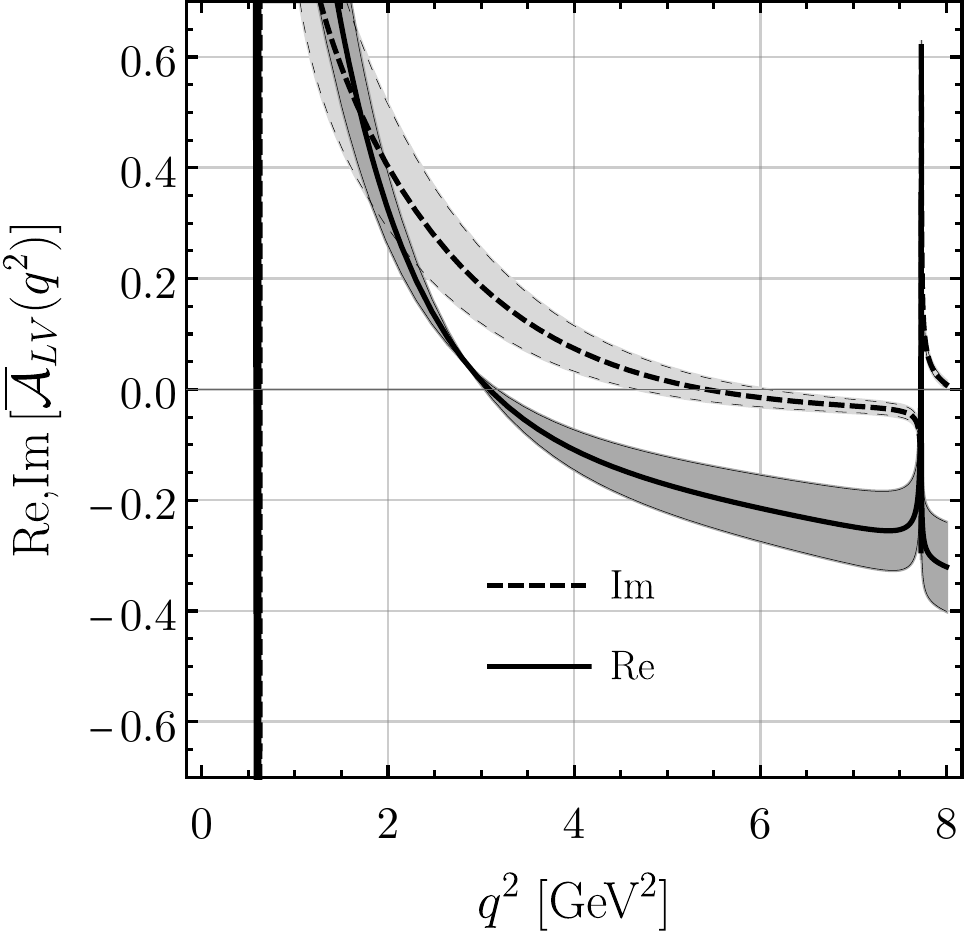}
\caption{\small The real [solid, darker] and imaginary [dashed, lighter] part
  of the combined LP and NLP result for $\oL{\calA}_{LV}$ in the default
  1- and 2-Res model for $q = s$ [left] and $q = d$ [right], respectively.
  The grey band shows the variation $r_\text{LP} = 0.2 \pm 0.2$.
}
\label{fig:amp-ALV-all}
\end{figure}

The final result of $\oL{\calA}_{LV}$, including all contributions
(LP and NLP, $A$-type and $B$-type), is shown in 
\reffig{fig:amp-ALV-all} for the default 1(2)-Res model of
the NLP nonlocal $B$-type contributions. The variation 
$r_\text{LP} = 0.2 \pm 0.2$ gives rise to the bands for the real and 
imaginary parts. The zero crossing of the real part for $q = s$ is 
shifted to $q_0^2 \approx 4\GeV^2$ when compared to the LP result of 
$\oL{\calA}_{LV}$ in \reffig{fig:LP-amp-mu-dep}. For $q = d$, 
the zero-crossing occurs at the lower value $q_0^2 \approx 2.5\GeV^2$.

%
%
\subsection[$\Bqtollgamma$ observables]
{\boldmath  $\Bqtollgamma$ observables}
\label{sec:Bqllgam-obs}

From the amplitude \refeq{eq:ampl-FFs} we obtain the two-fold 
$\oL{B}_q\to\gamma\ell\bar{\ell}$ decay-rate distribution
\begin{align}
  \label{eq:d2Gamma-ISR}
  \frac{d^2\oL{\Gamma}}{dq^2\, d\!\cos\theta_\ell}\Big|_{|\text{SD}|^2} &
  = \oL{a}(q^2) + \oL{b}(q^2) \cos\theta_\ell + \oL{c}(q^2) \cos^2\!\theta_\ell 
\end{align}
in terms of the kinematic variables $q^2 \in [4 m_\ell^2,\, m_{B_q}^2]$
and $\cos\theta_\ell \in [-1,\, 1]$, where $\theta_\ell$ is the angle
between the $\oL{B}_q$-meson direction and the lepton momentum in the
dilepton center-of-mass frame. The $q^2$-dependent angular coefficients
are expressed in terms of the transversity amplitudes as follows:
\begin{align}
  \label{eq:ang-obs-a}
  \oL{a} &
  = \Gamma_0 \sqrt{\lambda}^3  q^2 \beta_\ell \times \Big[
      (2 - \beta_\ell^2) \left(
         |\oL{\calA}_{\parallel V}|^2 + |\oL{\calA}_{\perp V}|^2 \right)
    + \beta_\ell^2 \left(
         |\oL{\calA}_{\parallel A}|^2 + |\oL{\calA}_{\perp A}|^2 \right) \Big],
\\
  \label{eq:ang-obs-b}
  \oL{b} &
  = \Gamma_0 \sqrt{\lambda}^3  q^2 \beta_\ell \times 4\, \beta_\ell\,
  \text{Re} \Big[ \oL{\calA}_{\parallel V}^{} \oL{\calA}_{\perp A}^\ast
                + \oL{\calA}_{\parallel A}^{} \oL{\calA}_{\perp V}^\ast \Big] ,
\\
  \label{eq:ang-obs-c}
  \oL{c} &
  = \Gamma_0 \sqrt{\lambda}^3  q^2 \beta_\ell \times \beta_\ell^2
    \Big[ |\oL{\calA}_{\parallel V}|^2 + |\oL{\calA}_{\perp V}|^2
        + |\oL{\calA}_{\parallel A}|^2 + |\oL{\calA}_{\perp A}|^2 \Big] ,
\end{align}
and
\begin{align}
  \Gamma_0 & \equiv \frac{\alE^3 |\normEW|^2}{2^{11} \pi^4 m_{B_q}^5} ,
&
  \sqrt{\lambda} & \equiv m_{B_q}^2 - q^2 ,
&
  \beta_\ell & \equiv \sqrt{1 - \frac{4 m_\ell^2}{q^2}}.
\end{align}
Eq.~(\ref{eq:d2Gamma-ISR}) refers to the square of the so-called
structure-dependent (SD) amplitude. We provide the lepton-mass suppressed
parts of the double-differential width due to the FSR amplitude of $\Op_{10}$
and its interferences with the SD amplitude in \refapp{app:FSR}. The differential
decay width and the normalized lepton forward-backward asymmetry are
\begin{align}
  \label{eq:decay-rate}
  \frac{d\oL{\Gamma}}{dq^2} &
  = 2 \left[ \oL{a}(q^2) + \frac{\oL{c}(q^2)}{3} \right]  + \ldots ,
&
  \oL{A}_\text{FB}(q^2) &
  = \frac{\oL{b}(q^2)}{d\oL{\Gamma}/dq^2} + \ldots ,
\end{align}
where the dots denote lepton-mass suppressed terms from the FSR
contribution. The angular observables $\oL{a}(q^2)$ and $\oL{c}(q^2)$
differ only by lepton-mass effects, such that their difference
\begin{align}
  \oL{a}(q^2) - \oL{c}(q^2) &
  \propto 2 (1 - \beta_\ell^2)
    \left(|\oL{\calA}_{\parallel V}|^2 + |\oL{\calA}_{\perp V}|^2 \right)
  \propto \frac{m_\ell^2}{q^2}
    \left(|\oL{\calA}_{\parallel V}|^2 + |\oL{\calA}_{\perp V}|^2 \right)
\end{align}
is lepton-mass suppressed. The lepton-mass suppression renders the
experimental measurement of these parts of the two-fold differential
decay width challenging. Moreover, the lepton-mass suppressed FSR
contributions in \refapp{app:FSR} cannot be disentangled from them
and introduce a dependence on $C_{10}$, which is absent in $\oL{a}(q^2)
- \oL{c}(q^2)$. Further, the FSR contribution introduces a nonanalytic
$\cos\theta_\ell$ dependence as can be seen in \eqref{eq:d2G-SDxFSR}
and \eqref{eq:d2G-FSR2}. In consequence there are the two main
observables, $d\oL{\Gamma}/dq^2$ and $\oL{A}_\text{FB}(q^2)$, in the
part of the phase space where $4 m_\ell^2 \ll q^2$.

The two-fold decay-rate distribution 
\begin{align}
  \label{eq:d2Gamma-ISR-CP}
  \frac{d^2\Gamma}{dq^2\, d\!\cos\theta_\ell}\Big|_{|\text{SD}|^2} &
  = a(q^2) - b(q^2) \cos\theta_\ell + c(q^2) \cos^2\!\theta_\ell 
\end{align}
for the CP-conjugated decay $B_q\to\gamma\ell\bar{\ell}$ is given by the quantities $a$, $b$ and $c$ without bars, which are
obtained from \refeq{eq:ang-obs-a}--\refeq{eq:ang-obs-c} by using the
CP-conjugated transversity amplitudes $\calA_{\perp \chi}$ and
$\calA_{\para \chi}$ ($\chi = V, A$). The CP-transformation properties
 \refeq{eq:ampl-CP} of $\calA_{\perp \chi}$ imply a minus sign in
front of $b(q^2)$ in correspondence with other $B\to V\ell\bar\ell$
decays ($V = K^*, \phi,\ldots$)~\cite{Kruger:1999xa}. 

The combination of the two-fold decay-rate distributions of the 
decay \refeq{eq:d2Gamma-ISR}
and the CP-conjugated decay \refeq{eq:d2Gamma-ISR-CP} allows for two
CP-averaged and two CP-asymmetric observables. We consider the CP-averaged
rate and the normalized lepton-forward-backward asymmetry as well as the 
CP rate asymmetry, defined by
\begin{align}
  \frac{d\BR}{dq^2} & 
  = \frac{\tau_{B_q}}{2} \frac{d[\oL{\Gamma} + \Gamma]}{dq^2} ,
&
  \AFB(q^2) & 
  = \frac{[\oL{b} + b](q^2)}
         {d[\oL{\Gamma} + \Gamma]/dq^2} ,
& 
  \ACP(q^2) & 
  = \frac{d[\oL{\Gamma} - \Gamma]/dq^2}
         {d[\oL{\Gamma} + \Gamma]/dq^2} .
\end{align}
We do not study the CP-asymmetry of the normalized lepton-forward-backward
asymmetry defined as $\mathcal{A}^\text{CP}_\text{FB} = [\oL{b} - b](q^2)
/(d[\Gamma + \oL{\Gamma}]/dq^2)$. We note that an untagged
sample actually depends on the CP asymmetry of the lepton forward-backward
asymmetry $\mathcal{A}^\text{CP}_\text{FB}(q^2) \propto d^2[\oL{\Gamma} + \Gamma]
\propto [\oL{b} - b](q^2)$. On the other hand the measurement of the
lepton forward-backward asymmetry $\AFB(q^2) \propto d^2[\oL{\Gamma} - \Gamma]
\propto [\oL{b} + b](q^2)$ requires tagging.

\begin{figure}
\centering
  \includegraphics[width=0.40\textwidth]{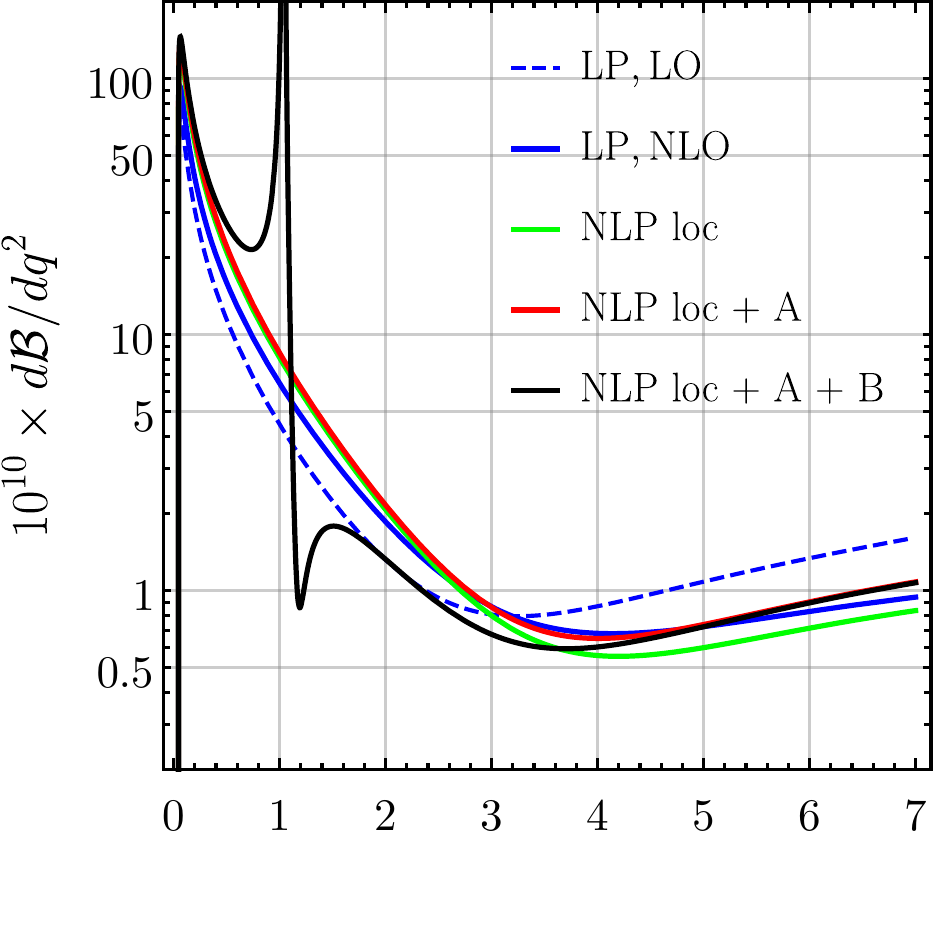}
  \hskip 0.05\textwidth
  \includegraphics[width=0.395\textwidth]{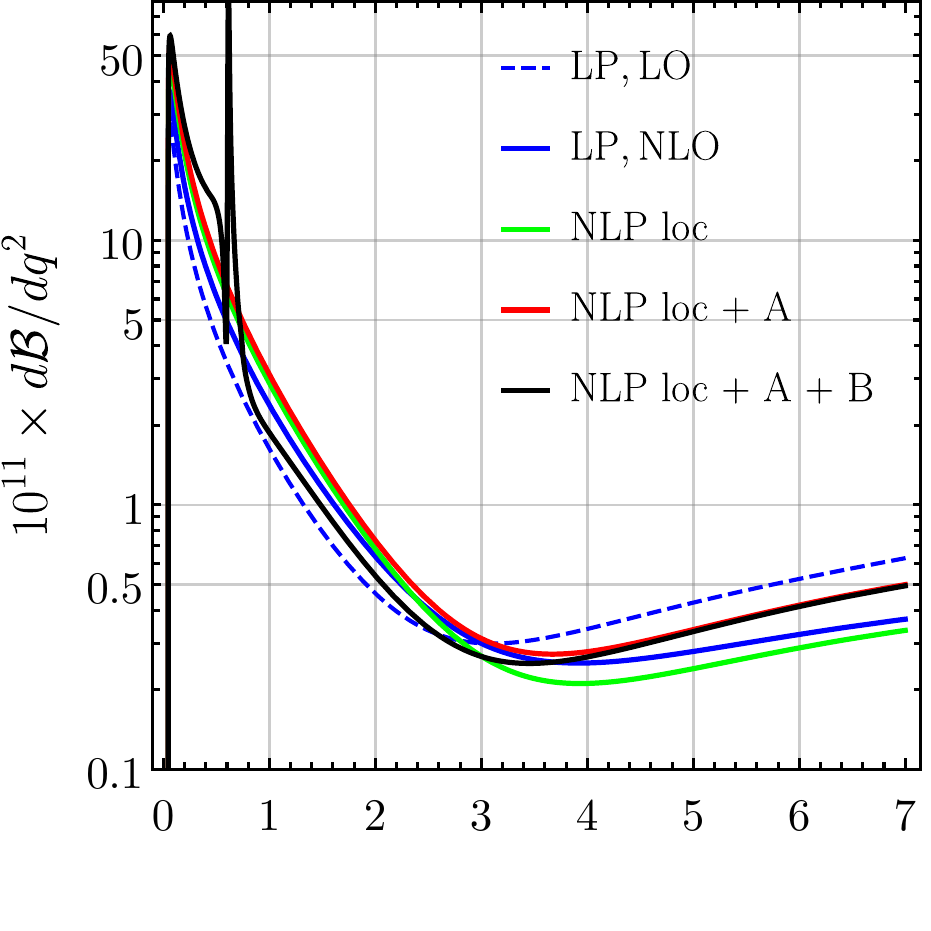}
\\[-0.5cm]
  \includegraphics[width=0.40\textwidth]{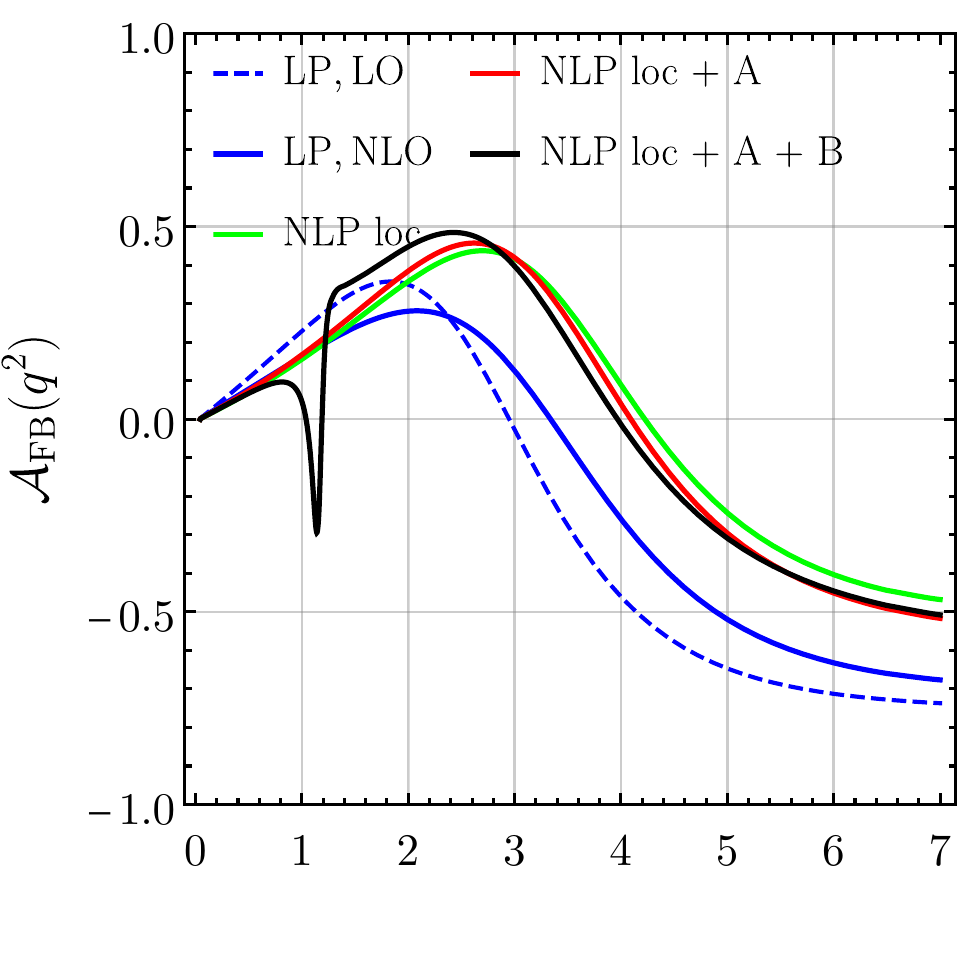}
  \hskip 0.05\textwidth
  \includegraphics[width=0.40\textwidth]{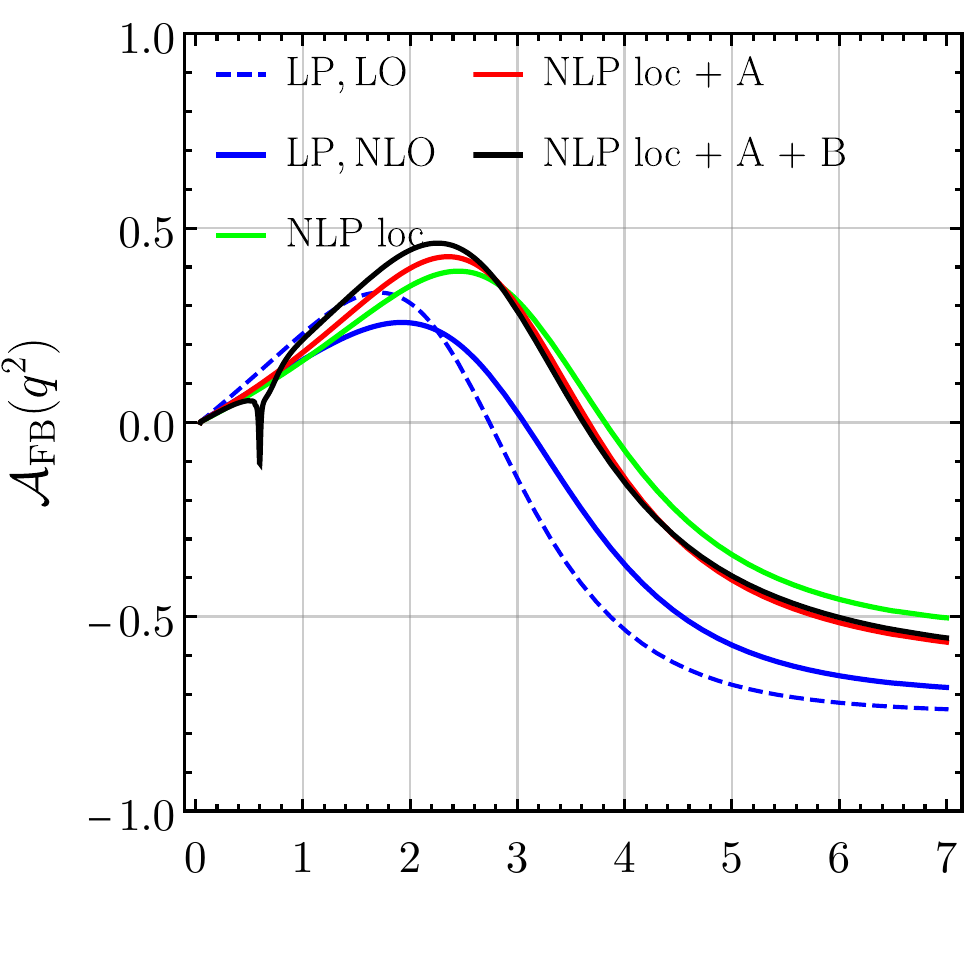}
\\[-0.5cm]
  \includegraphics[width=0.40\textwidth]{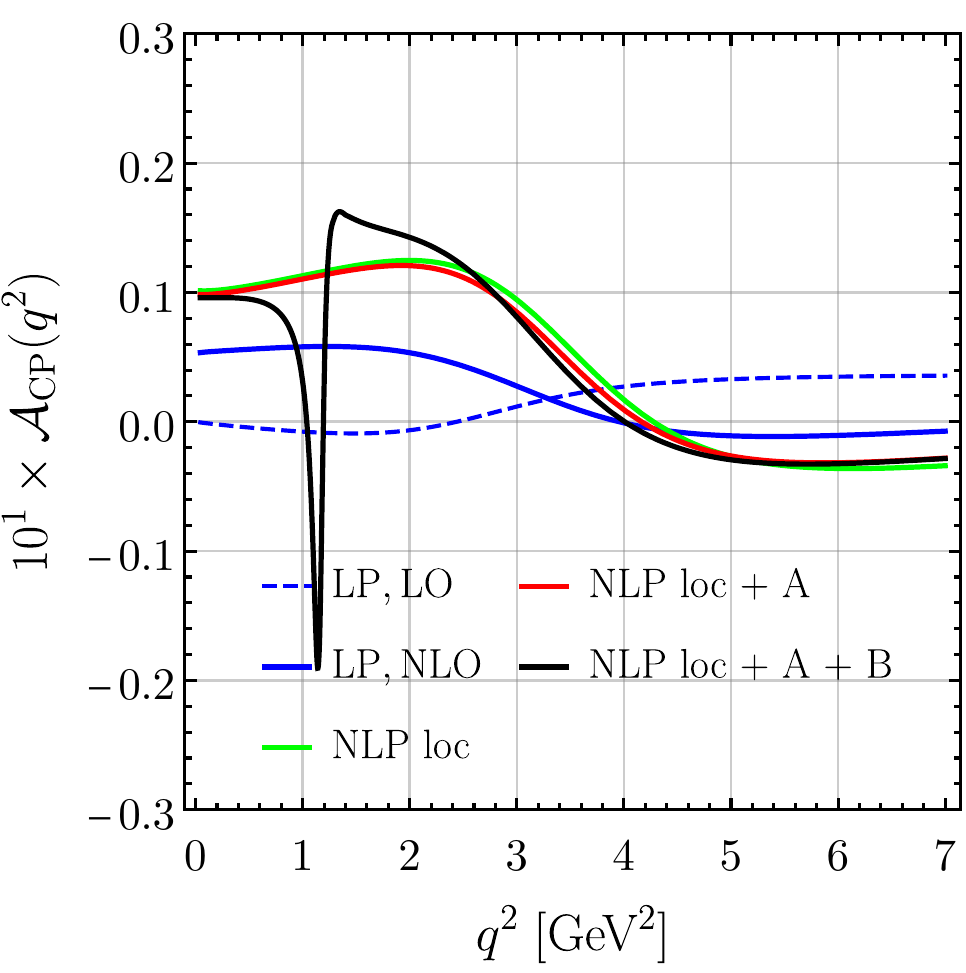}
  \hskip 0.05\textwidth
  \includegraphics[width=0.40\textwidth]{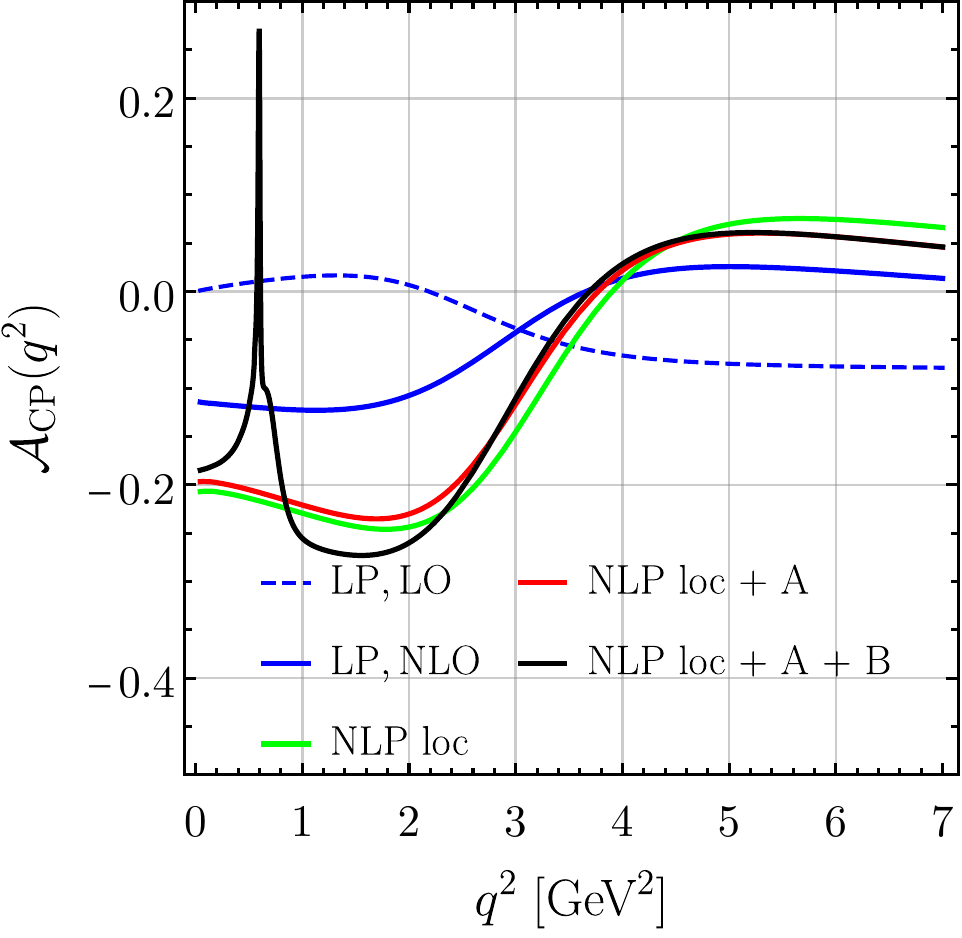}
\caption{\small The $d\BR/dq^2$ [upper], $\AFB(q^2)$ [middle] and 
  $\ACP(q^2)$ [lower] distributions 
for $B_s\to\gamma\mu\bar\mu$ [left] and
  $B_d\to\gamma\mu\bar\mu$ [right] when adding successively various 
  contributions:
  only LP [blue] at LO QCD [dashed] and NLO QCD [solid]; + local
  NLP corrections [green]; + nonlocal $A$-type [red]; 
  + nonlocal $B$-type with lowest resonances [black] NLP
  corrections. The parameter $r_\text{LP} = 0.2$.
}
\label{fig:Br-AFB-cen}
\end{figure}

The $q^2$ dependencies of the branching fraction $d\BR/dq^2$, $\AFB(q^2)$
and $\ACP(q^2)$ for $B_s\to\gamma\mu\bar\mu$ and $B_d\to\gamma\mu\bar\mu$
are shown in \reffig{fig:Br-AFB-cen} when including successively the various
higher-order QCD and NLP corrections discussed in \refsec{sec:FF-factorization}.
We recall that the calculation is valid in the ``low-$q^2$'' region $q^2
\lesssim 6\GeV^2$ or equivalently $E_\gamma \gtrsim 2.1 \GeV$
($y \gtrsim 0.8$). In the figure, the NLP corrections have been separated
into  1) purely local (``loc''), 2) nonlocal $A$-type (``A'') as given in
\eqref{eq:soft-FF-A-model} and 3) nonlocal $B$-type (``B'') as given in
\eqref{eq:soft-FF-B-model}. At LP, the NLO QCD corrections -- see ``LP, LO''
vs. ``LP, NLO'' -- decrease (increase) $d\BR/dq^2$ for $q^2 \gtrsim 3.0\GeV^2$
($q^2 \lesssim 3.0\GeV^2$) and shift the position of the zero-crossing of
$\AFB(q^2)$ by about $\Delta q^2_0 \approx + 0.5\GeV^2$ towards higher values.
Further, the NLO QCD corrections substantially change the CP asymmetry
$\ACP(q^2)$. The CP asymmetry is always tiny for $q = s$, because it is
suppressed by the Cabibbo angle, but for $q = d$ it can be reach 
$-10\%$ locally below the zero-crossing at $q^2 \gtrsim 3.5\GeV^2$. 

The local NLP corrections lead to a decrease of about $(15 -20)\%$ of
$d\BR/dq^2$ for $q^2 \gtrsim 3.0\GeV^2$, i.e. in the region where the
individual LP contributions $i = 7A\text{eff},\, 7B\text{eff},\, 9\text{eff}$
cancel, and give a large shift $\Delta q^2_0 \approx + 0.8\GeV^2$ to the 
zero of the $\AFB(q^2)$. The inclusion of $A$-type nonlocal corrections
(``NLP loc$\,+\,$A'') almost cancels the effect of ``NLP loc'' in $d\BR/dq^2$,
but not so in $\AFB(q^2)$ and $\ACP(q^2)$ for the chosen input $r^q_{fT} = 0.3$
and $r_\text{LP} = 0.2$, leaving $\ACP(q^2)$ for $q=d$ at about $-20\%$
at very low $q^2$ and $+5 \%$ at higher $q^2$. 
The NLP nonlocal $B$-type corrections for the default model with only lowest
resonances affects only the very-low $q^2$ region, with strong impact
from the $\phi(1020)$ on $d\BR/dq^2$ for $q = s$. The lowest resonances
also affect the $\AFB(q^2)$ and $\ACP(q^2)$ locally in $q^2$.
In total, the zero-crossing $q_0^2$ is significantly increased by NLP 
contributions compared to ``LP, NLO''. The difference in
$q_0^2$ between $B_s$ and $B_d$ decays of about $0.5\GeV^2$ is 
due to a different value of $\lamB{q}$, but also non-negligible 
contributions $\propto \lamCKM{u}{q}$ for $q = d$.

At very low $q^2$ the distributions shown in \reffig{fig:Br-AFB-cen} 
are strongly affected locally by the resonances from the $B$-type 
contributions, and therefore prone to the uncertainty in modelling 
the corresponding form factor. In the spirit of global parton-hadron
duality, we investigate whether such effects become averaged once
integrated over sufficiently large bins in $q^2$. The $q^2$ bins are chosen
in the low-$q^2$ region, with upper bin boundary $q^2_\text{max} = 6 \GeV^2$ 
to avoid large impacts of charmonium resonances.\footnote{The bin $q^2\in
[4 m_\mu^2,\, 8.64]\GeV^2$ is shown for comparison with \cite{Guadagnoli:2017quo}.}
The lower bin boundaries are chosen as $q^2_\text{min} = \{4 m_\mu^2,\,1.0,\,
2.0,\, 3.0,\, 4.0\} \GeV^2$, excluding $q^2_\text{min} = 1.0\GeV^2$ for $q = s$
because it is at the center of the $\phi(1020)$ peak region.\footnote{See
\cite{Kozachuk:2017mdk} for predictions in this $q^2$ bin.} The results for
$\BR(B_s\to\gamma\mu\bar\mu)$ and $\BR(B_d\to\gamma\mu\bar\mu)$  are listed
in \reftab{tab:Br-cen} for the same approximations of the various curves as in
\reffig{fig:Br-AFB-cen}.

\begin{table}
\centering
\renewcommand{\arraystretch}{1.5}
\begin{tabular}{|c|cc|ccc|cccc|}
\hline
  $q^2$ bin           
& \multicolumn{2}{|c|}{LP}
& \multicolumn{3}{|c|}{NLP}
& \multicolumn{4}{|c|}{uncertainty of ``NLP all''}
\\
  $[\GeV^2]$ & LO & NLO & loc & loc$\,+\,$A & all
& $\mu_{h,hc}$ & $\lamB{q},\, \hsigNB{1}{q}$ & $r_\text{LP}$ & total
\\
\hline \hline
  \multicolumn{10}{|c|}{$B_s\to\gamma\mu\bar\mu$}
\\
\hline
  $[4 m_\mu^2,\, 6.0]$  & 2.32 & 2.96 & 3.81 & 4.03 & 12.43 & ${}_{-0.56}^{+0.11}$ & ${}_{-1.42}^{+3.56}$ & ${}_{-1.19}^{+1.39}$ & ${}_{-1.93}^{+3.83}$ \\
  $[2.0,\, 6.0]$        & 0.40 & 0.34 & 0.31 & 0.36 &  0.30 & ${}_{-0.04}^{+0.01}$ & ${}_{-0.08}^{+0.21}$ & ${}_{-0.11}^{+0.14}$ & ${}_{-0.14}^{+0.25}$ \\
  $[3.0,\, 6.0]$        & 0.30 & 0.22 & 0.19 & 0.22 &  0.21 & ${}_{-0.03}^{+0.01}$ & ${}_{-0.07}^{+0.18}$ & ${}_{-0.08}^{+0.10}$ & ${}_{-0.10}^{+0.20}$ \\
  $[4.0,\, 6.0]$        & 0.22 & 0.15 & 0.12 & 0.15 &  0.15 & ${}_{-0.02}^{+0.01}$ & ${}_{-0.05}^{+0.14}$ & ${}_{-0.05}^{+0.07}$ & ${}_{-0.08}^{+0.16}$ \\
\hline
  $[4 m_\mu^2,\, 8.64]$ & 2.77 & 3.24 & 4.05 & 4.34 & 12.74 & ${}_{-0.60}^{+0.14}$ & ${}_{-1.50}^{+3.85}$ & ${}_{-1.31}^{+1.54}$ & ${}_{-2.08}^{+4.15}$ \\
\hline \hline
  \multicolumn{10}{|c|}{$B_d\to\gamma\mu\bar\mu$}
\\
\hline
  $[4 m_\mu^2,\, 6.0]$  &  8.77 & 11.31 & 14.07 & 15.54 & 18.43 & ${}_{-1.17}^{+0.43}$ & ${}_{-5.67}^{+17.46}$ & ${}_{-4.33}^{+5.08}$ & ${}_{-7.23}^{+18.19}$ \\
  $[1.0,\, 6.0]$        &  2.39 &  2.52 &  2.60 &  3.01 &  2.34 & ${}_{-0.28}^{+0.13}$ & ${}_{-0.69}^{+ 1.71}$ & ${}_{-0.77}^{+0.97}$ & ${}_{-1.07}^{+ 1.97}$ \\
  $[2.0,\, 6.0]$        &  1.52 &  1.25 &  1.13 &  1.43 &  1.29 & ${}_{-0.16}^{+0.06}$ & ${}_{-0.39}^{+ 1.13}$ & ${}_{-0.43}^{+0.56}$ & ${}_{-0.60}^{+ 1.26}$ \\
  $[3.0,\, 6.0]$        &  1.17 &  0.83 &  0.71 &  0.96 &  0.92 & ${}_{-0.11}^{+0.06}$ & ${}_{-0.32}^{+ 1.00}$ & ${}_{-0.32}^{+0.42}$ & ${}_{-0.47}^{+ 1.09}$ \\
  $[4.0,\, 6.0]$        &  0.86 &  0.56 &  0.49 &  0.68 &  0.67 & ${}_{-0.08}^{+0.05}$ & ${}_{-0.25}^{+ 0.76}$ & ${}_{-0.24}^{+0.31}$ & ${}_{-0.35}^{+ 0.83}$ \\
\hline
  $[4 m_\mu^2,\, 8.64]$ & 10.53 & 12.38 & 15.03 & 16.98 & 19.85 & ${}_{-1.33}^{+0.33}$ & ${}_{-6.07}^{+19.10}$ & ${}_{-4.85}^{+5.74}$ & ${}_{-7.88}^{+19.95}$ \\
\hline
\end{tabular}
\renewcommand{\arraystretch}{1.0}
\caption{\label{tab:Br-cen} \small The $q^2$-integrated branching fractions
  $10^9    \times \BR(B_s\to\gamma\mu\bar\mu)$ and
  $10^{11} \times \BR(B_d\to\gamma\mu\bar\mu)$ in four / five bins in the
  various approximations shown in \reffig{fig:Br-AFB-cen}.  ``NLP all''
  corresponds to ``NLP loc + A + B'' in \reffig{fig:Br-AFB-cen}. The last
  four columns list the uncertainties due to 1) renormalization scale
  uncertainties, 2) $B$-meson LCDA parameters and 3) the continuum fraction
  $r_\text{LP}$ as well as ``total'' when adding them in quadrature.
}
\end{table}

The block of the first three corrections in \reftab{tab:Br-cen}, ``LP, LO''
through ``NLP, loc'', are found within systematic approximations and
assumptions, and can be considered as model-independent. Omitting for
a moment the nonlocal $B$-type contributions and comparing ``NLP loc$\,+\,$A''
in the first two bins of $\BR(B_s\to\gamma\mu\bar\mu)$ shows that the photon
pole at $q^2\to 4 m_\mu^2$ in the bin $q^2 \in [4 m_\mu^2,\, 2.0\GeV^2]$ contributes almost 90\% of the rate. In this approximation the rate
is of the order $4 \cdot 10^{-9}$, comparable to the branching fraction
$\BR(B_s\to\mu\bar\mu)$ of the non-radiative mode. When adding the nonlocal
$B$-type contributions, however, the $\phi(1020)$ contribution dominates
over these ``short-distance'' corrections and enhances the rate by a factor of three to $12 \cdot 10^{-9}$, see ``NLP all'', and is responsible
for about 70\% of the signal in the bin $[4 m_\mu^2,\, 6.0\GeV^2]$. The
photon pole constitutes also about 80\% of the rate of
$\BR(B_d\to\gamma\mu\bar\mu)$ in $q^2\in [4 m_\mu^2,\, 6.0\GeV^2]$ when
omitting the nonlocal $B$-type contributions due to the $\rho$ and $\omega$
resonances, which however have only a small impact compared to the $\phi$
in $\BR(B_s\to\gamma\mu\bar\mu)$, as expected from the discussion 
of duality violation in \refsec{sec:duality}. They enhance the rate by
a factor $1.2$ to about $1.8 \cdot 10^{-10}$, which is comparable to the
SM prediction of $\BR(B_d\to\mu\mu) \approx 1 \cdot 10^{-10}$.
The comparison of the first bin $[4 m_\mu^2,\,6.0\GeV^2]$ to the 
second bin shows that the branching fractions are reduced by a 
factor of about $40$ and $8$ for $q = s$ and $q = d$, respectively,
once the photon pole
contribution is excluded, making them an order of magnitude smaller than
for the corresponding non-radiative rare decays, $B_q \to \mu\bar\mu$.
Note, however, that for $\ell = e$ the $B_q \to e\bar{e}$ rates are strongly
helicity suppressed by $(m_e/m_\mu)^2$, whereas the rates of $B_q \to \gamma
e\bar{e}$ are identical to $B_q \to \gamma \mu\bar{\mu}$ in the SM up to
phase-space effects for very low $q^2$.

Comparison of the value $\BR(B_s\to\gamma\mu\bar\mu) = (8.4 \pm 1.3) \cdot
10^{-9}$ in the $q^2$ bin $[4 m_\mu^2,\, 8.64 \GeV^2]$ given in
\cite{Guadagnoli:2017quo} with \reftab{tab:Br-cen} shows that our prediction 
is about a factor 1.5 larger and within the given errors there is
some tension. Here we used the QCD factorization approach to compute the
$A$- and $B$-type contributions,
allowing us also to include at LP the NLO QCD corrections, contrary to
\cite{Guadagnoli:2017quo}. In addition, we include in our model of the
nonlocal NLP $A$- and $B$-type contributions also a continuum contribution. 
On the other hand we do not include a model for charmonium resonances (but
rather stay away from them by restricting $q^2 \lesssim 6$~GeV$^2$),
which could be responsible for some differences when $q^2$ approaches 
$8 \GeV^2$. Similar comments apply to \cite{Kozachuk:2017mdk} who predict
$\BR(B_s\to\gamma\mu\bar\mu) = 7.79 \cdot 10^{-9}$ in the bin $q^2 \in
[4 m_\mu^2,\, 6.0\GeV^2]$ and $\BR(B_d\to\gamma\mu\bar\mu) = (1.02 \pm 0.16)
\cdot 10^{-11}$ in the bin $q^2 \in [1.0,\, 6.0]\GeV^2$, roughly a factor of 1.6 and 2.3, respectively,  smaller than our results in \reftab{tab:Br-cen}.
We attribute these differences to the difference between the QCD
factorization computation of the $B\to \gamma^*$ form factors and the more model-dependent approaches used in \cite{Guadagnoli:2017quo, Kozachuk:2017mdk}, and differences in the numerical values of the 
$T_1^{B_q \to V}$ form factors in the parameterization of the 
nonlocal NLP $B$-type contribution relative to \cite{Kozachuk:2017mdk}.

\begin{figure}
\centering
  \includegraphics[width=0.40\textwidth]{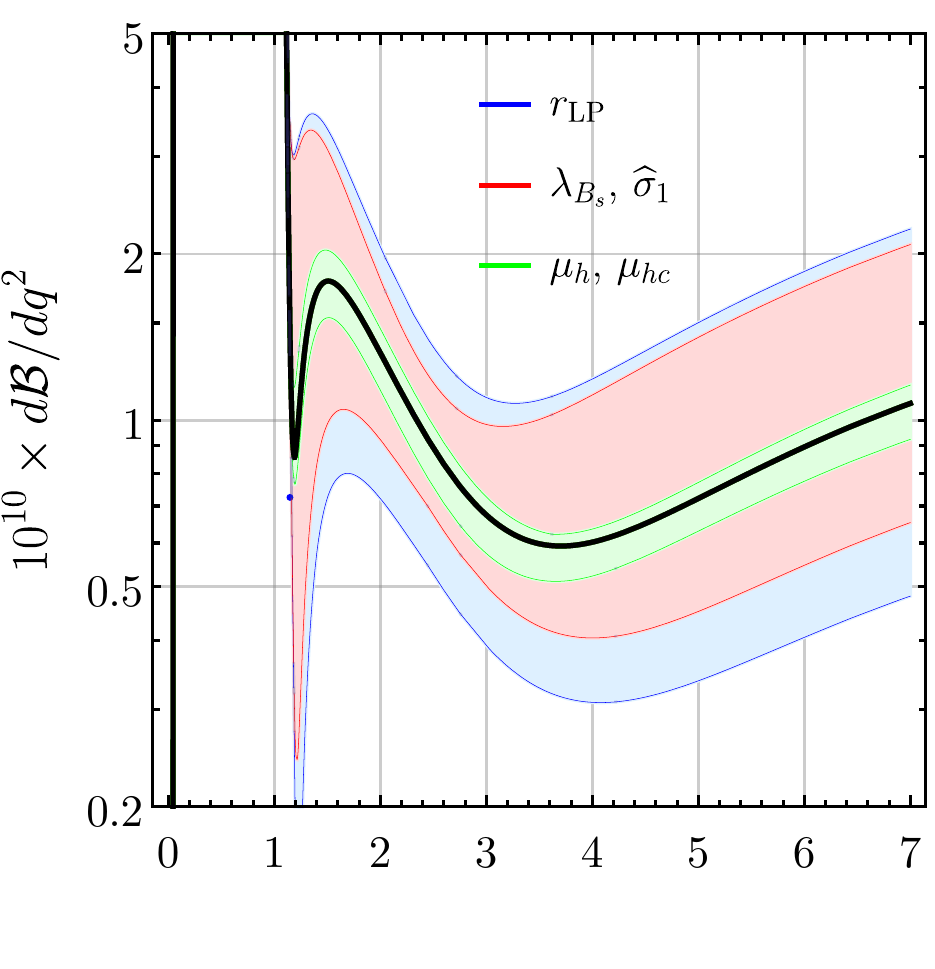}
  \hskip 0.05\textwidth
  \includegraphics[width=0.40\textwidth]{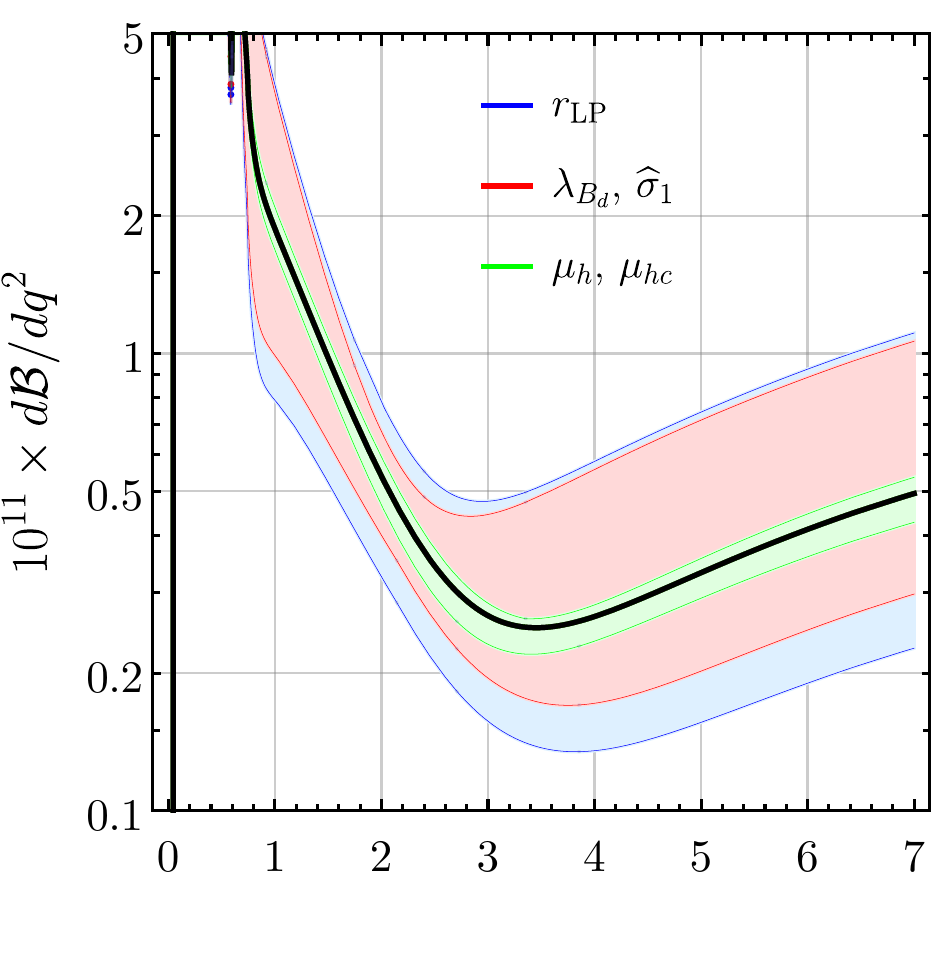}
\\[-0.52cm]
  \includegraphics[width=0.40\textwidth]{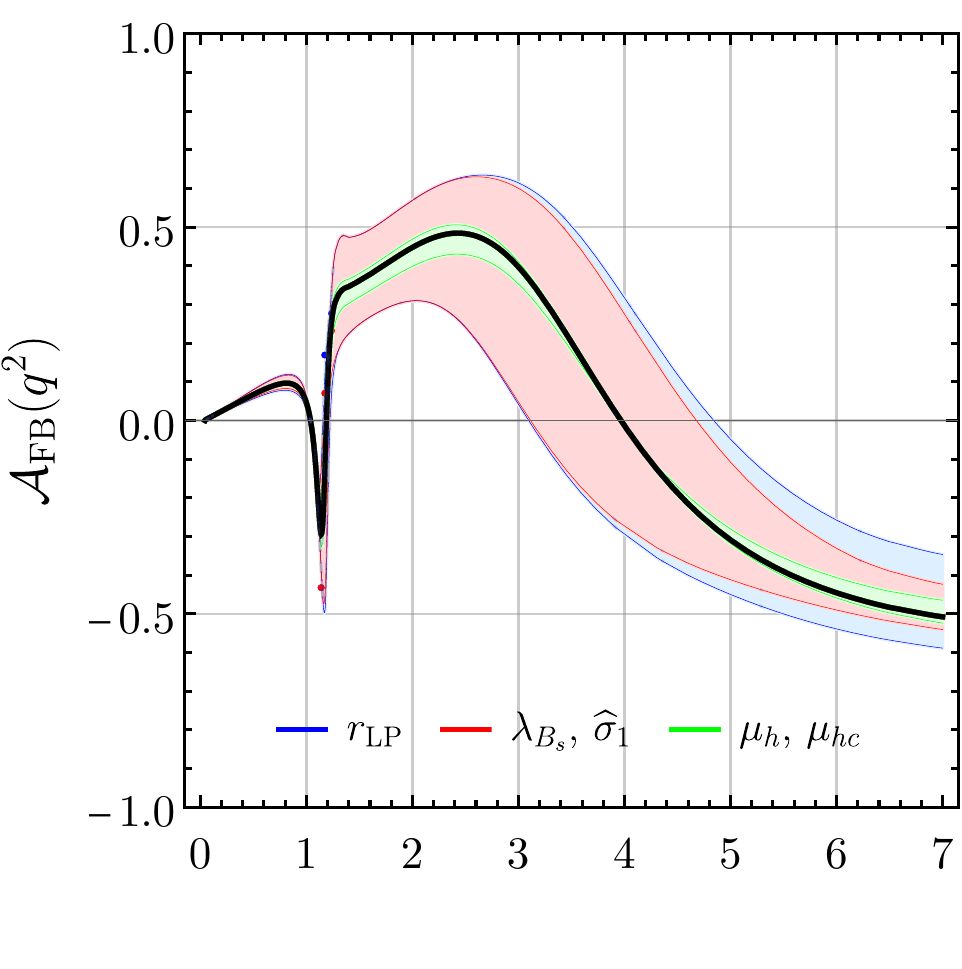}
  \hskip 0.05\textwidth
  \includegraphics[width=0.40\textwidth]{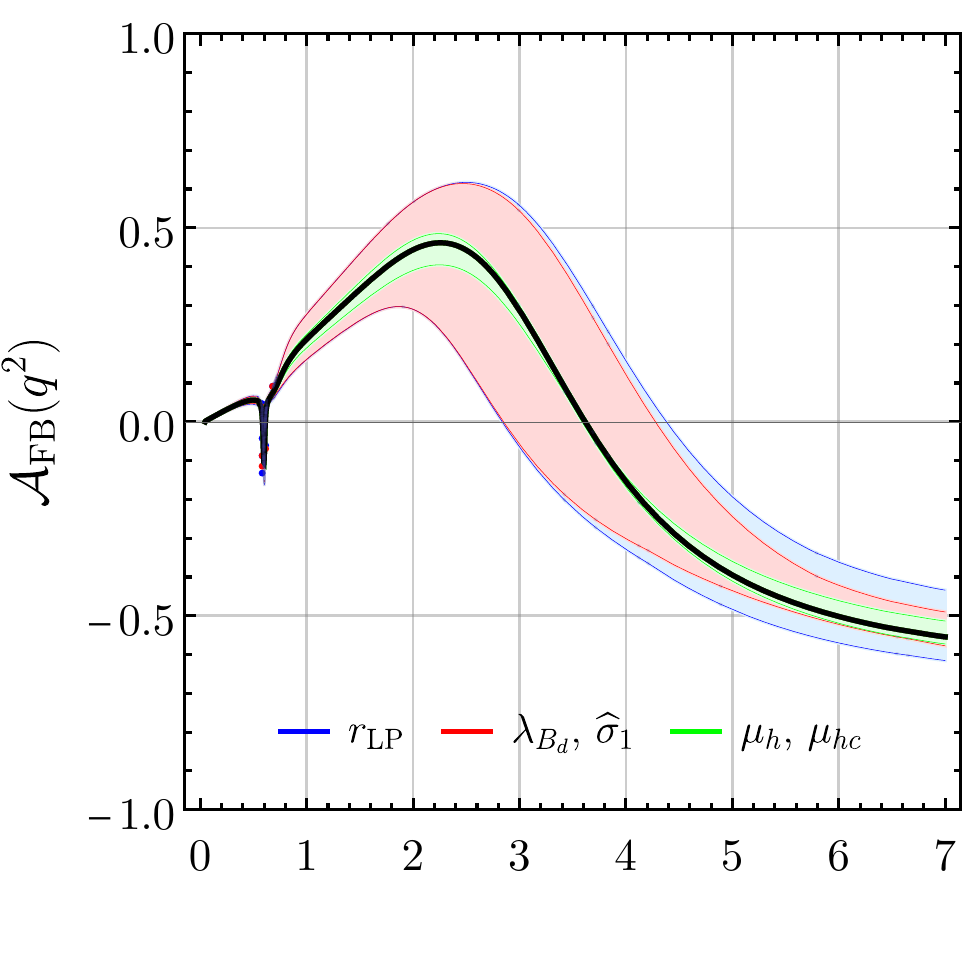}
\\[-0.52cm]
  \includegraphics[width=0.40\textwidth]{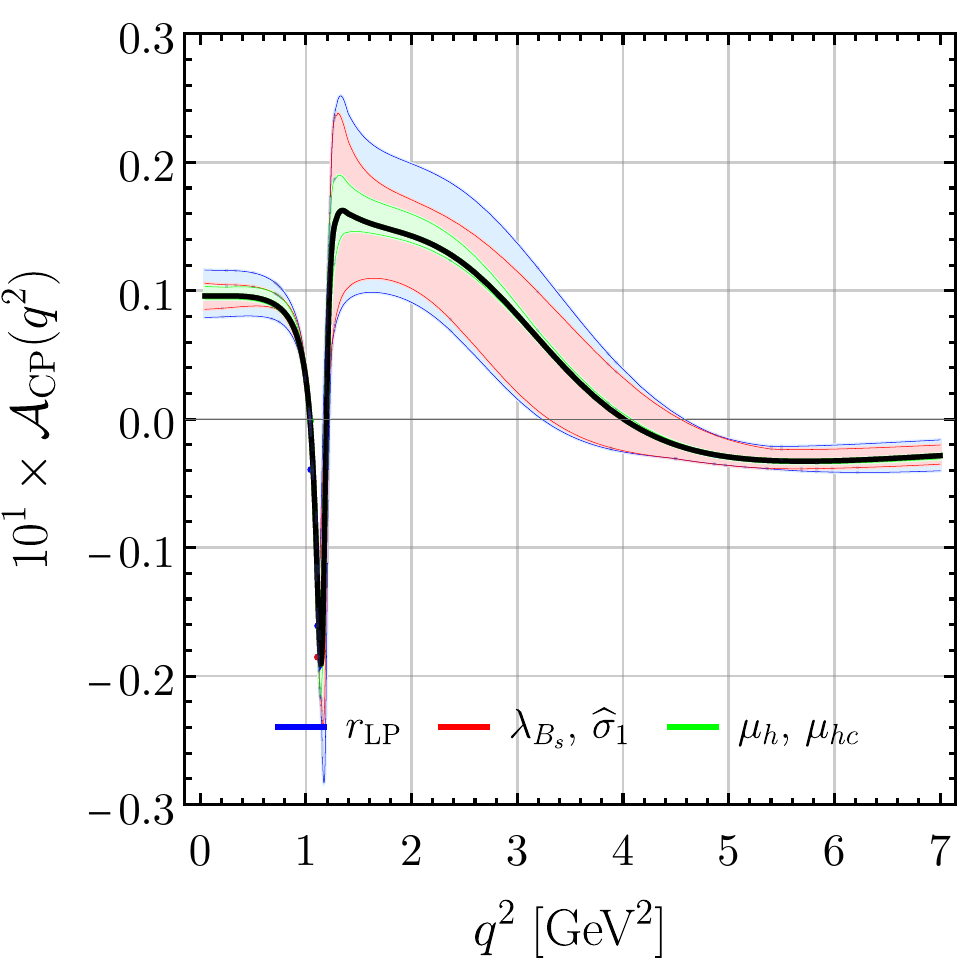}
  \hskip 0.05\textwidth
  \includegraphics[width=0.40\textwidth]{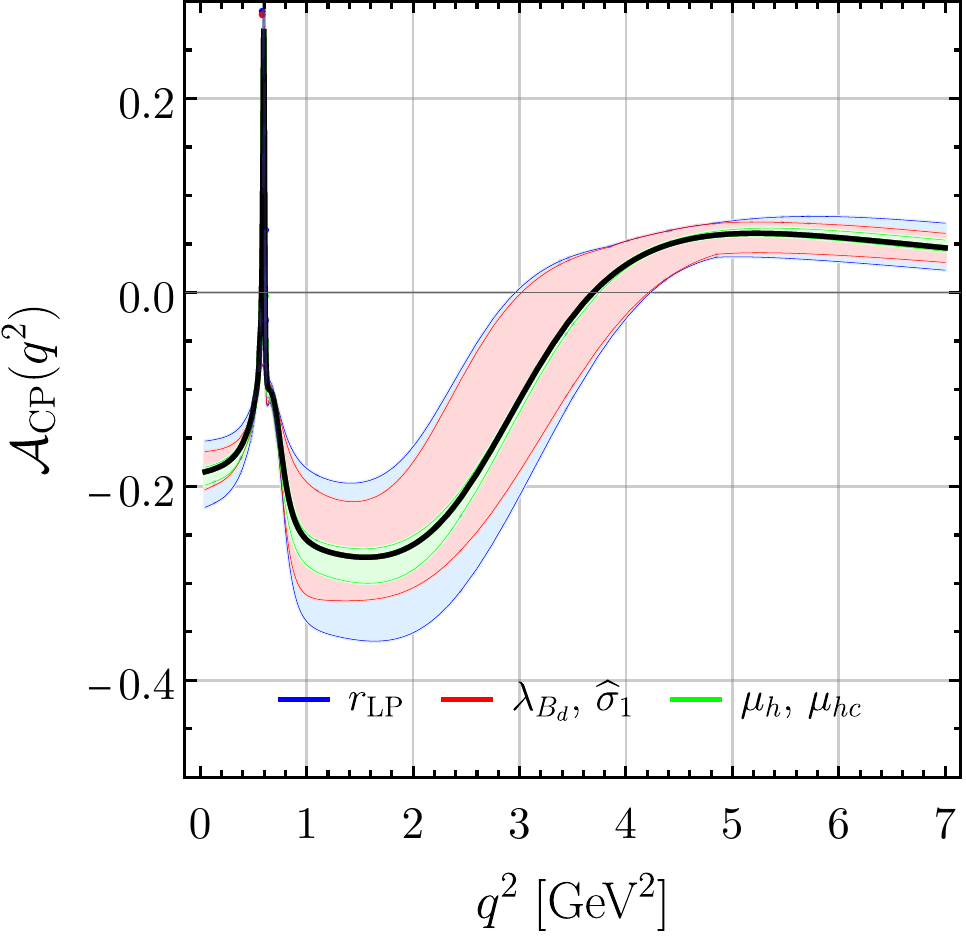}
\caption{\small 
  Uncertainty budget for $d\BR/dq^2$ [upper], $\AFB(q^2)$ [middle] and
  $\ACP(q^2)$ [lower] for $q = s$ [left] and $q = d$ [right] when
  adding successively in quadrature the uncertainties due to renormalization
  scales ($\mu_{h,hc}$) [green], $B$-meson LCDA parameters ($\lamB{q},\,
  \hsigNB{q}{1}$) [red] and the continuum fraction ($r_\text{LP}$) [blue].
}
  \label{fig:Br-unc} 
\end{figure}

We calculate the theoretical uncertainties from 1) scale variation,
2) $B$-meson LCDA parameters and 3) the modelling of the nonlocal $A$-
and $B$-type contributions, as discussed for the amplitudes in
\refsec{sec:Bqllgam-amp}. They are listed for the best approximation
``NLP, all'' in \reftab{tab:Br-cen} and shown in \reffig{fig:Br-unc}
for the $q^2$- distributions $d\BR/dq^2$, $\AFB(q^2)$ and $\ACP(q^2)$.
The default 1(2)-Res model is used and the uncertainties are added
successively in quadrature.

The largest uncertainty is due to the LCDA parameters, amounting to 
about ${}^{+70}_{-30}\,$\% for $B_s\to \gamma\mu\bar\mu$ in the 
bin $q^2 \in [2.0,\,  6.0] \GeV^2$ (see \reftab{tab:Br-cen}),
in particular the first inverse moment of the $B$-meson LCDA, $\lamB{q}$,
since roughly the LP is proportional to $1/\lamB{q}$.\footnote{The
dependence on the renormalization scales $\mu_{h,hc}$ reduces when
including NLO QCD corrections to the LP amplitude. At NLP there would
be a strong uncancelled $\mu_h$ dependence of $\oL{m}_b C_7^\text{eff}$
in the model of the nonlocal $A$- and $B$-type contributions in
\eqref{eq:soft-FF-A-model} and \eqref{eq:soft-FF-B-model}. We neglect
this particular scale dependence, which is a consequence of our chosen
model, and find that the remaining $\mu_h$ dependence of the local NLP
corrections is small. The scale uncertainties is sub-leading, less than
$15\,$\%, compared to the parametric uncertainties.} It is even larger
for $B_d\to \gamma\mu\bar\mu$ in that $q^2$ bin, because here the
relative variation of $\lamB{d}= (350 \pm 150)\MeV$ is larger and the
minimal value in the variation is $200\MeV$ to be compared to
$\lamB{s} = (400 \pm 150)\MeV$.
The variation $r_\text{LP} = 0.2 \pm 0.2$ of the continuum fraction in the
form factors model for the nonlocal $A$- and $B$-type contributions
contributes another $\pm (35 - 45)\,$\% to the uncertainty when going to 
$q_\text{min}^2 \gtrsim 2.0 \GeV^2$, beyond the region of the lowest
resonances. The large uncertainty of the power-suppressed contributions
does not come as a surprise given that the continuum is modelled as a
fraction of up to 40\% of the LP amplitude. 

\begin{figure}
\centering
  \includegraphics[width=0.41\textwidth]{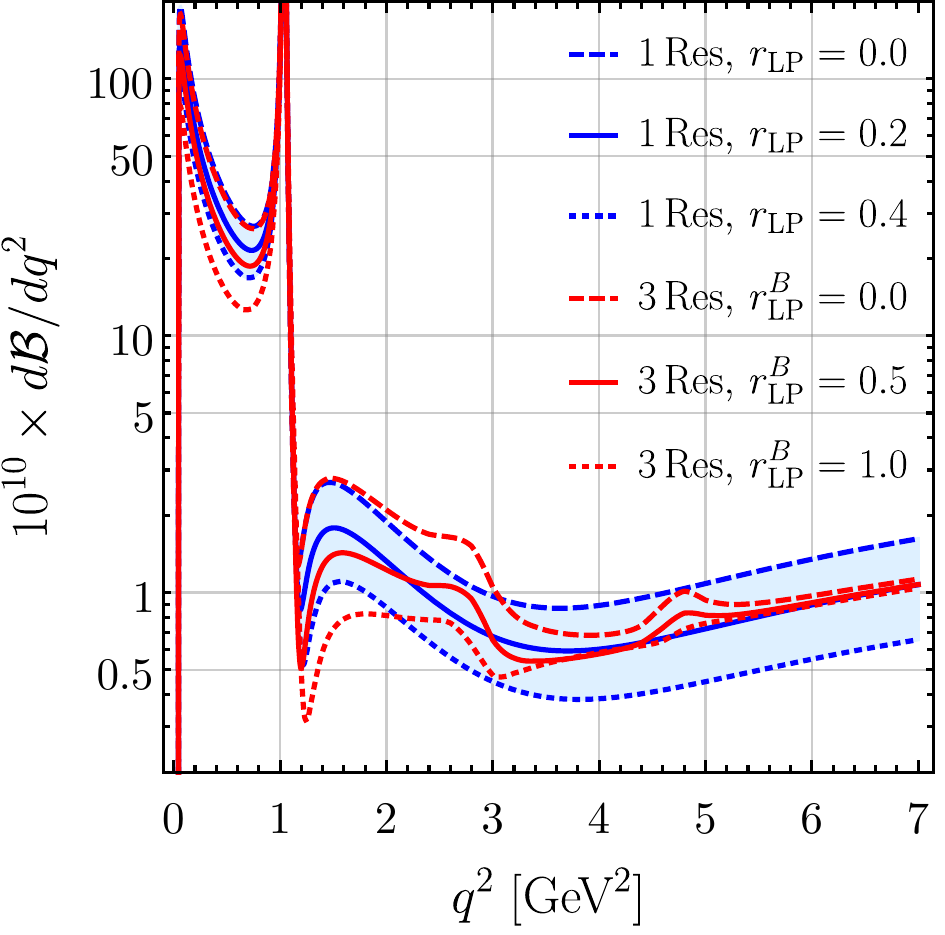}
  \hskip 0.05\textwidth
  \includegraphics[width=0.41\textwidth]{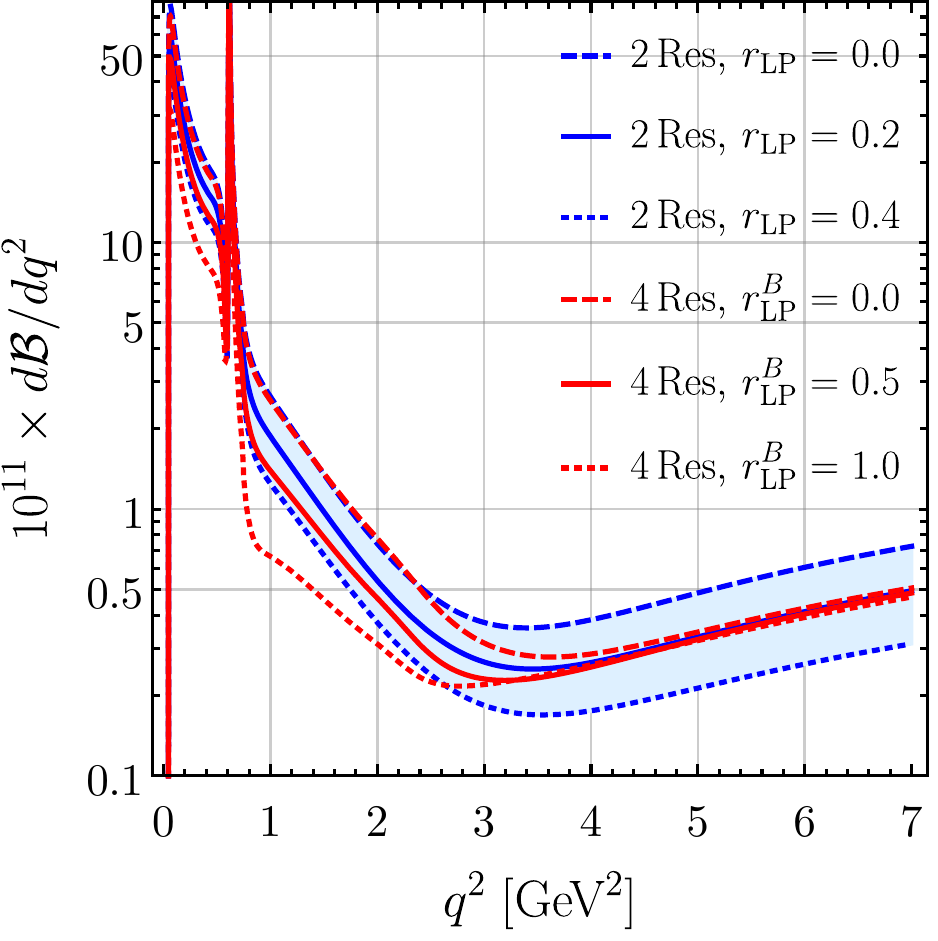}
\\[0.4cm]
  \includegraphics[width=0.41\textwidth]{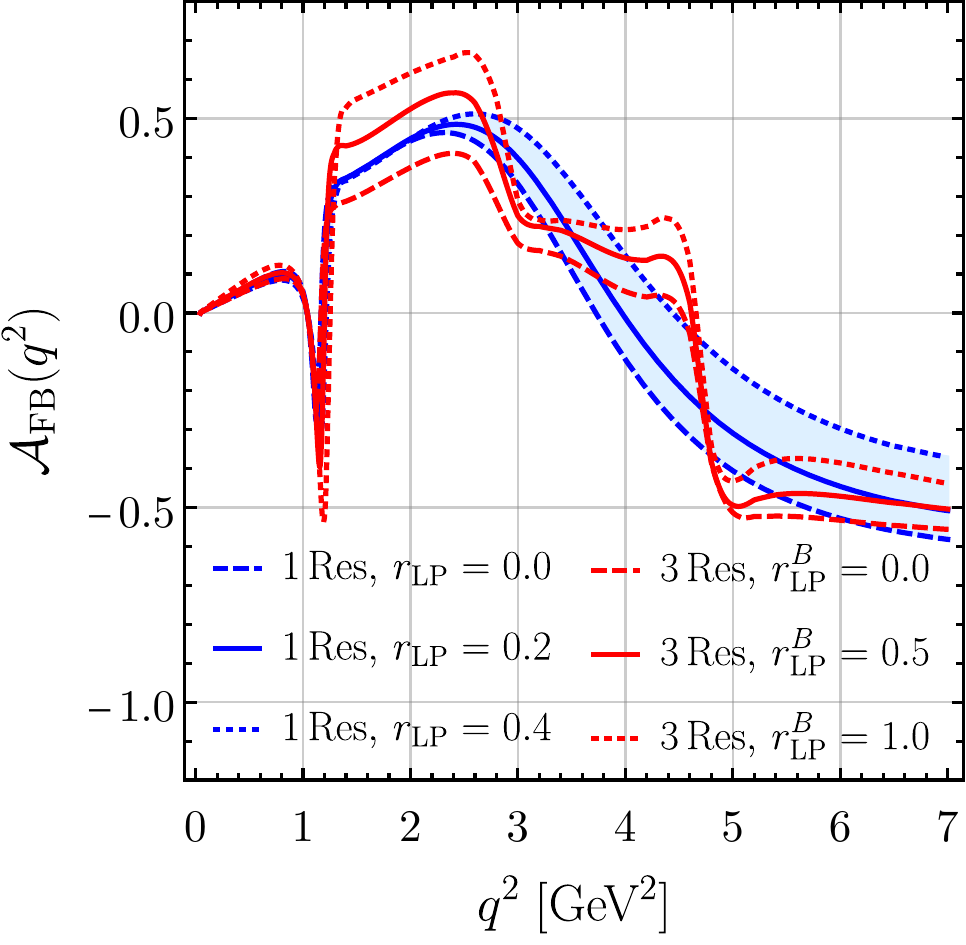}
  \hskip 0.05\textwidth
  \includegraphics[width=0.41\textwidth]{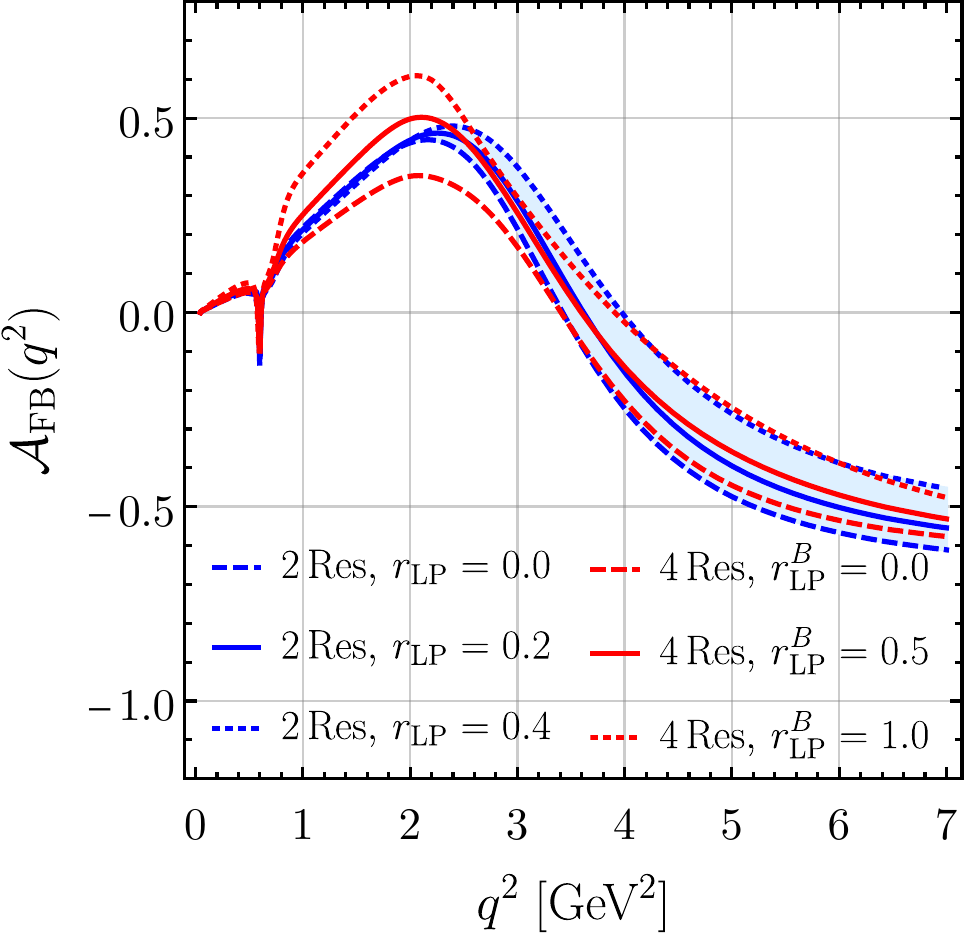}
\caption{\small 
  Comparison of the 1(2)-Res model with a continuum of $r_\text{LP} =
  0.2 \pm 0.2$ [blue] with the 3(4)-Res model for different values $r_\text{LP}^B$
  with fixed $r_\text{LP}^A = 0.2$ [red] for $d\BR/dq^2$ [upper] and
  $\AFB(q^2)$ [lower] in $B_s\to\gamma\mu\bar\mu$ [left] and
  $B_d\to\gamma\mu\bar\mu$ [right].
}
  \label{fig:Br-1+cont_vs_3} 
\end{figure}

The nonlocal $B$-type corrections affect the $q^2$-differential distributions
particularly in the very-low  $q^2$ domain as shown at the amplitude level in
\reffig{fig:NLP-B-type}. There the default 1(2)-Res model was compared to the
3(4)-Res model for various values of the continuum fraction $r_\text{LP}^B$.
This comparison is extended to the observables $d\BR/dq^2$ and $\AFB(q^2)$
in \reffig{fig:Br-1+cont_vs_3}. The 1(2)-Res model with a continuum parameter
$r_\text{LP}^A = r_\text{LP}^B = 0.2 \pm 0.2$ (blue) is compared to the 3(4)-Res
model for various values $r_\text{LP}^B$ and fixed $r_\text{LP}^A = 0.2$ (red).
The variation of $r_\text{LP}$ in the 1-Res model spans a band that includes
in the case of $d\BR[B_s\to\gamma\mu\bar\mu]/dq^2$ the effect of $\phi(1680)$
in the 3-Res model for values $r_\text{LP}^B \approx 0.5$ and for the
$\phi(2170)$ even for all values $r_\text{LP}^B \approx [0.0,\, 1.0]$. For 
$\AFB(q^2)$ the higher resonances are locally not completely covered by this
band, but oscillate around it, such that suitable binning would indeed average
out the resonances. The comparison of the 2-Res with the 4-Res model in
$B_d\to\gamma\mu\bar\mu$ shows a small impact of the higher resonances in all
distributions as expected from the discussion of the corresponding amplitudes.

\begin{figure}
\centering
  \includegraphics[width=0.40\textwidth]{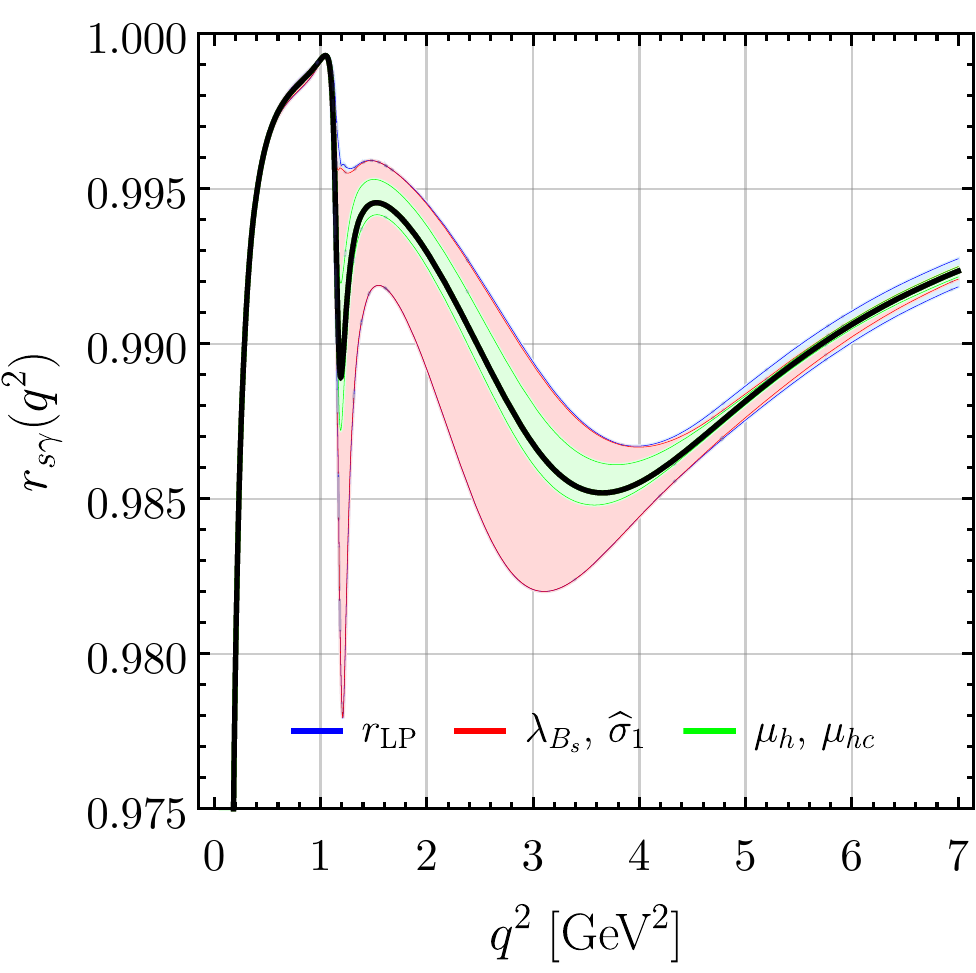}
  \hskip 0.05\textwidth
  \includegraphics[width=0.40\textwidth]{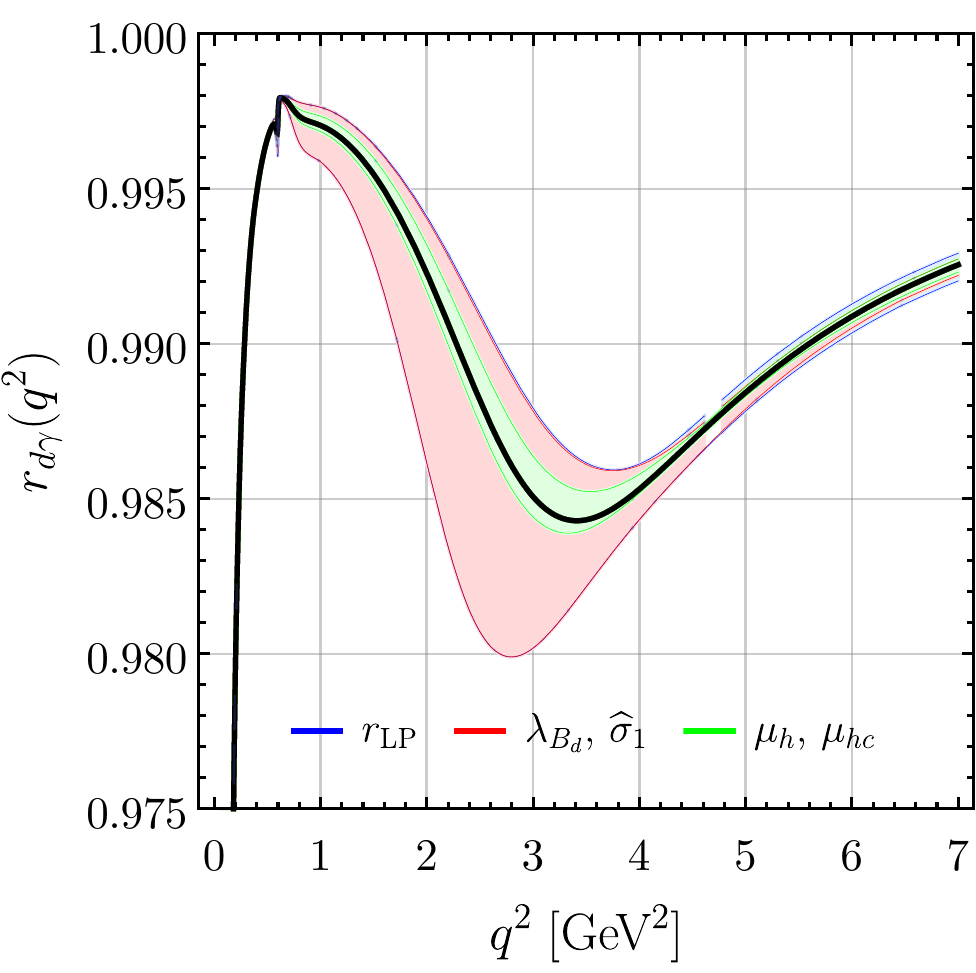}
\caption{\small 
  The ratio $r_{q\gamma}$ for $B_s\to \gamma\ell\bar\ell$ [left] and
  $B_d\to \gamma\ell\bar\ell$ [right] with uncertainty budget as
  in \reffig{fig:Br-unc}.
}
  \label{fig:LFV-r} 
\end{figure}

The ratios of observables with different lepton flavours provide interesting
tests of lepton-flavour universality. The deviations from unity are tiny in
the SM, and hadronic uncertainties cancel to large extent. Here we briefly
discuss the ratio 
\begin{align}
  r_{q\gamma}(q^2) &
  = \frac{d\BR[B_q\to \gamma\mu\bar\mu]/dq^2}{d\BR[B_q\to \gamma e\bar{e}]/dq^2}
& 
  (q & = d,s) .
\end{align}
As  shown in \reffig{fig:LFV-r}, these ratios take values between $(0.98-1.00)$
for both $q = s, d$,  except for $q^2 \lesssim 0.4 \GeV^2$. Within the framework
of our calculation, lepton-mass dependence arises only from $\beta_\ell$ through
the phase-space measure, not from the amplitudes themselves. In 
\reftab{tab:R_q-gamma} we give the ratio
\begin{align}
  R_{q\gamma} &
  = {\displaystyle \int_{q^2_\text{min}}^{q^2_\text{max}} dq^2 
     \frac{d\BR[B_q\to \gamma\mu\bar\mu]}{dq^2}} \Bigg/
    {\displaystyle \int_{q^2_\text{min}}^{q^2_\text{max}} dq^2
     \frac{d\BR[B_q\to \gamma e\bar{e}]}{dq^2}}
& 
  (q & = d,s)
\end{align}
for selected bins in $q^2$. $R_{q\gamma}$ is always very close to 
unity except when $q^2_\text{min}$
is below $1\GeV^2$. The considered sources of uncertainties cancel to very
high degree, and in view of the remaining tiny errors one should keep in mind
that further not included higher order QCD and QED as well as NLP corrections
can contribute at this level.

\begin{table}
\centering
\renewcommand{\arraystretch}{1.5}
\begin{tabular}{|c|cc|cc|}
\hline
  $q^2$ bin           
& \multicolumn{2}{|c|}{$B_s\to\gamma\ell\bar\ell$}
& \multicolumn{2}{|c|}{$B_d\to\gamma\ell\bar\ell$}
\\
  $[\GeV^2]$ 
& cen & total
& cen & total
\\
\hline 
  $[4 m_\mu^2,\, 6.0]$  & 0.9716 &  ${}_{-0.0137}^{+0.0081}$ & 0.9263 & ${}_{-0.0093}^{+0.0046}$ \\
  $[1.0,\, 6.0]$        & ---    &  ---                      & 0.9913 & ${}_{-0.0028}^{+0.0013}$ \\
  $[2.0,\, 6.0]$        & 0.9883 &  ${}_{-0.0021}^{+0.0012}$ & 0.9878 & ${}_{-0.0022}^{+0.0011}$ \\
  $[3.0,\, 6.0]$        & 0.9875 &  ${}_{-0.0009}^{+0.0007}$ & 0.9874 & ${}_{-0.0009}^{+0.0005}$ \\
  $[4.0,\, 6.0]$        & 0.9883 &  ${}_{-0.0006}^{+0.0005}$ & 0.9885 & ${}_{-0.0006}^{+0.0004}$ \\
\hline
  $[4 m_\mu^2,\, 8.64]$ & 0.9721 &  ${}_{-0.0130}^{+0.0077}$ & 0.9308 & ${}_{-0.0078}^{+0.0037}$ \\
\hline
\end{tabular}
\renewcommand{\arraystretch}{1.0}
\caption{\label{tab:R_q-gamma}
  \small The $q^2$-integrated lepton-flavour universality ratios 
  $R_{q\gamma}$ for $B_s\to\gamma\ell\bar\ell$ [left] and
  $B_d\to\gamma\ell\bar\ell$ [right] in several $q^2$ bins for
  the approximation``NLP all''. The total uncertainty is
  obtained as for \reftab{tab:Br-cen}.
}
\end{table}

So far we discussed the so-called instantaneous distributions at the 
time of the production of the neutral $B_q$ meson without accounting
for $B_q$-$\oL{B}_q$ mixing. The time evolution
due to mixing between production and decay further complicates the
predictions of experimentally measurable observables. Thus, besides
tagged and untagged there can be time-integrated or time-resolved
measurements, all of which require a specific theoretical treatment
of the time-dependence. In particular time-resolved measurements allow
to measure effective lifetimes as well as direct and mixing-induced
CP-asymmetries.

The modifications due to time-dependence might not be negligible 
for $B_s$ mesons, but
are certainly small compared to the current parametric 
uncertainties. The 
$\Bstollgamma$ observable that is most likely to be measured first 
is the time-integrated branching ratio
\begin{align}
  \frac{d\langle \BR \rangle}{dq^2} &
  = \frac{1}{2} \int_0^\infty \! dt \; \left(
    \frac{d\Gamma[\oL{B}_q(t) \to \gamma\ell\bar\ell]}{dq^2} +
    \frac{d\Gamma[B_q(t) \to \gamma\ell\bar\ell])}{dq^2} \right),
\end{align}
hence we focus on this to explore the phenomenological impacts
of the $B_q$-meson width difference. 
At the time $t = 0$
of the production of the $B_q$ meson the time-dependent decay rate
$d\Gamma[\oL{B}_q(t = 0) \to \gamma\ell\bar\ell]/dq^2 = d\oL{\Gamma}/dq^2$
is given by the instantaneous one in \eqref{eq:decay-rate} and
correspondingly for the CP-conjugated decay
$d\Gamma[B_q(t=0) \to \gamma\ell\bar\ell]/dq^2 = d\Gamma/dq^2$. In the
case of $B_s \to \gamma\ell\bar\ell$, the time-dependence can be
included for the time-integrated branching fraction by the simple
replacement 
\begin{align}
  \label{eq:Bs-time-dep}
  |\oL{\calA}_{\parallel \chi}|^2 &
  \to \frac{1}{1 + y_s} \,|\oL{\calA}_{\parallel \chi}|^2 ,
&
  |\oL{\calA}_{\perp \chi}|^2 &
  \to \frac{1}{1 - y_s} \,|\oL{\calA}_{\perp \chi}|^2 , 
&
  (\chi & = V, A)
\end{align}
in $\oL{a}(q^2)$ and $\oL{c}(q^2)$, where $y_s \equiv 
\Delta\Gamma_s/(2\Gamma_s)$.
This holds in the SM where the tiny $\lamCKM{u}{s}$ term can be 
neglected and no other weak phases than those in $V_{tb}^{} V_{ts}^*$ contribute.
The width difference of $B_s$ is already well measured to $y_s = 0.065\pm 0.003$
\cite{Amhis:2019ckw}. For $B_d\to\gamma\ell\bar\ell$, however, the $\lamCKM{u}{d}$
term cannot be neglected and the
formula for the time-integrated rate is slightly more involved. 
On the other hand, the upper bound on the width difference 
in the ballpark of $|y_d| \lesssim 0.005$ makes width effects 
negligible anyway. We computed the time-integrated
branching fractions for the bins in \reftab{tab:Br-cen} and find 
the following changes of the central values for 
$B_s \to \gamma\mu\bar\mu$: 
\begin{align}
  && &&
  10^9 \times \BR_{[q_\text{min}^2,\, q_\text{max}^2]} & \to 
  10^9 \times \langle \BR \rangle_{[q_\text{min}^2,\, q_\text{max}^2]}
\notag\\
  q^2 & \in [4 m_\mu^2,\, 6.0] \GeV^2 & : && 12.43 & \to 12.51 
\nonumber \\
  q^2 & \in [2.0,\, 6.0] \GeV^2       & : && 0.300 & \to 0.304 
\nonumber \\
  q^2 & \in [3.0,\, 6.0] \GeV^2       & : && 0.207 & \to 0.212 
\nonumber \\
  q^2 & \in [4.0,\, 6.0] \GeV^2        & : && 0.146 & \to 0.150 
\end{align}
These changes correspond to factors $1.006, 1.013, 1.024$ and 
$1.027$ with respect to the instantaneous branching ratios. 
The changes are even smaller than the $\mathcal{O}(y_s) \approx \mp 
\,6.5\%$ 
corrections of the individual transversity amplitudes squared 
(\ref{eq:Bs-time-dep}) due to cancellation between the two 
in the branching fraction. Such a cancellation 
should be expected, since at LP the two transversity amplitudes 
are equal, and in this case the width difference affects the 
sum of the two amplitudes by the factor $1/(1-y_s^2)$, 
which deviates from unity only by  $\mathcal{O}(y_s^2)$.
Overall, the width difference results in tiny modifications 
of the instantaneous rate compared to parametric 
and theoretical uncertainties. The
time-dependence of other observables can be worked out as well 
if required, but we refrain from doing so here.

%
%
%

\section{Summary and conclusions}
\label{sec:concl}

In this paper we provided the first analysis of the rare 
radiative $\Bqtollgamma$ decays in the kinematic 
regime of large photon energy with QCD factorization techniques, 
which construct a systematic expansion in inverse powers 
of large photon energy and heavy quark mass.
As in the case of $\Butolvgamma$, a two step matching on \SCETI{} and 
\SCETII{} leads to a systematic decoupling of hard and hard-collinear
degrees of freedom. At leading power (LP) in $1/E_\gamma$, $1/m_b$ 
this allows us to include the sizeable $\mathcal{O}(\alpha_s)$ QCD 
corrections, which have previously been omitted. 
The neutral current radiative decays are more involved than
$\Butolvgamma$, since the real photon is not only emitted
from the constituent quarks of the initial $B$-meson 
($A$-type emission), but also directly from an 
operator in the weak EFT ($B$-type emission). Power-counting
shows that the $B$-type emission also contributes at LP.

Further, we analyze the next-to-leading power (NLP) corrections 
and provide all local corrections, which can be calculated 
unambiguously. We also investigate the nonlocal corrections 
to $A$- and $B$-type emission from subleading-power form factors. 
A new feature compared to factorization calculations of 
$\Butolvgamma$ is the NLP form factor in the time-like region, 
which is affected by resonances in the region of very small 
dilepton mass, for which we adopted a parameterization in 
terms of resonances and a continuum term. 
We note that the well-known partial cancellation of the contributions
from the operators $\Op_9$ and $\Op_7$ to the transversity/helicity 
amplitudes, which is also responsible for the zero at $q_0^2$ in the
lepton-forward-backward asymmetry, makes the NLP corrections 
very important. This together with the resonance contamination 
in the very low $q^2$ region, the restriction to 
$q^2 \lesssim 6$~GeV$^2$ to avoid the charmonium resonance, 
and the large uncertainty from the 
$B$-meson LCDA parameters leads to the conclusion that  
precise calculations of rare radiative $\Bqtollgamma$ decay 
observables are difficult in practice.

The analytical results allow us to provide predictions in the 
Standard Model for the $q^2$-differential and $q^2$-integrated 
CP-averaged branching fraction ($d\BR/dq^2$), lepton-forward-backward asymmetry
($\AFB$), rate CP-asymmetry $(\ACP)$, and lepton-flavour ratios. 
We quantified the dominant uncertainties from renormalization scales,
the parametric dependencies from the $B$-meson light-cone distribution
amplitude (LCDA) as well as the modelling of nonlocal NLP corrections.
These studies show that the rate for the $B_s$ decay mode is 
strongly increased by the $\phi$ resonance due to the $B$-type contribution
of the operator $\Op_7$ at very low $q^2 \lesssim  2.0\GeV^2$, but becomes small 
once restricting to $q^2 > 2.0\GeV^2$. Our main results are summarized in 
Table~\ref{tab:Br-cen} and Figures~\ref{fig:Br-unc} and \ref{fig:Br-1+cont_vs_3}.
Here we quote the $B_q\to \gamma\mu\bar\mu$ branching fractions 
\begin{align}
  \langle \BR \rangle_{[4 m_\mu^2,\, 6.0]} & 
  = \big( 12.51^{+3.83}_{-1.93} \big) \cdot 10^{-9} , &
  \langle \BR \rangle_{[2.0,\, 6.0]}       &
  = \big(  0.30^{+0.25}_{-0.14} \big) \cdot 10^{-9} , &
  (q & = s)
\\
  \BR_{[4m_\mu^2,\, 6.0]} & = \big( 1.84^{+1.82}_{-0.72} \big) \cdot 10^{-10} , &
  \BR_{[2.0,\, 6.0]}      & = \big( 0.13^{+0.13}_{-0.06} \big) \cdot 10^{-10} , &
  (q & = d)
\end{align}
in two $q^2$-bins, of which for $q=s$ the first is dominated by 
resonances. In the
case of $B_s\to\gamma\mu\bar\mu$ the $\langle \BR \rangle$ denotes the
time-integrated branching fraction, which accounts for the non-vanishing
width difference of the $B_s$ system.
Compared to previous estimates \cite{Guadagnoli:2017quo, Kozachuk:2017mdk},
the branching fractions from the QCD factorization calculation 
performed here are roughly a factor of two larger. We attribute 
these differences to the difference between the QCD factorization 
computation of the form factors and the more model-dependent and 
less complete parameterizations used in previous work as well as
different form factor input in the nonlocal NLP $B$-type contributions
used in \cite{Kozachuk:2017mdk}. 
The presence of the additional real photon lifts the helicity suppression of the
purely leptonic $\Bqtoll$, which yields similar rates for final states with
electrons and muons. The rate CP asymmetry $\ACP$ for $q = s$ is tiny, but
reaches between  $-20$\% and $+5$\%  locally in $q^2$ for $q = d$.

\vskip 0.5cm

\noindent{\em Note added:} When this paper was completed, 
Ref.~\cite{Khodjamirian:2020hob} appeared, which presents the first theoretical
estimate of $\lamB{s}$. The QCD sum rule calculation for this quantity and
$\lamB{s}/\lamB{d}$ is in very good agreement with the values used here.

%
%
\subsubsection*{Acknowledgements}

CB thanks Robert Szafron for discussions. 
The work of MB is supported by the DFG Sonderforschungsbereich/Transregio 110
``Symmetries and the Emergence of Structure in QCD''.
YMW acknowledges support from the National Youth Thousand Talents Program, the
Youth Hundred Academic Leaders Program of Nankai University, the National
Natural Science Foundation of China with Grant No. 11675082~and 11735010, and the
Natural Science Foundation of Tianjin with Grant No. 19JCJQJC61100.

%
%
%

\appendix

\section{Definitions and conventions}
\label{app:defs}

Throughout we use the common definitions
\begin{align}
   g_{\mu\nu}^\perp &
  \equiv g_{\mu\nu} - \frac{\nbb^\mu \nnn^\nu}{2} - \frac{\nnn^\mu \nbb^\nu}{2} ,
&
  \veps_{\mu\nu}^\perp &
  \equiv \veps_{\mu\nu\alpha\beta} \, \frac{\nbb^\alpha \nnn^\beta}{2}
  = \veps_{\mu\nu\alpha\beta} \, \nbb^\alpha v^\beta ,
\end{align}
with $\veps_{0123} = -1$ and the conventions
\begin{align}
  \mbox{Tr}[\gamma_\mu \gamma_\nu \gamma_\alpha \gamma_\beta \gamma_5] &
  = - 4 i \veps_{\mu\nu\alpha\beta} ,
&
  \sigma_{\mu\nu} \gamma_5 &
  = \frac{i}{2} \veps_{\mu\nu\alpha\beta} \, \sigma^{\alpha\beta} ,
&
  \sigma_{\mu\nu}
  = \frac{i}{2} [\gamma_\mu, \gamma_\nu] .
\end{align}

%
%
%

\section{Details on final-state radiation}
\label{app:FSR}

Final-state radiation (FSR) is governed by the
$B_q$-to-vacuum hadronic matrix elements $S_\alpha^{(i)}$ appearing
in \refeq{eq:ampl-Ti},
\begin{align}
  S_\alpha^{(i)} &
  = \bra{0} \oL{q}\, \gamma_\alpha P_L\, b \ket{\oL{B}_q}
  = -\frac{i}{2} f_{B_q} p_\alpha, &
  i & = 9, 10 ,
\\
  S_\alpha^{(i)} &
  = \frac{2 \oL{m}_b}{m_{B_q}^2} \bra{0} \oL{q}\, i \sigma_{\alpha\beta} p^\beta P_R\, b \ket{\oL{B}_q}
  = 0 , &
  i & = 7,
\\
  S_\alpha^{(i)} &
  = \frac{(4\pi)^2 \,i}{m_{B_q}^2} \int \! d^4 x \, e^{i p x} \,
    \bra{0} \text{T} \lbrace j_{f\alpha}(x),\, \Op_i(0) \rbrace \ket{\oL{B}_q} , &
  i & = 1,\ldots,6,8 ,
\end{align}
which are linear in the $B_q$-meson momentum $S_\alpha^{(i)} \propto p_\alpha$.
They are contracted with the vector and axial-vector leptonic rank-two tensors
\begin{align}
  L^{\mu\nu}_{V(A)} &
  = i Q_\ell \, \bar{u}(p_\ell) \Big[
      \gamma^\mu \frac{\slashed p_\ell + \slashed k + m_\ell}
                         {(p_\ell + k)^2 - m_\ell^2} \gamma^\nu (\gf)
    - \gamma^\nu (\gf)  \frac{\slashed p_\oL{\ell} + \slashed k - m_\ell}
                         {(p_\oL{\ell} + k)^2 - m_\ell^2} \gamma^\mu
    \Big] v (p_\oL{\ell})
\end{align}
with the property $p_\nu L^{\mu\nu}_V  = 0$ and
$p_\nu L^{\mu\nu}_A \propto f_{B_q}\,  m_\ell/m_{B_q}$.
In consequence the FSR contributions vanish except the one from $\Op_{10}$,
which is helicity suppressed and proportional to $f_{B_q}$. The FSR amplitude
then takes the form
\begin{align}
  \label{eq:ampl-FSR-Op10}
  \oL{\calA}|_\text{FSR} &
  = i e \frac{\alE}{4\pi} \normEW
    \frac{m_\ell\, \oL{\calA}_\text{FSR}\, \eps^\star_\mu\, p_\nu}
         {\sqrt{\lambda} (1 - \beta_\ell^2 \cos^2\!\theta_\ell)}
    \big[\bar{u}(p_\ell)\! \left( 
    - i \, \sigma^{\mu\nu} \right)\! \gamma_5\, v(p_\oL{\ell}) \big]
\end{align}
in terms of the kinematic variables $(q^2,\, \cos\theta_\ell)$, where
$\oL{\calA}_\text{FSR} \equiv 4\, C_{10} Q_\ell f_{B_q}$.

The additional contributions to the two-fold differential decay width from
the so-called structure-dependent (SD) contributions in~\refeq{eq:d2Gamma-ISR}
consist of the interference of SD$\,\times\,$FSR
\begin{equation}
  \label{eq:d2G-SDxFSR}
\begin{aligned}
\frac{d^2\oL{\Gamma}}{dq^2\, d\!\cos\theta_\ell} \Big|_{\text{SD}\times\text{FSR}} &
  = \Gamma_0 \sqrt{\lambda}^3  q^2 \beta_\ell \times
    \frac{m_{B_q} (1 - \beta_\ell^2)}{\sqrt{\lambda}\, (1 - \beta_\ell^2 \cos^2\!\theta_\ell)}
\\ & \times (- \sqrt{2})
   \re \Big[ \Big( \oL{\calA}_{\perp V}
                 + \beta_\ell \cos\theta_\ell \, \oL{\calA}_{\parallel A}
             \Big) \oL{\calA}_\text{FSR}^* \Big]\,,
\end{aligned}
\end{equation}
and the term  proportional $|$FSR$|^2$
\begin{equation}
  \label{eq:d2G-FSR2}
\begin{aligned}
\frac{d^2\oL{\Gamma}}{dq^2\, d\!\cos\theta_\ell} \Big|_{|\text{FSR}|^2} &
  = \Gamma_0 \sqrt{\lambda}^3  q^2 \beta_\ell \times
    \frac{m_{B_q}^2 (1 - \beta_\ell^2)}{4\,\lambda^2(1 - \beta_\ell^2 \cos^2\!\theta_\ell)^2}
     |\oL{\calA}_\text{FSR}|^2
\\ & \times
    \Big[m_{B_q}^4 + q^4 - 2 (1 - \beta_\ell^2) q^2 m_{B_q}^2
         - \beta_\ell^2 \cos^2\!\theta_\ell (m_{B_q}^4 + q^4)
    \Big] ,
\end{aligned}
\end{equation}
where for the CP-conjugated decay the transversity amplitudes in
\eqref{eq:d2G-SDxFSR} should be replaced according to \eqref{eq:ampl-CP}.
Both are lepton-mass suppressed as can be seen by the overall factor
$(1 - \beta_\ell^2)$ and become important only for very low
$q^2 \sim 4 m_\ell^2 \ll 1$~GeV${}^2$. Further they have a nonanalytic
dependence on $\cos\theta_\ell$ given by the factor $(1 - \beta_\ell^2
\cos^2\!\theta_\ell)^{-n}$ with $n = 1, 2$, respectively, which becomes
most pronounced in regions of the phase space when $q^2 \gg 4 m_\ell^2$
and hence $\beta_\ell \to 1$. Then the collinear divergence in the limit
$\cos\theta_\ell \to \pm 1$ is regulated by the finite lepton mass
$m_\ell \neq 0$. The nonanalytic $\cos\theta_\ell$ dependence prevents
a simple polynomial dependence on $\cos\theta_\ell$ as given
in~\refeq{eq:d2Gamma-ISR}, but due to the lepton-mass suppression, this
is of concern only for small regions in the phase space, such that the
phenomenologically most important observables without lepton-mass
suppression remain the differential decay width and the
lepton forward-backward asymmetry.

%
%
%
\section{\boldmath $B_q$-meson LCDA: RG evolution and model}
\label{app:B-LCDA}

The factorization approach leads to a convolution of the jet function with 
the nonperturbative $B$-meson LCDA $\phi_+(\omega; \mu)$ in the LP amplitudes
\refeq{eq:ampLPfinal}. In $A$-type insertions, at NLO in QCD, the convolution
\refeq{eq:jet-func-conv} is related to the inverse and first two logarithmic
moments, whereas in $B$-type insertions the knowledge of the functional form
of $\phi_+(\omega; \mu)$ is required even at LO in QCD. Apart from theoretical
constraints, little is known about $\phi_+$, because of the absence of stringent
phenomenological constraints. In fact, the radiative charged-current decay
$\Butolvgamma$ in the same kinematical region as $\Bqtollgamma$ is considered
as the prime decay to determine the inverse ($\lamB{u}$) and the first
logarithmic ($\sigNB{u}{1}$) moments \cite{Beneke:2011nf, Braun:2012kp,
Wang:2016qii, Wang:2018wfj, Beneke:2018wjp}, and has been investigated by the
Belle Collaboration~\cite{Gelb:2018end}. There is a preference for small values
of $\lamB{u,d} \sim 0.2\GeV$ from hadronic two-body decays $B \to \pi\pi,\,
\pi\rho,\, \rho\rho$ \cite{Bell:2009fm, Beneke:2009ek} in the framework of
QCD factorization. QCD sum rules yield $\lamB{q} \sim 0.46(11)\GeV$ \cite{Braun:2003wx}.

Such values usually refer to a particular renormalization scale. Throughout we
set $\mu_0 = 1\GeV$ as the initial scale, and RG evolution is 
used to evolve $\phi_+$ to the hard-collinear scale $\mu_{hc}$, accounting for 
higher-order QCD corrections. The RG equations for $\phi_+(\omega; \mu)$ and
its moments involve a convolution in the momentum variable $\omega$,
which can be avoided at the leading-logarithmic order when going to dual
(position) space \cite{Bell:2013tfa}, where the corresponding nonperturbative
function $\eta_+(s; \mu) = U_+(s; \mu,  \mu_0) \eta_+(s; \mu_0)$ has autonomous
scaling for each value of $s$. 

We resort to the three-parameter model \cite{Beneke:2018wjp}
\begin{align}
  \label{eq:B-LCDA-BBJW'18}
  \phi_+(\omega; \mu_0) &
  = \frac{\Gamma(\beta)}{\Gamma(\alpha)} \frac{\omega}{\omega_0^2} 
    e^{-\omega/\omega_0} \, U(\beta-\alpha,\, 3 - \alpha,\, \omega/\omega_0) ,
\end{align}
which permits to solve the transformation from dual to momentum space
analytically, and in consequence also an analytic solution of the RG equation.
The three parameters $\omega_0$, $\alpha$ and $\beta$ are assumed to determine
$\phi_+(\omega)$ at the scale $\mu_0$ and the RG evolution is performed as
given in \cite{Beneke:2018wjp}. Further
\begin{align}
  \lamB{q} &
  = \frac{\alpha - 1}{\beta - 1} \omega_0 ,
\end{align}
which allows to associate $\omega_0$ to $\lamB{q}$, whereas $\alpha$ and $\beta$
determine the hatted logarithmic moments defined as
\begin{align}
  \label{eq:def-B-LCDA-moments-hat}
  \hsigNB{q}{n} &
   = \int_0^\infty d\omega \,\frac{\lamB{q}}{\omega} 
     \ln^n \frac{\lamB{q} e^{-\gamma_E}}{\omega} \phi_+(\omega) .
\end{align}
The three-parameter model  reduces to the exponential model \cite{Grozin:1996pq}
for $\alpha = \beta$, which depends only a single parameter $\lamB{q} = \omega_0$,
the inverse moment.

In the numerical analysis, we choose as default the exponential model,
i.e. $\alpha = \beta$. This fixes the logarithmic moments, as for 
example
\begin{align}
  \label{eq:B-lcda-exp-model-sig_1}
  \sigNBq{1} & 
  = \gamma_E + \ln\frac{\mu_0}{\lamB{q}} , &
  \hsigNBq{1} & = 0 ,
\\
  \label{eq:B-lcda-exp-model-sig_2}
  \sigNBq{2} & 
  = \gamma_E^2 + \frac{\pi^2}{6} + \left(
         2 \gamma_E + \ln\frac{\mu_0}{\lamB{q}} \right)
         \ln\frac{\mu_0}{\lamB{q}} , &
  \hsigNBq{2} & = \frac{\pi^2}{6} .
\end{align}
To calculate the uncertainty from the $B$-LCDA, besides the variation of
$\lamB{q}$ (i.e. $\omega_0$), we also vary $\hsigNB{q}{1} = (0.0 \pm 0.7)$,
$\hsigNB{q}{2} = (0 \pm 6)$ to estimate the further model-dependence, using
the three-parameter model \refeq{eq:B-LCDA-BBJW'18}. We use the following
$(\alpha,\, \beta)$ tuples together with $\omega_0$:\\[0.2cm]
\begin{center}
\renewcommand{\arraystretch}{1.5}
\begin{tabular}{|c|ccc|crrrr|}
\hline
& $\omega_0$ & $\alpha$ & $\beta$ 
& $\lamB{q}$
& $\hsigNB{q}{1}$ & $\hsigNB{q}{2}$ 
& $\sigNB{q}{1}$  & $\sigNB{q}{2}$
\\
\hline
  \multirow{2}*{$B_d$}
&  $0.295611\GeV$ & 1.22294 & 1.18830 & $0.35\GeV$ & $-0.70$ & $-6.00$ & 0.93 & $-5.63$ \\
&  $0.590920\GeV$ & 1.42192 & 1.71234 & $0.35\GeV$ &   0.70  &   6.00  & 2.33 &  10.93  \\
\hline
  \multirow{2}*{$B_s$}
& $0.337841\GeV$ & 1.22294 & 1.18830 & $0.40\GeV$ & $-0.70$ & $-6.00$ & 0.68 & $-6.03$ \\
& $0.675338\GeV$ & 1.42192 & 1.71234 & $0.40\GeV$ &   0.70  &   6.00  & 2.08 &   9.82  \\
\hline
\end{tabular}
\renewcommand{\arraystretch}{1.0}
\end{center}

%
%

\renewcommand{\refname}{R\lowercase{eferences}}

\addcontentsline{toc}{section}{References}

\bibliographystyle{JHEP}

\small

\bibliography{bibliography}

\providecommand{\href}[2]{#2}\begingroup\raggedright\begin{thebibliography}{10}

\bibitem{Bobeth:2013uxa}
C.~Bobeth, M.~Gorbahn, T.~Hermann, M.~Misiak, E.~Stamou and M.~Steinhauser,
  \emph{{$B_{s,d} \to \ell^+ \ell^-$ in the Standard Model with Reduced
  Theoretical Uncertainty}},
  \href{http://dx.doi.org/10.1103/PhysRevLett.112.101801}{\emph{Phys. Rev.
  Lett.} {\bfseries 112} (2014) 101801},
  [\href{https://arxiv.org/abs/1311.0903}{{\ttfamily 1311.0903}}].

\bibitem{Beneke:2017vpq}
M.~Beneke, C.~Bobeth and R.~Szafron, \emph{{Enhanced electromagnetic correction
  to the rare $B$-meson decay $B_{s,d} \to \mu^+ \mu^-$}},
  \href{http://dx.doi.org/10.1103/PhysRevLett.120.011801}{\emph{Phys. Rev.
  Lett.} {\bfseries 120} (2018) 011801},
  [\href{https://arxiv.org/abs/1708.09152}{{\ttfamily 1708.09152}}].

\bibitem{Beneke:2019slt}
M.~Beneke, C.~Bobeth and R.~Szafron, \emph{{Power-enhanced leading-logarithmic
  QED corrections to $B_q \to \mu^+\mu^-$}},
  \href{http://dx.doi.org/10.1007/JHEP10(2019)232}{\emph{JHEP} {\bfseries 10}
  (2019) 232}, [\href{https://arxiv.org/abs/1908.07011}{{\ttfamily
  1908.07011}}].

\bibitem{Bazavov:2017lyh}
A.~Bazavov et~al., \emph{{$B$- and $D$-meson leptonic decay constants from
  four-flavor lattice QCD}},
  \href{http://dx.doi.org/10.1103/PhysRevD.98.074512}{\emph{Phys. Rev.}
  {\bfseries D98} (2018) 074512},
  [\href{https://arxiv.org/abs/1712.09262}{{\ttfamily 1712.09262}}].

\bibitem{Aaij:2017vad}
{\scshape LHCb} collaboration, R.~Aaij et~al., \emph{{Measurement of the
  $B^0_s\to\mu^+\mu^-$ branching fraction and effective lifetime and search for
  $B^0\to\mu^+\mu^-$ decays}},
  \href{http://dx.doi.org/10.1103/PhysRevLett.118.191801}{\emph{Phys. Rev.
  Lett.} {\bfseries 118} (2017) 191801},
  [\href{https://arxiv.org/abs/1703.05747}{{\ttfamily 1703.05747}}].

\bibitem{Sirunyan:2019xdu}
{\scshape CMS} collaboration, A.~M. Sirunyan et~al., \emph{{Measurement of
  properties of B$^0_\mathrm{s}\to\mu^+\mu^-$ decays and search for
  B$^0\to\mu^+\mu^-$ with the CMS experiment}},
  \href{http://dx.doi.org/10.1007/JHEP04(2020)188}{\emph{JHEP} {\bfseries 04}
  (2020) 188}, [\href{https://arxiv.org/abs/1910.12127}{{\ttfamily
  1910.12127}}].

\bibitem{Aaboud:2018mst}
{\scshape ATLAS} collaboration, M.~Aaboud et~al., \emph{{Study of the rare
  decays of $B^0_s$ and $B^0$ mesons into muon pairs using data collected
  during 2015 and 2016 with the ATLAS detector}},
  \href{http://dx.doi.org/10.1007/JHEP04(2019)098}{\emph{JHEP} {\bfseries 04}
  (2019) 098}, [\href{https://arxiv.org/abs/1812.03017}{{\ttfamily
  1812.03017}}].

\bibitem{Eilam:1996vg}
G.~Eilam, C.-D. Lu and D.-X. Zhang, \emph{{Radiative dileptonic decays of $B$
  mesons}}, \href{http://dx.doi.org/10.1016/S0370-2693(96)01491-8}{\emph{Phys.
  Lett.} {\bfseries B391} (1997) 461--464},
  [\href{https://arxiv.org/abs/hep-ph/9606444}{{\ttfamily hep-ph/9606444}}].

\bibitem{Aliev:1996ud}
T.~M. Aliev, A.~Ozpineci and M.~Savci, \emph{{$B_q \to \gamma \ell^+\ell^-$
  decays in light cone QCD}},
  \href{http://dx.doi.org/10.1103/PhysRevD.55.7059}{\emph{Phys. Rev.}
  {\bfseries D55} (1997) 7059--7066},
  [\href{https://arxiv.org/abs/hep-ph/9611393}{{\ttfamily hep-ph/9611393}}].

\bibitem{Guadagnoli:2017quo}
D.~Guadagnoli, M.~Reboud and R.~Zwicky, \emph{{$B_s^0 \to \gamma \ell^+ \ell^-$
  as a test of lepton flavor universality}},
  \href{http://dx.doi.org/10.1007/JHEP11(2017)184}{\emph{JHEP} {\bfseries 11}
  (2017) 184}, [\href{https://arxiv.org/abs/1708.02649}{{\ttfamily
  1708.02649}}].

\bibitem{Kozachuk:2017mdk}
A.~Kozachuk, D.~Melikhov and N.~Nikitin, \emph{{Rare FCNC radiative leptonic
  $B_{s,d} \to \gamma \ell^+\ell^-$ decays in the standard model}},
  \href{http://dx.doi.org/10.1103/PhysRevD.97.053007}{\emph{Phys. Rev.}
  {\bfseries D97} (2018) 053007},
  [\href{https://arxiv.org/abs/1712.07926}{{\ttfamily 1712.07926}}].

\bibitem{Aditya:2012im}
Y.~G. Aditya, K.~J. Healey and A.~A. Petrov, \emph{{Faking $B_s \to
  \mu^+\mu^-$}},
  \href{http://dx.doi.org/10.1103/PhysRevD.87.074028}{\emph{Phys. Rev.}
  {\bfseries D87} (2013) 074028},
  [\href{https://arxiv.org/abs/1212.4166}{{\ttfamily 1212.4166}}].

\bibitem{Dettori:2016zff}
F.~Dettori, D.~Guadagnoli and M.~Reboud, \emph{{$B^{0}_{s} \to
  \mu^{+}\mu^{-}\gamma$ from $B^{0}_{s} \to \mu^{+}\mu^{-}$}},
  \href{http://dx.doi.org/10.1016/j.physletb.2017.02.048}{\emph{Phys. Lett.}
  {\bfseries B768} (2017) 163--167},
  [\href{https://arxiv.org/abs/1610.00629}{{\ttfamily 1610.00629}}].

\bibitem{DescotesGenon:2002ja}
S.~Descotes-Genon and C.~T. Sachrajda, \emph{{Universality of nonperturbative
  QCD effects in radiative B decays}},
  \href{http://dx.doi.org/10.1016/S0370-2693(03)00173-4}{\emph{Phys. Lett.}
  {\bfseries B557} (2003) 213--223},
  [\href{https://arxiv.org/abs/hep-ph/0212162}{{\ttfamily hep-ph/0212162}}].

\bibitem{Dubnicka:2018gqg}
S.~Dubni\v{c}ka, A.~Z. Dubni\v{c}kov{\'a}, M.~A. Ivanov, A.~Liptaj,
  P.~Santorelli and C.~T. Tran, \emph{{Study of $B_s \to \gamma\ell^+\ell^-$
  decays in covariant quark model}},
  \href{http://dx.doi.org/10.1103/PhysRevD.99.014042}{\emph{Phys. Rev.}
  {\bfseries D99} (2019) 014042},
  [\href{https://arxiv.org/abs/1808.06261}{{\ttfamily 1808.06261}}].

\bibitem{Korchemsky:1999qb}
G.~P. Korchemsky, D.~Pirjol and T.-M. Yan, \emph{{Radiative leptonic decays of
  $B$ mesons in QCD}},
  \href{http://dx.doi.org/10.1103/PhysRevD.61.114510}{\emph{Phys. Rev.}
  {\bfseries D61} (2000) 114510},
  [\href{https://arxiv.org/abs/hep-ph/9911427}{{\ttfamily hep-ph/9911427}}].

\bibitem{DescotesGenon:2002mw}
S.~Descotes-Genon and C.~T. Sachrajda, \emph{{Factorization, the light cone
  distribution amplitude of the $B$ meson and the radiative decay $B \to \gamma
  \ell \bar\nu_\ell$}},
  \href{http://dx.doi.org/10.1016/S0550-3213(02)01066-0}{\emph{Nucl. Phys.}
  {\bfseries B650} (2003) 356--390},
  [\href{https://arxiv.org/abs/hep-ph/0209216}{{\ttfamily hep-ph/0209216}}].

\bibitem{Lunghi:2002ju}
E.~Lunghi, D.~Pirjol and D.~Wyler, \emph{{Factorization in leptonic radiative
  $B \to \gamma \ell \bar\nu_\ell$ decays}},
  \href{http://dx.doi.org/10.1016/S0550-3213(02)01032-5}{\emph{Nucl. Phys.}
  {\bfseries B649} (2003) 349--364},
  [\href{https://arxiv.org/abs/hep-ph/0210091}{{\ttfamily hep-ph/0210091}}].

\bibitem{Bosch:2003fc}
S.~W. Bosch, R.~J. Hill, B.~O. Lange and M.~Neubert, \emph{{Factorization and
  Sudakov resummation in leptonic radiative $B$ decay}},
  \href{http://dx.doi.org/10.1103/PhysRevD.67.094014}{\emph{Phys. Rev.}
  {\bfseries D67} (2003) 094014},
  [\href{https://arxiv.org/abs/hep-ph/0301123}{{\ttfamily hep-ph/0301123}}].

\bibitem{Beneke:2011nf}
M.~Beneke and J.~Rohrwild, \emph{{B meson distribution amplitude from $B \to
  \gamma \ell \bar\nu_\ell$}},
  \href{http://dx.doi.org/10.1140/epjc/s10052-011-1818-8}{\emph{Eur. Phys. J.}
  {\bfseries C71} (2011) 1818},
  [\href{https://arxiv.org/abs/1110.3228}{{\ttfamily 1110.3228}}].

\bibitem{Braun:2012kp}
V.~M. Braun and A.~Khodjamirian, \emph{{Soft contribution to $B\to \gamma \ell
  \bar\nu_\ell$ and the $B$-meson distribution amplitude}},
  \href{http://dx.doi.org/10.1016/j.physletb.2012.11.047}{\emph{Phys. Lett.}
  {\bfseries B718} (2013) 1014--1019},
  [\href{https://arxiv.org/abs/1210.4453}{{\ttfamily 1210.4453}}].

\bibitem{Wang:2016qii}
Y.-M. Wang, \emph{{Factorization and dispersion relations for radiative
  leptonic $B$ decay}},
  \href{http://dx.doi.org/10.1007/JHEP09(2016)159}{\emph{JHEP} {\bfseries 09}
  (2016) 159}, [\href{https://arxiv.org/abs/1606.03080}{{\ttfamily
  1606.03080}}].

\bibitem{Wang:2018wfj}
Y.-M. Wang and Y.-L. Shen, \emph{{Subleading-power corrections to the radiative
  leptonic $B \to \gamma \ell \bar\nu_\ell$ decay in QCD}},
  \href{http://dx.doi.org/10.1007/JHEP05(2018)184}{\emph{JHEP} {\bfseries 05}
  (2018) 184}, [\href{https://arxiv.org/abs/1803.06667}{{\ttfamily
  1803.06667}}].

\bibitem{Beneke:2018wjp}
M.~Beneke, V.~M. Braun, Y.~Ji and Y.-B. Wei, \emph{{Radiative leptonic decay
  $B\to \gamma \ell \nu_\ell$ with subleading power corrections}},
  \href{http://dx.doi.org/10.1007/JHEP07(2018)154}{\emph{JHEP} {\bfseries 07}
  (2018) 154}, [\href{https://arxiv.org/abs/1804.04962}{{\ttfamily
  1804.04962}}].

\bibitem{Beneke:2001at}
M.~Beneke, T.~Feldmann and D.~Seidel, \emph{{Systematic approach to exclusive
  $B \to V \ell^+ \ell^-$, $V \gamma$ decays}},
  \href{http://dx.doi.org/10.1016/S0550-3213(01)00366-2}{\emph{Nucl. Phys.}
  {\bfseries B612} (2001) 25--58},
  [\href{https://arxiv.org/abs/hep-ph/0106067}{{\ttfamily hep-ph/0106067}}].

\bibitem{Beneke:2004dp}
M.~Beneke, T.~Feldmann and D.~Seidel, \emph{{Exclusive radiative and
  electroweak $b \to d$ and $b \to s$ penguin decays at NLO}},
  \href{http://dx.doi.org/10.1140/epjc/s2005-02181-5}{\emph{Eur. Phys. J.}
  {\bfseries C41} (2005) 173--188},
  [\href{https://arxiv.org/abs/hep-ph/0412400}{{\ttfamily hep-ph/0412400}}].

\bibitem{Wang:2013rfa}
W.-Y. Wang, Z.-H. Xiong and S.-H. Zhou, \emph{{Complete Analyses on the
  Short-Distance Contribution of $B_s \to \gamma \ell^+\ell^-$ in the Standard
  Model}}, \href{http://dx.doi.org/10.1088/0256-307X/30/11/111202}{\emph{Chin.
  Phys. Lett.} {\bfseries 30} (2013) 111202},
  [\href{https://arxiv.org/abs/1303.0660}{{\ttfamily 1303.0660}}].

\bibitem{Bosch:2002bv}
S.~W. Bosch and G.~Buchalla, \emph{{The Double radiative decays $B \to \gamma
  \gamma$ in the heavy quark limit}},
  \href{http://dx.doi.org/10.1088/1126-6708/2002/08/054}{\emph{JHEP} {\bfseries
  08} (2002) 054}, [\href{https://arxiv.org/abs/hep-ph/0208202}{{\ttfamily
  hep-ph/0208202}}].

\bibitem{Beneke:2009az}
M.~Beneke, G.~Buchalla, M.~Neubert and C.~Sachrajda, \emph{{Penguins with Charm
  and Quark-Hadron Duality}},
  \href{http://dx.doi.org/10.1140/epjc/s10052-009-1028-9}{\emph{Eur. Phys. J.
  C} {\bfseries 61} (2009) 439--449},
  [\href{https://arxiv.org/abs/0902.4446}{{\ttfamily 0902.4446}}].

\bibitem{Chetyrkin:1996vx}
K.~G. Chetyrkin, M.~Misiak and M.~M{\"u}nz, \emph{{Weak radiative $B$-meson
  decay beyond leading logarithms}},
  \href{http://dx.doi.org/10.1016/S0370-2693(97)00324-9}{\emph{Phys. Lett.}
  {\bfseries B400} (1997) 206--219},
  [\href{https://arxiv.org/abs/hep-ph/9612313}{{\ttfamily hep-ph/9612313}}].

\bibitem{Bobeth:1999mk}
C.~Bobeth, M.~Misiak and J.~Urban, \emph{{Photonic penguins at two loops and
  $m_t$ dependence of $Br(B \to X_s \ell^+ \ell^-)$}},
  \href{http://dx.doi.org/10.1016/S0550-3213(00)00007-9}{\emph{Nucl. Phys.}
  {\bfseries B574} (2000) 291--330},
  [\href{https://arxiv.org/abs/hep-ph/9910220}{{\ttfamily hep-ph/9910220}}].

\bibitem{Bobeth:2003at}
C.~Bobeth, P.~Gambino, M.~Gorbahn and U.~Haisch, \emph{{Complete NNLO QCD
  analysis of $\bar B \to X_s \ell^+ \ell^-$ and higher order electroweak
  effects}}, \href{http://dx.doi.org/10.1088/1126-6708/2004/04/071}{\emph{JHEP}
  {\bfseries 04} (2004) 071},
  [\href{https://arxiv.org/abs/hep-ph/0312090}{{\ttfamily hep-ph/0312090}}].

\bibitem{Huber:2005ig}
T.~Huber, E.~Lunghi, M.~Misiak and D.~Wyler, \emph{{Electromagnetic logarithms
  in $\bar B \to X_s \ell^+ \ell^-$}},
  \href{http://dx.doi.org/10.1016/j.nuclphysb.2006.01.037}{\emph{Nucl. Phys.}
  {\bfseries B740} (2006) 105--137},
  [\href{https://arxiv.org/abs/hep-ph/0512066}{{\ttfamily hep-ph/0512066}}].

\bibitem{Melikhov:2004mk}
D.~Melikhov and N.~Nikitin, \emph{{Rare radiative leptonic decays $B_{d,s} \to
  \gamma \ell^+\ell^-$}},
  \href{http://dx.doi.org/10.1103/PhysRevD.70.114028}{\emph{Phys. Rev.}
  {\bfseries D70} (2004) 114028},
  [\href{https://arxiv.org/abs/hep-ph/0410146}{{\ttfamily hep-ph/0410146}}].

\bibitem{Grinstein:2000pc}
B.~Grinstein and D.~Pirjol, \emph{{Long distance effects in $B \to V \gamma$
  radiative weak decays}},
  \href{http://dx.doi.org/10.1103/PhysRevD.62.093002}{\emph{Phys. Rev.}
  {\bfseries D62} (2000) 093002},
  [\href{https://arxiv.org/abs/hep-ph/0002216}{{\ttfamily hep-ph/0002216}}].

\bibitem{Bauer:2000yr}
C.~W. Bauer, S.~Fleming, D.~Pirjol and I.~W. Stewart, \emph{{An Effective field
  theory for collinear and soft gluons: Heavy to light decays}},
  \href{http://dx.doi.org/10.1103/PhysRevD.63.114020}{\emph{Phys. Rev.}
  {\bfseries D63} (2001) 114020},
  [\href{https://arxiv.org/abs/hep-ph/0011336}{{\ttfamily hep-ph/0011336}}].

\bibitem{Beneke:2005gs}
M.~Beneke and D.~Yang, \emph{{Heavy-to-light $B$ meson form-factors at large
  recoil energy: Spectator-scattering corrections}},
  \href{http://dx.doi.org/10.1016/j.nuclphysb.2005.11.027}{\emph{Nucl. Phys.}
  {\bfseries B736} (2006) 34--81},
  [\href{https://arxiv.org/abs/hep-ph/0508250}{{\ttfamily hep-ph/0508250}}].

\bibitem{Asatrian:2001de}
H.~H. Asatrian, H.~M. Asatrian, C.~Greub and M.~Walker, \emph{{Two loop virtual
  corrections to $B \to X_s \ell^+ \ell^-$ in the standard model}},
  \href{http://dx.doi.org/10.1016/S0370-2693(01)00441-5}{\emph{Phys. Lett.}
  {\bfseries B507} (2001) 162--172},
  [\href{https://arxiv.org/abs/hep-ph/0103087}{{\ttfamily hep-ph/0103087}}].

\bibitem{Asatryan:2001zw}
H.~H. Asatryan, H.~M. Asatrian, C.~Greub and M.~Walker, \emph{{Calculation of
  two loop virtual corrections to $b \to s \ell^+ \ell^-$ in the standard
  model}}, \href{http://dx.doi.org/10.1103/PhysRevD.65.074004}{\emph{Phys.
  Rev.} {\bfseries D65} (2002) 074004},
  [\href{https://arxiv.org/abs/hep-ph/0109140}{{\ttfamily hep-ph/0109140}}].

\bibitem{Seidel:2004jh}
D.~Seidel, \emph{{Analytic two loop virtual corrections to $b \to d \ell^+
  \ell^-$}}, \href{http://dx.doi.org/10.1103/PhysRevD.70.094038}{\emph{Phys.
  Rev.} {\bfseries D70} (2004) 094038},
  [\href{https://arxiv.org/abs/hep-ph/0403185}{{\ttfamily hep-ph/0403185}}].

\bibitem{Buras:2002tp}
A.~J. Buras, A.~Czarnecki, M.~Misiak and J.~Urban, \emph{{Completing the NLO
  QCD calculation of $\bar{B} \to X_s \gamma$}},
  \href{http://dx.doi.org/10.1016/S0550-3213(02)00261-4}{\emph{Nucl. Phys.}
  {\bfseries B631} (2002) 219--238},
  [\href{https://arxiv.org/abs/hep-ph/0203135}{{\ttfamily hep-ph/0203135}}].

\bibitem{Grozin:1996pq}
A.~G. Grozin and M.~Neubert, \emph{{Asymptotics of heavy meson form-factors}},
  \href{http://dx.doi.org/10.1103/PhysRevD.55.272}{\emph{Phys. Rev.} {\bfseries
  D55} (1997) 272--290},
  [\href{https://arxiv.org/abs/hep-ph/9607366}{{\ttfamily hep-ph/9607366}}].

\bibitem{Beneke:2000wa}
M.~Beneke and T.~Feldmann, \emph{{Symmetry breaking corrections to heavy to
  light $B$ meson form-factors at large recoil}},
  \href{http://dx.doi.org/10.1016/S0550-3213(00)00585-X}{\emph{Nucl. Phys.}
  {\bfseries B592} (2001) 3--34},
  [\href{https://arxiv.org/abs/hep-ph/0008255}{{\ttfamily hep-ph/0008255}}].

\bibitem{Albrecht:2019zul}
J.~Albrecht, E.~Stamou, R.~Ziegler and R.~Zwicky, \emph{{Probing flavoured
  Axions in the Tail of $B_q \to \mu^+\mu^-$}},
  \href{https://arxiv.org/abs/1911.05018}{{\ttfamily 1911.05018}}.

\bibitem{Straub:2015ica}
A.~Bharucha, D.~M. Straub and R.~Zwicky, \emph{{$B\to V\ell^+\ell^-$ in the
  Standard Model from light-cone sum rules}},
  \href{http://dx.doi.org/10.1007/JHEP08(2016)098}{\emph{JHEP} {\bfseries 08}
  (2016) 098}, [\href{https://arxiv.org/abs/1503.05534}{{\ttfamily
  1503.05534}}].

\bibitem{Charles:1998dr}
J.~Charles, A.~Le~Yaouanc, L.~Oliver, O.~Pene and J.~Raynal, \emph{{Heavy to
  light form-factors in the heavy mass to large energy limit of QCD}},
  \href{http://dx.doi.org/10.1103/PhysRevD.60.014001}{\emph{Phys. Rev. D}
  {\bfseries 60} (1999) 014001},
  [\href{https://arxiv.org/abs/hep-ph/9812358}{{\ttfamily hep-ph/9812358}}].

\bibitem{Zyla:2020zbs}
{\scshape Particle Data Group} collaboration, P.~Zyla et~al., \emph{{Review of
  Particle Physics}}, \href{http://dx.doi.org/10.1093/ptep/ptaa104}{\emph{PTEP}
  {\bfseries 2020} (2020) 083C01}.

\bibitem{Aoki:2019cca}
{\scshape Flavour Lattice Averaging Group} collaboration, S.~Aoki et~al.,
  \emph{{FLAG Review 2019}},
  \href{https://arxiv.org/abs/1902.08191}{{\ttfamily 1902.08191}}.

\bibitem{Beneke:2014pta}
M.~Beneke, A.~Maier, J.~Piclum and T.~Rauh, \emph{{The bottom-quark mass from
  non-relativistic sum rules at NNNLO}},
  \href{http://dx.doi.org/10.1016/j.nuclphysb.2014.12.001}{\emph{Nucl. Phys.}
  {\bfseries B891} (2015) 42--72},
  [\href{https://arxiv.org/abs/1411.3132}{{\ttfamily 1411.3132}}].

\bibitem{Beneke:2016oox}
M.~Beneke, A.~Maier, J.~Piclum and T.~Rauh, \emph{{NNNLO determination of the
  bottom-quark mass from non-relativistic sum rules}},
  \href{http://dx.doi.org/10.22323/1.235.0035}{\emph{PoS} {\bfseries
  RADCOR2015} (2016) 035}, [\href{https://arxiv.org/abs/1601.02949}{{\ttfamily
  1601.02949}}].

\bibitem{Chakraborty:2014aca}
B.~Chakraborty, C.~T.~H. Davies, B.~Galloway, P.~Knecht, J.~Koponen, G.~C.
  Donald et~al., \emph{{High-precision quark masses and QCD coupling from
  $N_f=4$ lattice QCD}},
  \href{http://dx.doi.org/10.1103/PhysRevD.91.054508}{\emph{Phys. Rev.}
  {\bfseries D91} (2015) 054508},
  [\href{https://arxiv.org/abs/1408.4169}{{\ttfamily 1408.4169}}].

\bibitem{Colquhoun:2014ica}
B.~Colquhoun, R.~J. Dowdall, C.~T.~H. Davies, K.~Hornbostel and G.~P. Lepage,
  \emph{{$\Upsilon$ and $\Upsilon^{\prime}$ Leptonic Widths, $a_{\mu}^b$ and
  $m_b$ from full lattice QCD}},
  \href{http://dx.doi.org/10.1103/PhysRevD.91.074514}{\emph{Phys. Rev.}
  {\bfseries D91} (2015) 074514},
  [\href{https://arxiv.org/abs/1408.5768}{{\ttfamily 1408.5768}}].

\bibitem{Bussone:2016iua}
{\scshape ETM} collaboration, A.~Bussone et~al., \emph{{Mass of the $b$-quark
  and $B$-meson decay constants from $N_f=2+1+1$ twisted-mass lattice QCD}},
  \href{http://dx.doi.org/10.1103/PhysRevD.93.114505}{\emph{Phys. Rev.}
  {\bfseries D93} (2016) 114505},
  [\href{https://arxiv.org/abs/1603.04306}{{\ttfamily 1603.04306}}].

\bibitem{Gambino:2017vkx}
P.~Gambino, A.~Melis and S.~Simula, \emph{{Extraction of heavy-quark-expansion
  parameters from unquenched lattice data on pseudoscalar and vector
  heavy-light meson masses}},
  \href{http://dx.doi.org/10.1103/PhysRevD.96.014511}{\emph{Phys. Rev.}
  {\bfseries D96} (2017) 014511},
  [\href{https://arxiv.org/abs/1704.06105}{{\ttfamily 1704.06105}}].

\bibitem{Bazavov:2018omf}
{\scshape Fermilab Lattice, MILC, TUMQCD} collaboration, A.~Bazavov et~al.,
  \emph{{Up-, down-, strange-, charm-, and bottom-quark masses from four-flavor
  lattice QCD}},
  \href{http://dx.doi.org/10.1103/PhysRevD.98.054517}{\emph{Phys. Rev.}
  {\bfseries D98} (2018) 054517},
  [\href{https://arxiv.org/abs/1802.04248}{{\ttfamily 1802.04248}}].

\bibitem{Carrasco:2014cwa}
{\scshape European Twisted Mass} collaboration, N.~Carrasco et~al., \emph{{Up,
  down, strange and charm quark masses with N$_f$ = 2+1+1 twisted mass lattice
  QCD}}, \href{http://dx.doi.org/10.1016/j.nuclphysb.2014.07.025}{\emph{Nucl.
  Phys.} {\bfseries B887} (2014) 19--68},
  [\href{https://arxiv.org/abs/1403.4504}{{\ttfamily 1403.4504}}].

\bibitem{Alexandrou:2014sha}
C.~Alexandrou, V.~Drach, K.~Jansen, C.~Kallidonis and G.~Koutsou, \emph{{Baryon
  spectrum with $N_f=2+1+1$ twisted mass fermions}},
  \href{http://dx.doi.org/10.1103/PhysRevD.90.074501}{\emph{Phys. Rev.}
  {\bfseries D90} (2014) 074501},
  [\href{https://arxiv.org/abs/1406.4310}{{\ttfamily 1406.4310}}].

\bibitem{Lytle:2018evc}
{\scshape HPQCD} collaboration, A.~T. Lytle, C.~T.~H. Davies, D.~Hatton, G.~P.
  Lepage and C.~Sturm, \emph{{Determination of quark masses from
  $\mathbf{n_f=4}$ lattice QCD and the RI-SMOM intermediate scheme}},
  \href{http://dx.doi.org/10.1103/PhysRevD.98.014513}{\emph{Phys. Rev.}
  {\bfseries D98} (2018) 014513},
  [\href{https://arxiv.org/abs/1805.06225}{{\ttfamily 1805.06225}}].

\bibitem{Dowdall:2013tga}
{\scshape HPQCD} collaboration, R.~J. Dowdall, C.~T.~H. Davies, R.~R. Horgan,
  C.~J. Monahan and J.~Shigemitsu, \emph{{$B$-Meson Decay Constants from
  Improved Lattice Nonrelativistic QCD with Physical $u$, $d$, $s$, and $c$
  Quarks}}, \href{http://dx.doi.org/10.1103/PhysRevLett.110.222003}{\emph{Phys.
  Rev. Lett.} {\bfseries 110} (2013) 222003},
  [\href{https://arxiv.org/abs/1302.2644}{{\ttfamily 1302.2644}}].

\bibitem{Hughes:2017spc}
C.~Hughes, C.~T.~H. Davies and C.~J. Monahan, \emph{{New methods for $B$ meson
  decay constants and form factors from lattice NRQCD}},
  \href{http://dx.doi.org/10.1103/PhysRevD.97.054509}{\emph{Phys. Rev.}
  {\bfseries D97} (2018) 054509},
  [\href{https://arxiv.org/abs/1711.09981}{{\ttfamily 1711.09981}}].

\bibitem{Charles:2015gya}
J.~Charles et~al., \emph{{Current status of the Standard Model CKM fit and
  constraints on $\Delta F=2$ New Physics}},
  \href{http://dx.doi.org/10.1103/PhysRevD.91.073007}{\emph{Phys. Rev.}
  {\bfseries D91} (2015) 073007},
  [\href{https://arxiv.org/abs/1501.05013}{{\ttfamily 1501.05013}}].

\bibitem{Bona:2016dys}
{\scshape UTfit} collaboration, M.~Bona, \emph{{Unitarity Triangle analysis in
  the Standard Model from the UTfit collaboration}},
  \href{http://dx.doi.org/10.22323/1.282.0554}{\emph{PoS} {\bfseries ICHEP2016}
  (2016) 554}.

\bibitem{Beneke:1998rk}
M.~Beneke, \emph{{A Quark mass definition adequate for threshold problems}},
  \href{http://dx.doi.org/10.1016/S0370-2693(98)00741-2}{\emph{Phys. Lett. B}
  {\bfseries 434} (1998) 115--125},
  [\href{https://arxiv.org/abs/hep-ph/9804241}{{\ttfamily hep-ph/9804241}}].

\bibitem{Hambrock:2015wka}
C.~Hambrock, A.~Khodjamirian and A.~Rusov, \emph{{Hadronic effects and
  observables in $B\to \pi\ell^{+}\ell^{-}$ decay at large recoil}},
  \href{http://dx.doi.org/10.1103/PhysRevD.92.074020}{\emph{Phys. Rev.}
  {\bfseries D92} (2015) 074020},
  [\href{https://arxiv.org/abs/1506.07760}{{\ttfamily 1506.07760}}].

\bibitem{Kruger:1999xa}
F.~Kr{\"u}ger, L.~M. Sehgal, N.~Sinha and R.~Sinha, \emph{{Angular distribution
  and CP asymmetries in the decays $\bar B \to K^- \pi^+ e^- e^+$ and $\bar B
  \to \pi^- \pi^+ e^- e^+$}},
  \href{http://dx.doi.org/10.1103/PhysRevD.61.114028}{\emph{Phys. Rev. D}
  {\bfseries 61} (2000) 114028},
  [\href{https://arxiv.org/abs/hep-ph/9907386}{{\ttfamily hep-ph/9907386}}].

\bibitem{Amhis:2019ckw}
{\scshape HFLAV} collaboration, Y.~S. Amhis et~al., \emph{{Averages of
  $b$-hadron, $c$-hadron, and $\tau$-lepton properties as of 2018. Update 2020
  at \url{https://hflav-eos.web.cern.ch/hflav-eos/osc/PDG_2020/}}},
  \href{https://arxiv.org/abs/1909.12524}{{\ttfamily 1909.12524}}.

\bibitem{Khodjamirian:2020hob}
A.~Khodjamirian, R.~Mandal and T.~Mannel, \emph{{Inverse moment of the
  $B_{s}$-meson distribution amplitude from QCD sum rule}},
  \href{http://dx.doi.org/10.1007/JHEP10(2020)043}{\emph{JHEP} {\bfseries 10}
  (2020) 043}, [\href{https://arxiv.org/abs/2008.03935}{{\ttfamily
  2008.03935}}].

\bibitem{Gelb:2018end}
{\scshape Belle} collaboration, M.~Gelb et~al., \emph{{Search for the rare
  decay of $B^+ \to \ell^{\,+} \nu_{\ell} \gamma$ with improved hadronic
  tagging}}, \href{http://dx.doi.org/10.1103/PhysRevD.98.112016}{\emph{Phys.
  Rev.} {\bfseries D98} (2018) 112016},
  [\href{https://arxiv.org/abs/1810.12976}{{\ttfamily 1810.12976}}].

\bibitem{Bell:2009fm}
G.~Bell and V.~Pilipp, \emph{{$B^- \to \pi^- \pi^0/\rho^- \rho^0$ to NNLO in
  QCD factorization}},
  \href{http://dx.doi.org/10.1103/PhysRevD.80.054024}{\emph{Phys. Rev.}
  {\bfseries D80} (2009) 054024},
  [\href{https://arxiv.org/abs/0907.1016}{{\ttfamily 0907.1016}}].

\bibitem{Beneke:2009ek}
M.~Beneke, T.~Huber and X.-Q. Li, \emph{{NNLO vertex corrections to
  non-leptonic $B$ decays: Tree amplitudes}},
  \href{http://dx.doi.org/10.1016/j.nuclphysb.2010.02.002}{\emph{Nucl. Phys.}
  {\bfseries B832} (2010) 109--151},
  [\href{https://arxiv.org/abs/0911.3655}{{\ttfamily 0911.3655}}].

\bibitem{Braun:2003wx}
V.~M. Braun, D.~{\relax Yu}. Ivanov and G.~P. Korchemsky, \emph{{The B meson
  distribution amplitude in QCD}},
  \href{http://dx.doi.org/10.1103/PhysRevD.69.034014}{\emph{Phys. Rev.}
  {\bfseries D69} (2004) 034014},
  [\href{https://arxiv.org/abs/hep-ph/0309330}{{\ttfamily hep-ph/0309330}}].

\bibitem{Bell:2013tfa}
G.~Bell, T.~Feldmann, Y.-M. Wang and M.~W.~Y. Yip, \emph{{Light-Cone
  Distribution Amplitudes for Heavy-Quark Hadrons}},
  \href{http://dx.doi.org/10.1007/JHEP11(2013)191}{\emph{JHEP} {\bfseries 11}
  (2013) 191}, [\href{https://arxiv.org/abs/1308.6114}{{\ttfamily 1308.6114}}].

\end{thebibliography}\endgroup

\end{document}